# Giant optomechanical spring effect in plasmonic nano- and picocavities probed by surface-enhanced Raman scattering

Lukas A. Jakob[1], William M. Deacon[1], Yuan Zhang[2*], Bart de Nijs[1], Elena Pavlenko[1], Shu Hu[1], Cloudy Carnegie[1], Tomas Neuman[3], Ruben Esteban[3], Javier Aizpurua[3*], Jeremy J. Baumberg[1*]

[1] Nanophotonics Centre, Cavendish Laboratory, University of Cambridge, Cambridge CB3 0HE, UK.
[2] Henan Key Laboratory of Diamond Optoelectronic Materials and Devices, Key Laboratory of Material Physics, Ministry of Education, School of Physics and Microelectronics, Zhengzhou University, Zhengzhou 450052, China.
[3] Center for Material Physics (CSIC - UPV/EHU and DIPC), Paseo Manuel de Lardizabal 5, Donostia-San Sebastian Gipuzkoa 20018, Spain.

Correspondence to J.J.B. (jjb12@cam.ac.uk), Y.Z. (yzhuaudipc@zzu.edu.cn), J.A. (aizpurua@ehu.es)

## Abstract

**Molecular vibrations couple to visible light only weakly, have small mutual interactions, and hence are often ignored for non-linear optics. Here we show the extreme confinement provided by plasmonic nano- and pico-cavities can sufficiently enhance optomechanical coupling so that intense laser illumination drastically softens the molecular bonds. This optomechanical pumping regime produces strong distortions of the Raman vibrational spectrum related to giant vibrational frequency shifts from an optical spring effect which is hundred-fold larger than in traditional cavities. The theoretical simulations accounting for the multimodal nanocavity response and near-field-induced collective phonon interactions are consistent with the experimentally-observed non-linear behavior exhibited in the Raman spectra of nanoparticle-on-mirror constructs illuminated by ultrafast laser pulses. Further, we show indications that plasmonic picocavities allow us to access the optical spring effect in single molecules with continuous illumination. Driving the collective phonon in the nanocavity paves the way to control reversible bond softening, as well as irreversible chemistry.**

## Introduction

Molecular vibrations increasingly dominate electronic, thermal, and spin transport in a wide range of devices from photovoltaics[1–4] to molecular electronics[5] as well as being of fundamental interest. Vibrations also underpin label-free molecular sensing[6], harnessed with metal-induced plasmonic enhancements to overcome small Raman cross sections. Surface-Enhanced Raman Spectroscopy (SERS) is well-established for studying molecular vibrations[7,8], exciting the molecular ground state to the first vibrational level simultaneously with emission of a Stokes-shifted longer-wavelength photon. To enhance SERS signals, plasmonic nanostructures are designed to maximize the nanocavity optical field confinement in intense localized hot-spots in which molecules are immersed[9].

Recently it was shown that SERS can be described as molecular optomechanics, in which molecular vibrations and the optical nanocavity are highly coupled[10,11]. Despite their large nanocavity linewidths ($\kappa$), the ~200 nm³ effective mode volumes $V$ yield single-plasmon optomechanical couplings $g$ exceeding 3 meV, approaching room temperature thermal energies[12–14]. So far, optomechanical models for plasmonic cavities used descriptions based on cavity-QED, extended to account for plasmonic losses, and were often restricted to a single resonant photonic mode[13]. The Stokes scattering spectrum, $S(\omega_v) \propto g^2(1 + n_v)I_l$ at Raman shift $\omega_v$ did not vary in shape with pump laser intensity $I_l$ for excited vibrational population $n_v$.

Here, we show that this approximation is incomplete, and that a full multimodal treatment of the nanocavity[15] is needed to explain optical field-dependent softening of molecular vibrations seen when collectively driving molecules at higher powers. For the first time to the best of our knowledge, we report indications of a vibrational frequency shift associated with the optical spring effect, a novel effect in the context of molecular nanotechnology. The considerations raised here will also be important for optomechanics of phonons in thin crystals when integrated into plasmonic nanocavities, such as perovskites or 2D layered materials.

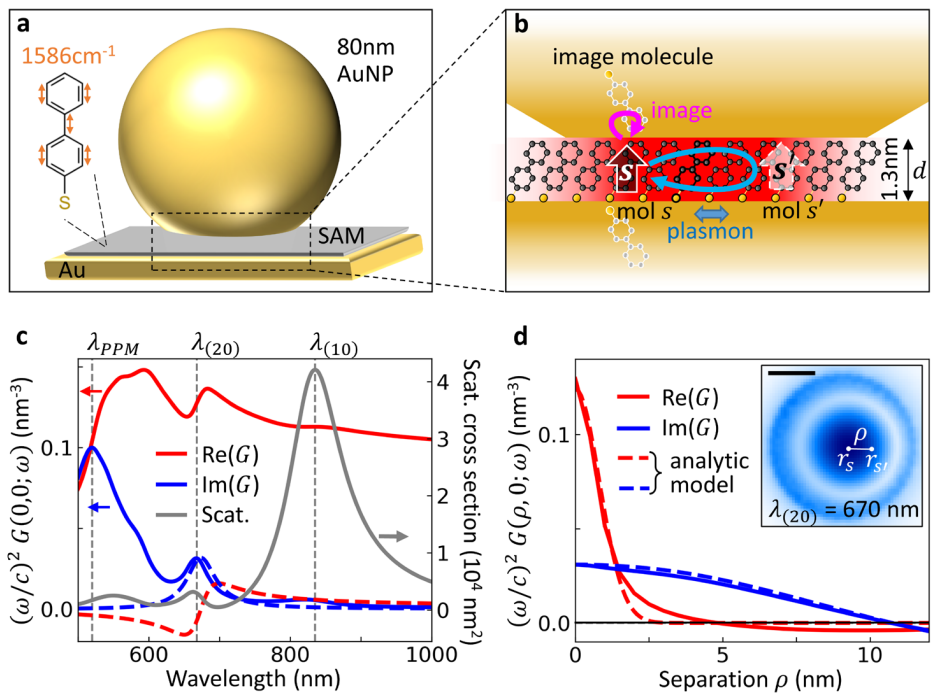

**Fig. 1 | Theory of nonlinear vibrational coupling in plasmonic nanocavities. a,** Schematic of 80 nm nanoparticle-on-mirror (NPoM) containing 1.3 nm-thick self-assembled monolayer (SAM) of biphenyl-4-thiol (BPT) molecules, showing benzene ring stretch at 1586 cm$^{-1}$. **b,** Nanogap supports localized plasmon modes (red). BPT molecular Raman dipoles ($s, s'$) interact via their image molecules (pink arrow) and through plasmon modes (blue arrow). **c,** Complex self-interaction Green's function (Re{$G$}=red, Im{$G$}=blue) for response produced by a vertical dipole in the gap centre. Scattering cross section (gray) shows dominant localized plasmonic (10) (lowest order) and (20) (second order) modes (vertical dashed) at $\lambda_{(10)} \approx 830$nm, $\lambda_{(20)} \approx 670$nm, and a peak in Im$G$ at $\lambda_{PPM} \approx 520$ nm identified as the plasmon pseudo-mode (PPM), originating from overlapping higher order modes. Dashed curves show $G$ when only a single-optical-mode is considered in the model. **d,** Two-point Green's function between spatially separated locations in the gap (separation $\rho$), at $\lambda_{(20)} = 670$ nm as obtained numerically (solid line) and with an analytic model based on image dipoles (dashed lines, see Supplementary Note S3.1). Inset shows midgap electric near-field at $\lambda_{(20)}$, with a scale bar 10 nm.

## Results

### Continuum-field description of plasmonic nanocavity

We start by theoretically considering a realistic plasmonic nanocavity which optimizes molecular optomechanical coupling. The molecules are embedded in a metal-insulator-metal nm-thick waveguide. This is easily realized using the nanoparticle-on-mirror geometry (NPoM), where a self-assembled monolayer (SAM) of active molecules is formed on a flat Au substrate before deposition of Au nanoparticles (NPs) on top (Fig. 1a)[16]. This places a few hundred molecules in the ~10 nm-wide optical field, highly-confined within the nanogap (thickness $d$~1.3 nm) between the flat bottom facet of the 80 nm-diameter Au nanoparticle and the Au surface underneath (Fig. 1b). More details on the simulated nanostructure are given in Supplementary Note S3. The resulting optical field enhancement factor $\mathrm{EF}>300$ gives SERS $\propto \mathrm{EF}^4$ which is hence increased by $\geq 10^{10}$ or more.[9,16]

Previous optomechanical models typically considered a dominant single cavity mode. However, if the full nanocavity plasmonic spectrum (grey line in Fig. 1c) with a more complex structure of resonances is included in a continuum-field model[15,17], new behaviors are predicted as compared to the single-mode description. Furthermore, within a general scheme of optomechanical dynamics, increasing the optical pumping predicts spectral changes of the Stokes scattering due to vibrational shifts induced by optomechanical interactions with the plasmonic modes. These interactions are enhanced in molecular self-assemblies by the coupling of different molecules leading to collective phonon modes[18]. The effects above can be obtained from the dynamics of the vibrational amplitude $\beta_s$ of the $v^{\mathrm{th}}$ vibrational mode of the $s^{\mathrm{th}}$ molecule:

$$\frac{\partial}{\partial t}\beta_s = -i\left[(\omega_v - \mathrm{Re}\, v_{ss}) - i\left(\frac{\gamma_v}{2} + \mathrm{Im}\, v_{ss}\right)\right]\beta_s + i\sum_{s'\neq s} v_{s's}\beta_{s'} \tag{1}$$

with $\omega_v, \gamma_v$ its vibrational frequency and decay rate, respectively. We observe that both the frequency and decay rate are modified by the term $v_{s's} = \left(S^+_{ss'}\right)^* + S^-_{ss'}$, given by the spectral density associated with the Stokes $\left(S^+_{ss'}\right)^*$ and the anti-Stokes $S^-_{ss'}$ scattering, defined as (see Supplementary Note S1.2)

$$S^{\pm}_{ss'} = \frac{1}{4\hbar\varepsilon_0}\left(\frac{\omega_l \mp \omega_v}{c}\right)^2 [\mathbf{p}_s(\omega_l)]^* \cdot \overleftrightarrow{G}(\mathbf{r}_s, \mathbf{r}_{s'}; \omega_l \mp \omega_v) \cdot \mathbf{p}_{s'}(\omega_l), \tag{2}$$

where $\hbar, \varepsilon_0, c$ are the reduced Planck constant, vacuum permittivity and speed of light. Here $\mathbf{p}_s(\omega_l) = \overleftrightarrow{\alpha}_v \mathbf{E}(\mathbf{r}_s, \omega_l)$ is the Raman dipole of the $s^{\mathrm{th}}$ molecule induced by the local electric field $\mathbf{E}(\mathbf{r}_s, \omega_l)$, excited by a laser of frequency $\omega_l$, acting on the molecule at position $\mathbf{r}_s$ with Raman polarizability of the $v^{\mathrm{th}}$ vibrational mode $\overleftrightarrow{\alpha}_v$. As observed in Eqn. (2), in addition to the local field enhancement, the Green's function of the plasmonic system, $\overleftrightarrow{G}(\mathbf{r}_s, \mathbf{r}_{s'}; \omega_l \mp \omega_v)$, at the Stokes, $\omega_l - \omega_v$, and anti-Stokes, $\omega_l + \omega_v$, frequencies is the key magnitude that governs the optomechanical interaction. The term $v_{s's}$ with $s = s'$ is explicitly separated in Eqn. (1) and describes the self-interaction of the Raman-induced dipoles for a single molecule, where the imaginary part $\mathrm{Im}\{v_{ss}\}$ leads to an increase of the decay rate (broadening the Raman lines by $2\mathrm{Im}\{v_{ss}\} \propto \mathrm{Im}\{G\}$), as reported in previous work[13,15], while the real part, $\mathrm{Re}\{v_{ss}\}$ leads to a reduction of the vibrational frequency of the mode (spectral shift of Raman lines, $\Delta\omega_v \propto \mathrm{Re}\{G\}$), corresponding to the optical spring effect in cavity optomechanics. In the case of many molecules, the terms $v_{s's}$ with $s \neq s'$ couple the different molecules, and the resulting collective response modifies the optomechanical effects, as explained below. In this work we explore the properties of and the evidence for this optical spring effect in our NPoM configuration. We note that the occurrence and magnitude of the observed effect is linked to the exact design of this nanostructure and its plasmonic cavity modes.

The self-interaction Green's function at the centre of the NPoM gap shows the relevant landscape of plasmonic modes (Fig. 1c). We focus on the contribution due to scattering by the NPoM (ignoring the contribution of the direct dipole-dipole interaction in the homogeneous medium, which avoids a divergence in the case of the self-interaction, and requires careful renormalization[19-21]). Importantly, Re$\{G\}$ for the full NPoM cavity is 10-fold larger than when a single-mode plasmonic cavity is considered (solid versus dashed red line, see Supplementary Note S6 for a discussion of the single-mode model), highlighting the importance of fully incorporating all plasmonic modes to correctly address the optomechanics. The spatial distribution of plasmonic modes impacts differently the real and imaginary parts of $G$ (Fig. 1d). While $\mathrm{Im}\{v_{ss'}\} \propto \mathrm{Im}\{G(\rho = |\mathbf{r}_{s'} - \mathbf{r}_s|)\}$ (blue solid line in Fig. 1d) extends across the whole facet and follows the near-field of the (20) mode at $\lambda_{(20)}$ = 670 nm (blue dashed in Fig. 1d), $\mathrm{Re}\{v_{ss'}\} \propto \mathrm{Re}\{G(\rho)\}$ is seen to be extremely short-ranged due to the self-interaction of highly localized dipole image charges in the gap (Fig. 1d, red line). This short-range interaction is important for the vibrational shifts of many molecules, is nearly spectrally independent (Fig. 1c, red solid line), and can be analytically derived from the coupling of image dipoles in the gap (Fig. 1d red dashed line, see Supplementary Note S3.1), giving a profile of width $\delta \approx 0.9d$ for gap size $d$.

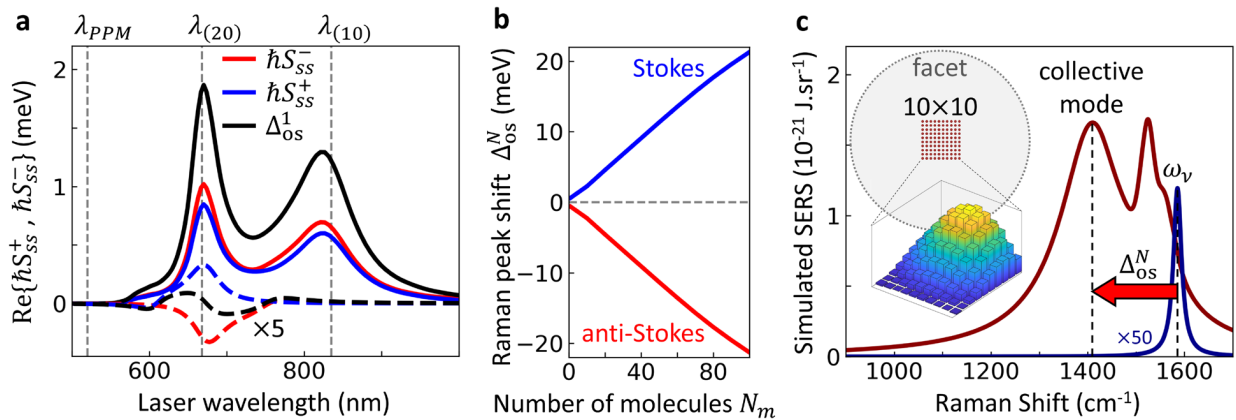

**Fig. 2 | Origin of optical spring effect in molecular optomechanics. a,** Optical spring effect *vs* laser wavelength for $\omega_{v1}$ = 1586 cm$^{-1}$ mode in the NPoM gap, showing contributions of the Stokes $S^+$ (blue) and anti-Stokes $S^-$ (red) optomechanical parameters to the total vibrational shift, $\Delta_{os}^1$, for a single molecule in the nanocavity centre. $S^+$ and $S^-$ scale linearly with laser intensity (shown here for $10^7$ μW μm$^{-2}$). Dashed curves show single-mode plasmonic cavity results. **b,** Dependence of shift in the fundamental collective Raman bright mode $\Delta_{os}^N$ with the number of molecules $N_m$ arranged in a lattice at the middle of gap (Stokes in blue, anti-Stokes in red) at 5x10$^7$ μW μm$^{-2}$. **c,** SERS emission from full multi-molecule model for the $\omega_{v1}$ = 1586 cm$^{-1}$ mode with 633 nm CW pump intensity of $10^5$ μW μm$^{-2}$ (blue, multipled by 50) and 5x10$^7$ μW μm$^{-2}$ (red). At the larger intensity the broad peak is down-shifted $\Delta_{os}^N \sim$170 cm$^{-1}$ from $\omega_v$ due to the dominant bright Raman collective phonon mode. Top inset shows the square array of 100 molecules spaced $\rho$ = 0.58 nm apart and centred in the facet (dashed, radius 16 nm). Bottom inset shows each molecular contribution to the fundamental bright collective phonon mode. Other parameters are specified in Supplementary Note S5.

## Optical spring effect and collective phonon modes

We examine first the vibrational shift experienced by a single molecule at the centre of the gap. The real part of the spectral density, $\text{Re}\{S_{ss}^{\pm}\}$, associated with the Stokes (+), and anti-Stokes (-) frequencies, determines the total frequency shift of each vibrational mode. The resulting values of $\text{Re}\{S_{ss}^{\pm}\}$ are shown in Fig. 2a. Compared to the single-mode model (dashed lines), the values of $\text{Re}\{S_{ss}^{\pm}\}$ are about ten-fold larger when considering the full plasmonic response (solid lines) because of the larger value of the real part of the Green's function (Fig. 1c). Further, in the single-mode model $\text{Re}\{S_{ss}^{\pm}\}$ both approach a maximum around the single plasmon cavity resonance (set to $\lambda_{(20)}$, see Supplementary Note S6) but with opposite sign. This results in a small total frequency shift from the optomechanical optical spring effect

$$\Delta_{\text{os}}^{1} = \text{Re}\{S_{ss}^{+} + S_{ss}^{-}\} \tag{3}$$

due to cancelation of frequency shifts from combining both Stokes and anti-Stokes contributions. In contrast, in the full plasmonic model $S_{ss}^{\pm}$ both approach a maximum around plasmon modes but have the same sign, which results in a much larger total frequency shift than in the single cavity mode description. We therefore predict a substantial optical spring effect in our NPoM configuration.

This significant optical spring effect in molecular self-assemblies is further collectively enhanced by coupling of molecules laterally separated by $\rho = |\mathbf{r}_{s'} - \mathbf{r}_{s}|$ arising from their image-dipole local Coulomb interactions. The term $v_{ss'}$ with $s \neq s'$ in Eqn. (1) describes the interaction between the Raman-induced dipoles, which leads to formation of collective phonon modes by superposing vibrations in individual molecules. In the mid-infrared, such collective phonons have been termed vibrational excitons[2,22–24], but here they are dynamically induced only by the laser driven Raman dipoles. We suggest the term 'vibrational exciton' in the literature is misleading since excitons refer to bound electron-hole pairs, while the collective vibrations here are analogous to localized phonons in a continuous material. We thus prefer the term 'molecular phonons'. This collective response results in the emergence of Raman-bright collective phonons across all $N_m$ molecules (which fit inside the NP facet and interact in the NPoM gap), together with other dark collective modes. The optical spring shift induced by the fundamental bright collective mode corresponds to a bond softening, and scales linearly with $N_m$ (Fig. 2b). This dependency is obtained by calculating the eigenmodes of equation (1) for an increasing number of molecules (see Supplementary Note S5.4) all with vibrational frequency $\omega_v = \omega_{v1} = 1586$ cm$^{-1}$. The shift for 100 molecules forming a dense square patch at the centre of the NPoM gap (inset Fig. 2c) is 32 times larger than when only the molecule in the corner is present (corresponding to $N$=1 in the figure) and 12 times larger than for a molecule in the gap centre.

We obtain the Raman spectra by applying the quantum regression theorem (Supplementary Note S1.2) to the patch of 100 molecules in the centre of the NPoM gap under continuous wave illumination. The obtained spectra (Fig. 2c) show a single narrow peak for weak illumination (blue line) at the vibrational energy $\omega_{v1}$ of the individual molecules, corresponding to the standard Raman line. In contrast, the collective SERS spectrum under strong pumping (red line) comprises a broad and strongly down-shifted line (by $\Delta_{\text{os}}^{N}$, arrow) associated with the fundamental collective bright mode (vertical dotted line), superposed on contributions from the remaining $N_m - 1$ weak near-unshifted modes. The latter lead to a relatively narrow line close to $\omega_{v1}$ (blue line). Thus, the model predicts a redistribution of the scattered Raman signal from a frequency near $\omega_{v1}$, towards the softened frequency of the broad bright molecular phonon mode. The individual molecular contributions to the broad bright phonon mode are displayed in the inset of Fig. 2c with a clear symmetric maximum at the central position of the gap.

## SERS saturation effect in plasmonic nanocavities

Most SERS experiments use continuous wave (CW) excitation, where the spring shifts remain small (as discussed further below). In order to probe these optical spring shifts in a nanocavity, high instantaneous powers are demanded, which are first here supplied through pulsed excitation, together with NPoM constructs that localize light fields to a very small volume (thus increasing the local fields and the Green's function). The use of laser pulses together with these inhomogeneous fields however smears out the power-dependent SERS spectrum predicted by our model (Fig. 2c), although the effects appear consistent with what is observed as a repeatable intensity-induced saturation of the sharp vibrational peak, as described below.

To explore the theoretical predictions, we use 0.5 ps laser pulses and concentrate on a strong SERS peak together with the region where the softened mode is expected to appear. For each of hundreds of NPoMs (depicted in Fig. 1a, see Methods for sample preparation), the plasmonic dark-field scattering spectra $DF(\lambda)$ and power-dependent SERS response $S(\omega_v, I_l)$ are characterized (Fig. 3). The plasmonic gap size, nanoparticle diameter, facet size, and gap contents control the spectral position of the main coupled plasmon $\lambda_{(10)} \sim 800$ nm. For the SAM of biphenyl-4-thiol (BPT) initially used, we concentrate on the strong $\omega_{v1}$= 1586 cm$^{-1}$ ring breathing mode producing Stokes emission from the pump pulse $\lambda_l$ at $\lambda_S^{-1} = \lambda_l^{-1} - \omega_{v1}/2\pi c$. Power series are recorded when blue-detuned from $\lambda_{(10)}$ ($\lambda_l$=633 nm, $\lambda_S$=704 nm), and compared with near-resonance emission ($\lambda_l$=700 nm, $\lambda_S$=787 nm $\sim \lambda_{(10)}$).

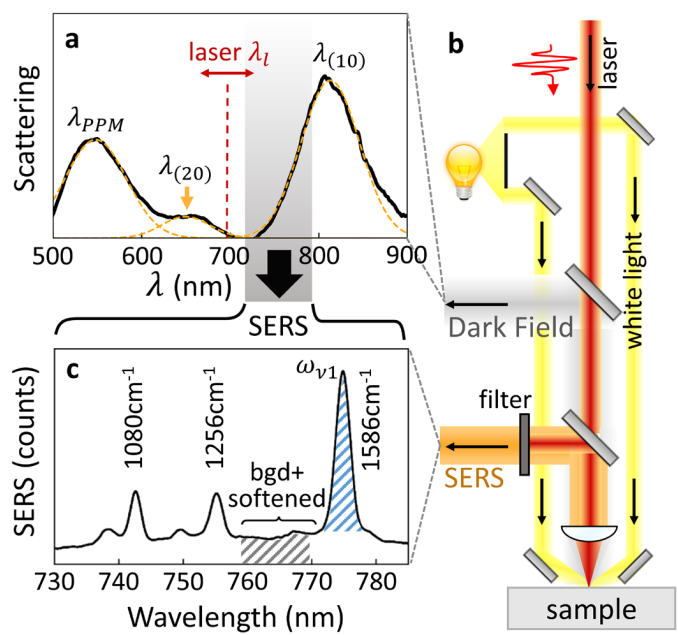

**Fig. 3 | Pulsed Raman scattering from plasmonic nanocavities. a,** Dark-field spectrum of typical 80 nm nanoparticle-on-mirror (NPoM) containing BPT molecular SAM. Pump laser (red dashed) is spectrally tunable, shaded region shows range of SERS emission, and individual plasmon modes are labelled. **b,** Pulsed SERS experiment combined with white-light dark-field scattering on individual NPoMs. Spectrally-tuned 0.5 ps pump pulses excite individual NPoM (with white light off), and laser is filtered from the collected emission. **c,** Pulsed Stokes SERS spectrum of BPT for three vibrational modes indicated. Blue shading shows 1586 cm$^{-1}$ peak area, grey shading shows region of softened mode + background (which is integrated for comparison).

Pulsed laser excitation easily causes permanent damage from the high peak fields through irreversible chemistry or gold surface melting[14,25–27], so we develop here a strategy that employs the shortest possible illumination times and measures many plasmonic nanocavities through fully automated experiments (see Methods). Pulses of duration 0.5 ps at 80 MHz repetition rate and average power ≤60 µW are focused with a x100 dark-field high-NA=0.9 objective to a sub-micron focus[14]. Dark-field images guide particle tracking software to concentrate only on well-formed NPoMs, monitor damage in real time, account for spectral aberrations, and minimize spatial drift. To reduce structural damage[14], the exposure time is scaled to keep the measurement fluence constant (10 µJ) while the average laser power ramps from 100 nW to 60 µW. To compare each NPoM $i$ with slightly different size and shape (limited by the precision of nanoscale fabrication) which varies their excitation and collection efficiencies, we normalize the results based on the integrated SERS counts at the lowest intensity $I_0$, using $\eta^i = S^i(\omega_{v1}, I_0)/I_0$. Given average $\bar{\eta} = \mathrm{mean}(\eta^i)$, we then normalize the in-coupled intensity for each NPoM as $I_{\mathrm{in}}^i = I_l \cdot \eta^i/\bar{\eta}$. This accounts for in-coupling efficiency so that each NPoM gives the same normalized SERS emission at the same low input intensity (see Supplementary Figure S19). Enabled by this experimental correction, we explicitly present here the entire dataset on hundreds of NPoMs to show the reproducibility of the effect across many nanostructures and avoid bias incurred when selecting individual particles. However, all observations can be confirmed by data on individual cavities as presented in Supplementary Note S11.

Despite the intensity averaging from using laser pulses (which smears out the spectral shifts and makes direct identification of the $\Delta_{\mathrm{os}}^N$ shift challenging), the evolution of the average SERS spectra for increasing $I_{\mathrm{in}}$ shows clear nonlinear changes (Figs. 4a-c, normalized by in-coupled intensity). A repeatable weakening of the original sharp vibrational peak is seen in the power-dependent SERS, while the region at lower wavenumber grows superlinearly, indicating the energy redistribution into collective modes as predicted by the optomechanical theory (Fig. 2c).

To quantitatively analyze this SERS saturation, we extract the integrated SERS areas $S^i$ from the peaked emission around the $\omega_{v1}$=1586 cm$^{-1}$ vibrational frequency (Fig. 3c, blue shaded). This is compared to the integrated emission from the softened peak region at lower wavenumbers (Fig. 3c, grey shaded), which however also contains a significant contribution from the SERS background. This ever-present background mainly arises from electronic Raman scattering (ERS) inside the Au[16]. At low powers ERS dominates this background, while for intense pulsed illumination it also contains the (smeared) softened modes from the broad collective phonon.

The integrated Raman peak emission $S^i$ shows a clear saturation with laser intensity, consistently for different vibrational peaks, NPoMs, pump wavelengths, and power series (Fig. 4d-f). To better show this, we also plot the SERS $\tilde{S}^i$ normalized to $I_{\mathrm{in}}^i$ (Fig. 4g-i). The nonlinearity in pulsed SERS is dramatic, with up to ten-fold suppression of linear scaling of the sharp vibrational peak area at the highest pulsed powers. To confirm the saturation behavior and eliminate damage or drift as the cause, power-dependent SERS measurements are repeated twice on each NPoM (Supplementary Fig. S23), showing that the saturation is reversible. Further power-cycling on each NPoM demonstrates that even for large SERS saturations, accompanying permanent damage of only 10% is seen. These observations contrast with literature reports on nonlinear SERS arising from irreversible damage under longer irradiation[28–30]. Vibrational pumping is also clearly observed in our data from the quadratic power dependence of the anti-Stokes emission (Supplementary Fig. S24), which then prevents meaningful temperature extraction (Supplementary Fig. S25).

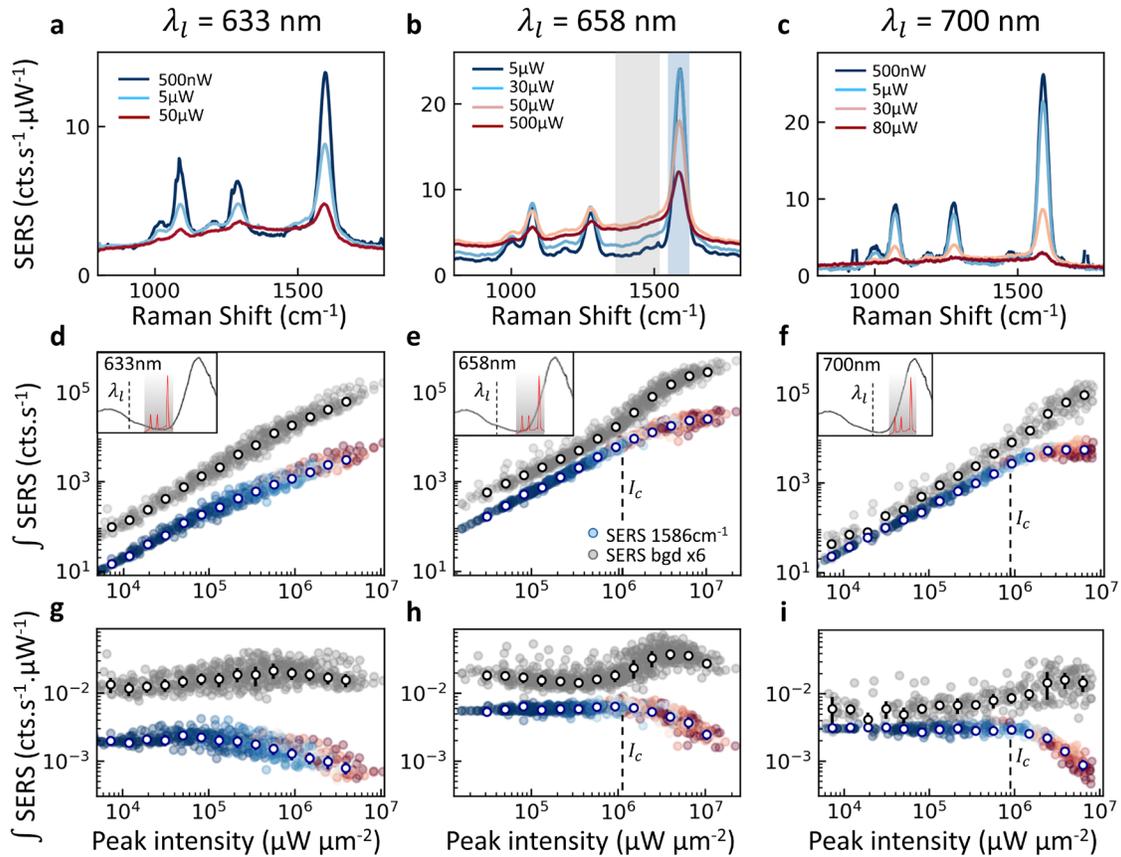

**Fig. 4 | Saturation of pulsed Raman scattering from many NPoMs. a-c,** Averaged power-normalized SERS spectra for increasing in-coupled average powers at different pump wavelengths $\lambda_l$. **d-f,** Integrated SERS emission from $\omega_{v1}$ = 1586 cm$^{-1}$ mode (colors show laser power) and integrated background + softened spectral region (grey, x6), excited by pulsed laser at $\lambda_l$ = 633, 658, 700 nm. Open points are averages of individual measurements in each power range. Insets show relative position of pump wavelength and plasmon resonances. **g-i,** Integrated SERS normalized by in-coupled peak intensity and integration time, error bars indicate their standard deviation. The critical laser intensity for saturation is marked as $I_c$ (indicated in the main text).

Examining the high power spectra shows that as the vibrational peaks saturate, SERS emission in the softened mode region at lower wavenumbers correspondingly increases (see Fig.5a, also evident in Fig.4e,h with spectral integration range indicated in grey in Fig.4b). This indeed points towards a redistribution of vibrational frequencies of many hundreds of cm$^{-1}$ (as predicted in Fig.2c), consistent with unprecedentedly large softening of the vibrational frequencies in the optical nanocavity. This effect is very different from dc bias-induced vibrational Stark shifts where sharp lines shift by ≲10 cm$^{-1}$. [31]

Additional experiments show that this SERS saturation is not specific to BPT, but is also seen for other molecules including naphthalene-thiol, triphenyl-thiol, and 4-mercaptobenzonitrile (Supplementary Figs. S26-S28), with similar saturation thresholds that depend on detuning (none have electronic transitions in the visible or near-infrared). We also find that molecule-metal hybridization is not a dominant influence, by substituting the first monolayer of atoms on the underlying Au mirror with Pd atoms. This changes the thiol binding but gives near-identical SERS saturation of BPT (Supplementary Fig. S29), showing that there is no power-dependent hybridization or charge transfer of the molecular orbitals. Similar behavior is seen for pump wavelengths of 785 nm (Supplementary Note S13). In the minimal-dose regime shown here, previously observed superlinear SERS increases at high power[14] are still seen for some detunings (small initial rise in Fig.4g) but at the higher powers that can now be achieved before bond-breaking, the SERS saturation is much more significant .

One other possible origin for these observations could be through the anharmonicity of the vibrational potentials, usually revealed through their thermal occupation. Transiently exciting molecules in solution[32–34] can give slightly red-shifted Raman peaks[35–38] arising from vibrational anharmonicity. However in our case, based on the reported biphenyl peak shifts with temperature[34,39,40], shifts of only $\Delta\omega_v < 30$ cm$^{-1}$ would result, much smaller than observed here (estimates in Supplementary Note S9). A model including anharmonicity cannot reproduce our results (Supplementary Figs. S15-18), instead suggesting that, even if contributing, a different interaction must be present with a much larger energy scale.

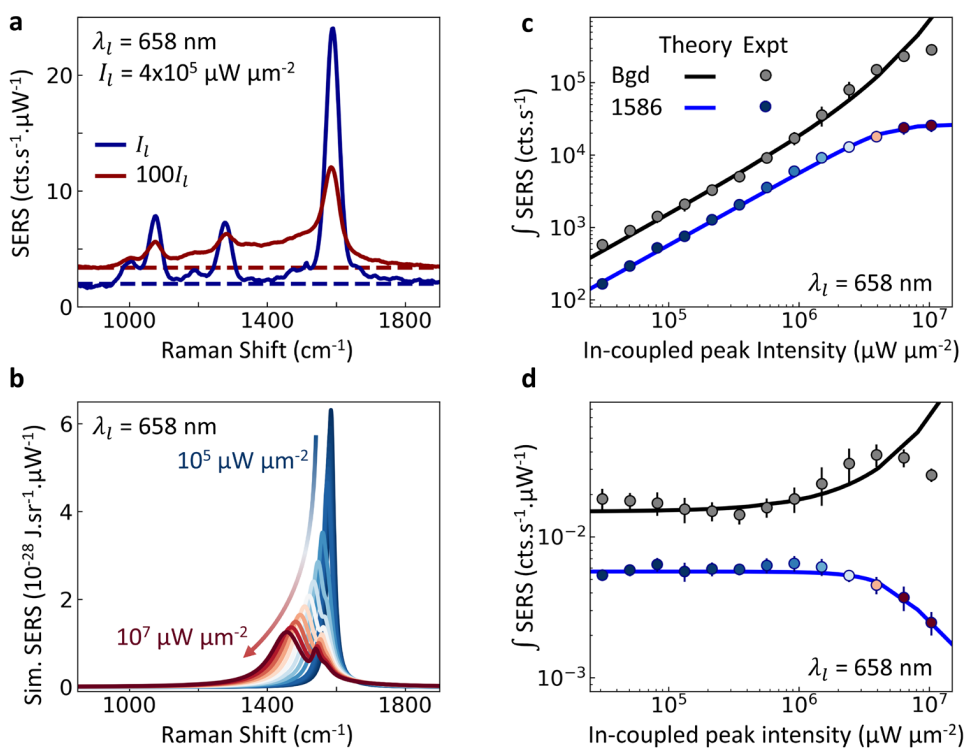

**Fig. 5 | Nonlinear vibrational coupling model vs expt. a,** Experimental power-normalized SERS spectra at low (blue) and high (red) powers. Constant ERS background is estimated by dashed lines. **b,** Corresponding theoretical results showing the SERS spectra vs CW illumination power, for 100 molecules arrayed around the gap centre. **c,d,** Extracted Raman integrated in the region around the 1586 cm$^{-1}$ peak (blue line/symbols) and in the softened+background region between 1350-1500 cm$^{-1}$ (grey line/symbols) for theory and experiment (SERS normalized by power in d, as in Fig.4g-i). Experimental data are averages of many particles with error bars indicating standard deviation of individual measurements. In **c,d**, scaling of in-coupled power from theory by 0.24 is used to match with experiment. The transfer in weight from the 1586 cm$^{-1}$ peak to lower wavenumbers arises from the redistribution of emission to the red-shifted lowest-energy bright Raman collective mode.

The correspondence between the spectral shifts predicted in theory and the experimentally identified saturation is analyzed in Fig. 5. The multimode optomechanical model indicates that above a critical pump power, a bright collective vibrational mode rapidly broadens and red-shifts linearly with power (Fig. 5b), producing a redistribution of vibrational frequencies that appears as a saturation of the originally-sharp SERS $v_1$ line and a superlinear increase in the softened + background region (Fig. 5c). We note that the saturation obtained with this model agrees well with the experiment (Fig. 5c,d), although the ERS and smearing mask the direct identification of shifts in Fig. 5a. In both theory and experiment, the total Raman yield integrated over all wavenumbers remains linear with power (Supplementary Figs. S8c and S24d, respectively), but is redistributed by the optomechanical interaction. To match the experimental and theoretical saturation it is only necessary to consider a slight scaling of the pump intensity. This scaling is not surprising since the calculations do not include all molecules in the gap due to the high computational cost (with different patches of molecules experiencing different intensity-dependent optomechanical coupling, Supplementary Fig. S9). Moreover, the exact spectral distribution of the collective Raman peaks and their relative weights will be influenced by the plasmon-mediated interaction of induced Raman dipoles (as noted above), as well as the specific configuration and orientation of molecules within the cavity. We note that although non-circular facets under the NPoM will shift plasmonic mode frequencies, the overall model still works in the same way.

We also can obtain a simplified analytical equation (see Supplementary Note S7) that predicts the critical illumination power needed to shift the Raman line outside its $\gamma_v$ = 20 cm$^{-1}$ low-power linewidth (where saturation starts),

$$I_c \, [10^6 \, \mu W \, \mu m^{-2}] \simeq a \, \frac{(2\pi c)^3 m_u \varepsilon_g \, d^3}{\eta_1' N_m \, \mathrm{EF}^2 R_v^2} e^{(1.1\rho/d)^2} \left[ \mathrm{Re}\left( \frac{\varepsilon_{Au} - \varepsilon_g}{\varepsilon_{Au} + \varepsilon_g} \right) \right]^{-1} \omega_v [\mathrm{cm}^{-1}] \gamma_v [\mathrm{cm}^{-1}] \quad (4)$$

with proportionality constant $a$ = 1.2x10$^6$ (Supplementary Note S7), gap permittivity $\varepsilon_g$ = 2.1, gap size $d$ and intermolecular spacing $\rho$ in nm, permittivity of Au at 658 nm $\varepsilon_{Au}$ = -13.5, [41] and $R_v \sim 960$ (in units of $\varepsilon_0 \text{Å}^2/\sqrt{\mathrm{amu}}$, where $R_v^2$ is the Raman activity of the 1586 cm$^{-1}$ line, and amu $m_u$ =1.7x10$^{-21}$ kg). The effective coupling to the collective Raman bright mode is accounted for by the factor $\eta_1(\rho) \approx \eta_1' \exp\{-(1.1\rho/d)^2\}$, here obtained as $\eta_1 \sim 0.12$ (at $\rho$ = 0.6 nm) from the discussion in Fig. 2b. We note this simplification provides a useful intuition and compares well with the rigorous result used for Fig. 5, while omitting the dependence on plasmonic resonances. With the field enhancement factor $\mathrm{EF}(\lambda_l) \sim 300$ of NPoM at the laser wavelength, for the 1586 cm$^{-1}$ mode, Eqn. (4) gives $I_c \sim$ 3x10$^6$ µW µm$^{-2}$ at 670 nm for $N_m \sim 100$ molecules, in good agreement with the experiments. This formula shows why nanocavities are essential to bring the optical powers into a viable domain that does not damage the sample, since $I_c$ scales cubically with gap size and inversely with the power enhancement $\mathrm{EF}^2$ (overall factors in excess of 10$^6$ compared to free space). Eqn. (4) also clearly shows close-packed molecules in SAMs ($\rho < d$) are needed to observe such collective effects (though molecular ordering is not required).

The fractional reduction in bond strength from the optomechanical interaction by light is then

$$\frac{\Delta \omega_v}{\omega_v} = b \, \eta_1 N_m \left( \frac{\mathrm{EF} \, R_v}{\omega_v} \right)^2 I_l \, , \quad (5)$$

where for $\omega_v$ in cm$^{-1}$ and $I_l$ in 10$^6$ µW µm$^{-2}$, $b$ = 2.23x10$^{-8}$. The relative lineshift can become even larger for low frequency lines, implying that irradiation even more strongly weakens rotations, librational and shearing deformations of molecules and macromolecules. As a result, for instance, enzymes might be optically-dressed and switched through this scheme of interactions (note proteins and lipids have already been placed in such plasmonic nanocavities[42]).

## Single-molecule optical spring shift in picocavities

Whilst pulsed excitation of nanocavities red-shifts and smears out SERS spectra (as Fig. 5a), direct visualization of the optical spring shifts is desirable. This requires CW illumination, but to reach observable spring shifts requires extra field enhancement. To show this we use NPs with integrated nanolenses (called SPARKs to enhance coupling efficiencies[43], Fig. 6a) which support picocavities inside them[44] (formed when single gold adatoms are pulled from the facet into the gap) that additionally confine light below the 1 nm scale[12,45] (Fig. 6b). Picocavities, as compared to nanocavities (which lack the atomic protrusion), give particularly strong optomechanical coupling and focus the optical fields down to the scale of an individual molecule[12,45].

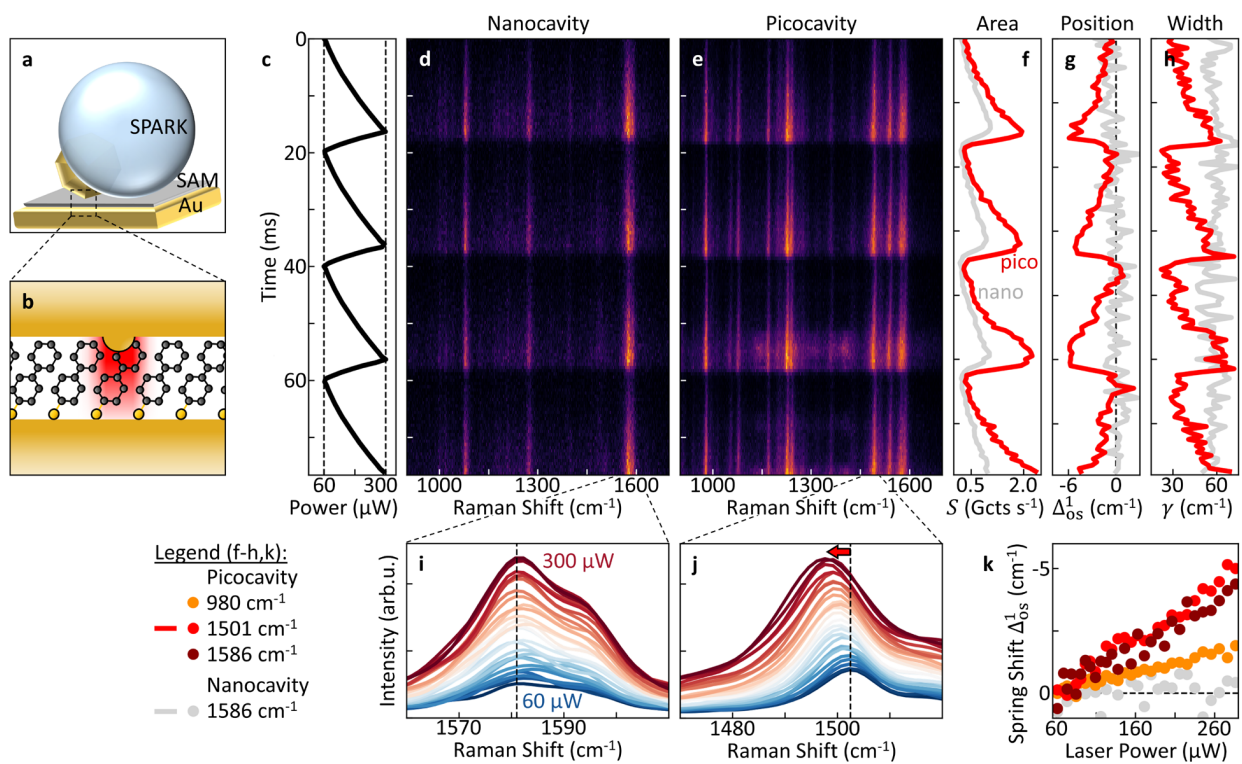

**Fig. 6 | Optical spring shift in picocavities. a,** Schematic of nanolens on NPoM (SPARK construct). **b,** Generation of a picocavity when Au atom moves onto facet, enhancing field at a single BPT molecule. **c,** Sawtooth modulation of 633 nm CW laser power from 60 to 300 µW at 50 Hz. **d,e,** Fast spectral scans (0.5 ms integration time) of Stokes emission from the SPARK nanocavity (d), and after formation of a picocavity (e). **f-h,** Extracted fits to 1586 cm$^{-1}$ line in nanocavity (grey) and 1501 cm$^{-1}$ picocavity line (red). Peak area (f) is linear in laser power, while optical spring effect in the picocavity leads to a repeatable shift in position (g) and broadening (h) of the vibrational line. **i,j,** Spectra of vibrational lines investigated in f-h averaged over 4 periods of laser modulation. Nanocavity line (i) shows constant width and position while picocavity line (j) shifts and broadens (colour gives laser power). **k,** Optical spring shift $\Delta_{os}^1$ dependence on laser intensity for several vibrational lines in nano- and picocavities. Each picocavity vibration experiences a different optical spring magnitude.

As previous work on picocavities has shown, SERS lines in picocavities fluctuate in position on multi-second timescales due to the single-molecule nature of the signal[12,45,46]. Hence, the meaningful extraction of the optical spring shift is experimentally challenging. Here, we greatly increase the speed at which power sweeps are performed, by $10^4$ compared to previous investigations[12], to avoid slower spectral wandering. A 633 nm, 300 µW CW laser is used to acquire Stokes spectra at kHz rates (Fig. 6c-e, see Methods). A sawtooth modulation of laser power from 60 to 300 µW at 50 Hz (Fig. 6c) repetitively probes the power dependence of the SERS in the nanocavity and picocavity within 20 ms, faster than any spectral drift (Fig. 6d,e, respectively). Fitting the vibrational lines with a Lorentzian peak gives the area, position, and linewidth *vs* laser power (Fig. 6f-h). For the picocavity lines, increasing illumination leads to a reversible red-shift in the vibrational energy as well as strong broadening by a factor of 2, while vibrations in the nanocavity remain unchanged. This behaviour can be clearly seen in power-dependent spectra averaged over four laser modulation cycles for the nano- and picocavity (Fig. 6i,j respectively). Extracting the position of different vibrational lines shows that all vibrational modes in the picocavity experience a different shift rate (Fig. 6k), but are always reversible. The strongest shifts of 5 cm$^{-1}$ are observed for the 1501 cm$^{-1}$ and 1586 cm$^{-1}$ lines in the picocavity (Fig. 6k).

This observation of a reversible peak shift with laser power in picocavities is found to be widely reproducible across many particles and picocavity events (further data provided in Supplementary Figure S30). A detectable spring shift was found in over two thirds of stable picocavity events recorded from 150 SPARKs, with undetectable shifts only for particles with weaker light in-coupling. A combined statistical analysis of all picocavities (as conducted above for nanocavities) is however not possible, since each picocavity is characterized by a different set of vibrational lines[45].

We now compare the observed shift for CW excitation of picocavities in SPARKs to the shifts in our pulsed experiments in NPoMs (Figs. 4,5). Eqn. (5) shows that the spring shift depends on the local field and the number of molecules (including correction factor $\eta_1$). Using the experimentally observed ratio of SERS from SPARK nanocavities and standard NPoMs, we estimate the field in the latter is 7-fold smaller.[43] Additionally, near-field focusing around the atom tip of picocavities enhances the local field further by ~3 times (which is consistent with observed experimental enhancements of SERS from picocavities of $3^4$~100). Collective vibrations lead to a 12-times larger spring shift for 100 molecules than a single molecule (see Fig. 2b). The CW laser has a much lower power than the pulsed peak laser power, $I_l(\text{CW})/I_l(\text{pulsed}) = 3\times10^{-4}$. Assuming that the Raman cross-section of molecules in the picocavity does not change (full calculation not yet possible), the pulsed nanocavity containing several molecules should give ≈ 100 times larger shift than a single molecule in a CW-pumped SPARK picocavity. Experimentally, we observe a shift >250 cm$^{-1}$ with laser pulses compared to ~5 cm$^{-1}$ in SPARK picocavities, giving a factor of ~50 consistent with the above approximations. For comparison, in SPARK nanocavities (Fig. 6i) the spring shift is below the instrument resolution due to the ~3 times lower optical fields. Importantly, we note that the analytical model in Eqn. (5) uses a simplified description of the plasmonic modes of the metal-dielectric nanostructure. Since the plasmonic modes of SPARKS are not yet known in detail (since darkfield scattering is obscured by reflections from the silica microlens), the above approximation can only confirm that the order of magnitude of the spectral shift observed is consistent with the optomechanical spring shift expected. Further, modifications of the Raman tensor induced by the picocavity might give different field enhancements and effective number of molecules involved in the picocavity-induced optical spring. However, the consistency of our estimates reinforces the optomechanical origin of the experimental results, while these picocavity observations directly evidence the repeatable spring shift from optomechanical coupling.

The observed vibrational shifts might also arise from vibrational anharmonicities under strong vibrational pumping in picocavities[12]. From our above model of the anharmonicity and linear scaling of this shift with laser intensity (Supplementary Note S9), we estimate the effect to be one order of magnitude smaller than the shift observed here. While not sufficient to fully discard contributions from vibrational anharmonicity, further detailed work on anharmonicity in picocavities is thus also needed.

## Discussion

The effect of the optomechanical interaction of the molecules, apparently observed here in plasmonic nanocavities, is to adiabatically decrease the bond strengths during a pulse (as the optical spring squeezes them). This effect is analogous to the Lamb shift in excitonic emitter-cavity coupling. It provides a new way to manipulate bonds which should be contrasted both with coherent control (based on electronic wavepacket excitation of light-absorbing molecules[47]), and with vacuum Rabi splitting (based on infrared light-matter coupling without any light present[48]). In the latter, strong coupling at mid-infrared frequencies causes energy to cycle between vibrational dipoles and photons, completely different from the optomechanical effect here where optical radiation pressure weakens the bonds themselves.

Our results indicate that in nanocavities, vibrational shifts much larger than the linewidths ($\gg$20 cm$^{-1}$) can be attained, comparable with the largest vibrational strong-coupling Rabi splitting[49]. The estimated $\Delta\omega_v$ implies fractional bond energy reductions of >10% (exceeding thermal energies and with minimal damage), corresponding to light-controlled weakening of the bond spring constants by $\sqrt{(\Delta\omega_v/\omega_v)}$~25%. Further pumping seems to reach light-induced dissolution of the collective bond, leading to irreversible bond breaking. Such effects are impossible to observe for molecules in solution[32–34] which are too far apart ($\gg d$) to coherently couple, and without the nearby metal surface they show negligible shifts.

The shifts also depend on the number of molecules hybridizing to give bright coherently-coupled phonon states. We thus explored mixed SAMs with 50% TPT and 50% BPT (as well as other fractions), and find that the power threshold is little affected (Supplementary Fig. S28). The similar vibrational spectra of these molecules thus suggests that hybridization also occurs between distinct but similar molecules.

We conclude by summarizing our results in the context of optomechanics. The large value of $\text{Re}\{G\}$ associated with the short-ranged interaction dramatically enhances the optical spring effect in such plasmonic gaps. Indeed, calculating the optical spring effect for single molecules in our NPoM gap as a function of laser wavelength (Fig. 2a) shows the light-induced spectral shift of the Raman line of a molecule, $\Delta_{os}^1 \propto \text{Re}\{G\}$, can be more than hundred-fold enhanced over that obtained in traditional dielectric cavities. This is a consequence of (i) fully including image charges from the transient Raman dipoles through the Green's function (tenfold enhancement), and (ii) $\text{Re}\{G\}$ being positive at both Stokes and anti-Stokes frequencies, so that the two contributions in $v_{ss} = (S_{ss}^+)^* + S_{ss}^-$ add up instead of largely cancelling, as occurring in single-mode optical resonances (another tenfold enhancement, Fig. 2a, Supplementary Note S6). We illustrate the difference in Figure 7a by plotting the ratio of the spring shift to the vibrational linewidth $\Gamma_{\text{tot}}$. The latter includes the optomechanically-induced broadening (or narrowing) of the vibrational losses (optomechanical damping), and is shown at laser intensities chosen so it equals half the vibrational losses, $|\gamma_v - \Gamma_{\text{tot}}(I_l)| = \gamma_v/2$, at the frequency that maximizes $\Gamma_{\text{tot}}(I_l)$. The color map behind shows the results for single-mode cavities as a function of the cavity losses $\kappa_c$ and detuning of the incident laser. The corresponding ratio for the NPoM (shown in the box) is significantly larger than for a single-mode cavity of similar losses, and is less dependent on the frequency of the laser.

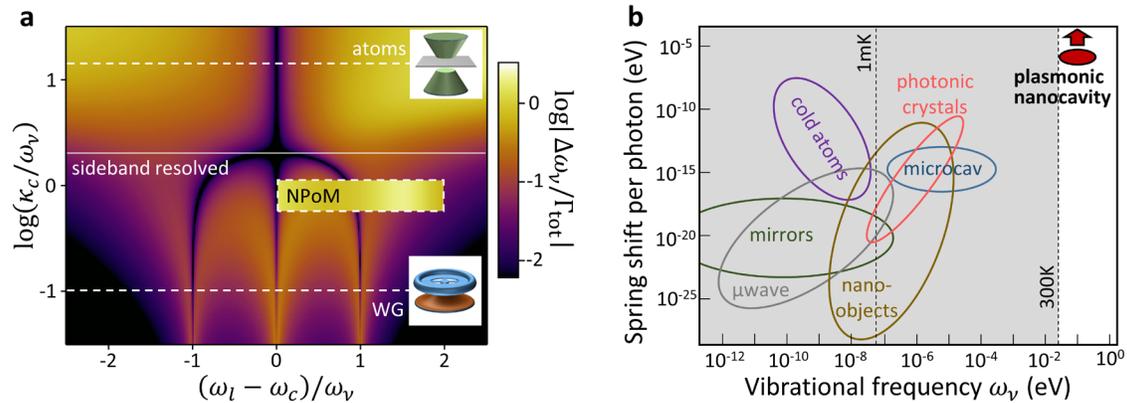

**Fig. 7 | Optical spring shifts in optomechanics. a,** Ratio of spring shift $\Delta\omega_v$ to the vibrational linewidth $\Gamma_{tot}$ mapped vs cavity linewidth $\kappa_c$ and detuning $\omega_l - \omega_c$ of laser $\omega_l$ from cavity $\omega_c$, both normalized to vibrational frequency $\omega_v$, for single-mode cavities. Typical regimes for cold atoms and whispering gallery (WG) resonances are shown dashed. For each $\kappa_c$, the lowest intensity is used at which $|\gamma_v - \Gamma_{tot}(I_l)| = \gamma_v/2$ over the detuning range. Box shows the results for the NPoM including the full multimode plasmonic response, as considered in the simulations, and with $\omega_c$ fixed at the NPoM dark-field resonance of 800 nm, giving ~10-fold enhancement. **b,** Comparison of spring shifts per cavity photon vs $\omega_v$ for a range of systems[50].

The net effect of the NPoM cavity according to Eqn. (1) is for the Raman peaks to linearly shift and broaden with laser power, even for a single molecule. The resulting spring shifts per photon in the cavity are correspondingly higher than other systems[50] (Fig. 7b), for vibrations in the ground state at 300 K, and offer further enhancements accessible from improved nanocavities or collective effects (arrow), even potentially exceeding the linewidths of cavity and vibrations.

The continuum-field optomechanical model for multimode plasmonic cavities reveals that ultrasmall mode-volume plasmonic nanocavities yield SERS emission from pumped molecular collective vibrations red-shifted by 10 to 100-fold more than in single-mode systems, leading to a redistribution of energy and a saturation of emission from the original sharp vibrational lines. This physical picture is commensurate with experimental observations of SERS saturation under pulsed illumination, and reversible SERS lineshifts under CW illumination of picocavities. Independently, the two experiments cannot unambiguously confirm the existence of the proposed optical spring shift, however both are in good agreement with the optomechanical model developed here. This also gives implications beyond simply limiting the maximum Raman yield from molecules (to $<10^{12}$ counts.s$^{-1}$).[16] In the plasmonic cavities, light transiently softens the bonds of molecules near metal interfaces, which may find use in optical catalysis of reactions and photodecomposition for recycling, as well as controlling molecular photodetectors and other molecular nanoscale optoelectronic devices. We also note that permanent damage to molecules (see Supplementary Note S12) occurs when bond softening starts, suggesting its mechanism may be related. Such plasmon-induced bond softening opens up fruitful possibilities to explore correlations of vibrations at room temperature, since $\hbar\Delta\omega_v > k_B T$. The convergence of molecular electronics, plasmonics, quantum emitters, and vibrational coherence gives opportunities for using quantum-correlated SERS to probe electronic transport, dissipation and switching. We emphasize that the results here operate not just for molecular layers, but also for 2D layered crystals such as transition metal dichalcogenides or graphene, and will lead to their drastically different optomechanical device operation when paired with plasmonic nanocavities.

## Methods

### Optomechanical simulations

A detailed description of the optomechanical theory employed to calculate SERS spectra is provided in the Supplementary Information.

### Sample preparation

The Au substrates are prepared via a template stripping method, which has been detailed elsewhere[51]. Briefly, atomically flat Au surfaces are produced by evaporating 100 nm of Au onto Si wafers at a rate of 0.5 Å/s. Small pieces of silicon are then glued to the wafer using epoxy, and the wafer slowly cooled from the 150°C curing temperature to room temperature. These silicon pieces can then be peeled off to reveal a smooth Au surface with rms roughness <0.2 nm. The SAM is prepared on the Au surface by immersion in a 1 mM analyte solution in anhydrous ethanol (>99.5%) for 22 h. The nanoparticles are purchased in suspension from BBI Solutions (80 nm, OD1, citrate capped). They are dropcast onto the SAM for 30 s before being rinsed with deionised water. The short time used for dropcasting ensures a low density across the sample so they can be individually observed in optical microscopy. Aggregation is prevented by citrate capping around the Au NPs.

### Pulsed SERS spectroscopy on NPoMs

In a custom-built, inverted darkfield microscope NPoMs are located automatically by particle tracking algorithms and centred by moving the sample stage. On each NPoM structure, Raman and dark-field spectra are acquired in quick succession. For pulsed Raman spectroscopy, a Spectra-Physics Maitai laser at 80 MHz is used to drive an optical parametric oscillator (Spectra-Physics Inspire), producing 100 fs pulses of tunable wavelength. For the required spectral resolution, pulses are filtered by a tunable bandpass filter (PhotonETC LLTF contrast) to 1.5 nm spectral bandwidth and 500 fs pulse duration. The light scattered by the sample is filtered by a Fianium Superchrome tunable longpass filter and detected by a Raman spectrometer. For power-dependent experiments, average laser power is ramped from 100 nW to 60 µW while adjusting integration times to keep constant fluence (100 s integration time for 100 nW, decreasing to 167 ms at 60 µW). We follow thousands of SERS spectra by examining hundreds of individual NPoMs for long periods of time. Correction for NPoM in-coupling efficiency is described in Supplementary Note S10.

The integrated SERS emission of the Raman lines is obtained by calculating the area underneath the spectra in a 40 cm$^{-1}$ wide window centred on the peak. The SERS background area is obtained by integrating a 150 cm$^{-1}$ wide window between the Raman lines (1350 cm$^{-1}$ to 1500 cm$^{-1}$). To compare the acquired spectra with theoretical calculations, the average in-coupled laser power is converted to the in-coupled peak power of the pulsed laser. For our laser pulses of 0.5 ps duration and 80 MHz repetition rate, an average power of 1 µW corresponds to a peak power of 2.5x10$^4$ µW and a peak intensity of 3.2x10$^4$ µW µm$^{-2}$.

## Fast Raman Power Sweeps on SPARKs

Fast Stokes spectra acquisition is carried out on Au nanoparticles with an integrated silica nanolens, termed "Superefficient plasmonic nanoarchitectures for Raman kinetics" (SPARK). SPARK constructs are produced by organosilica synthesis using Au NPs as seeds for nucleation and growth[43]. The SPARK particles are then deposited according to NPoM sample preparation described above. Such SPARK samples provide sufficient signal to allow the acquisition of SERS spectra with sub-ms integration times. To repetitively probe the power dependence of SERS spectra, a 633 nm continuous-wave laser is modulated with an acousto-optic modulator driven by a sawtooth voltage at 50 Hz supplied by a function generator. The power modulation is calibrated by measuring the minimum and maximum power with a power meter and characterising the shape of the modulated SERS background. Spectra are acquired with an Andor Newton 970BVF using the Fast Kinetic readout mode. An automated darkfield microscope is used to scan hundreds of nanoparticles and collect successive kinetic spectral scans for several minutes on each particle. The spectra are then screened for the generation of picocavities (characterised by the emergence of new, intense Raman lines) and the optical spring shift is analysed by fitting Lorentzian peaks to each individual vibrational line.

## Data availability
The figure data in this study are deposited in the Cambridge open data archive under DOI: 10.17863/CAM.95259.

## Acknowledgements
We thank Mikolaj Schmidt and Adrián Juan Delgado for fruitful discussions, and Marlous Kamp for synthesizing SPARK constructs. We acknowledge EPSRC grants EP/N016920/1, EP/L027151/1, NanoDTC EP/L015978/1, NSFC grant 12004344, NSFC-DPG grant 21961132023, Basque Government grant IT1526-22, grant PID2019-107432GB-I00 funded by MCIN/AEI/10.13039/501100011033/, and EU THOR 829067, POSEIDON 861950 and PICOFORCE 883703. LAJ acknowledges support from the Cambridge Trust and EPSRC award 2275079. B.d.N acknowledges support from the Winton Programme for the Physics of Sustainability, and the Royal Society in the form of a University Research Fellowship URF \R1\211162. CC thanks NPL for PhD funding.

## Author Contributions
WMD, EP, LAJ, BdN, SH, CC and JJB devised the experimental techniques and developed sample fabrication; WMD, EP, LAJ, JJB developed the spectral analysis; YZ, JA, RE, TN, JJB developed the simulations and models; all authors contributed to writing the manuscript.

## Conflicts of interest
The authors declare no conflicts of interest.

# References

1. Nikolis, V. C. *et al.* Strong light-matter coupling for reduced photon energy losses in organic photovoltaics. *Nat. Commun.* **10**, (2019).
2. Kennehan, E. R. *et al.* Using molecular vibrations to probe exciton delocalization in films of perylene diimides with ultrafast mid-IR spectroscopy. *Phys. Chem. Chem. Phys.* **19**, 24829–24839 (2017).
3. Tsoi, W. C. *et al.* In-situ monitoring of molecular vibrations of two organic semiconductors in photovoltaic blends and their impact on thin film morphology. *Appl. Phys. Lett.* **102**, 173302 (2013).
4. Nayak, P. K., Mahesh, S., Snaith, H. J. & Cahen, D. Photovoltaic solar cell technologies: analysing the state of the art. *Nat. Rev. Mater.* **4**, 269–285 (2019).
5. Solomon, G. C., Herrmann, C., Hansen, T., Mujica, V. & Ratner, M. A. Exploring local currents in molecular junctions. *Nat. Chem.* **2**, 223–228 (2010).
6. Saar, B. G. *et al.* Video-Rate Molecular Imaging in Vivo with Stimulated Raman Scattering. *Science* **330**, 1368–1370 (2010).
7. Sharma, B., Frontiera, R. R., Henry, A. I., Ringe, E. & Van Duyne, R. P. SERS: Materials, applications, and the future. *Materials Today* vol. 15 16–25 (2012).
8. Blackie, E. J., Le Ru, E. C. & Etchegoin, P. G. Single-molecule surface-enhanced raman spectroscopy of nonresonant molecules. *J. Am. Chem. Soc.* **131**, 14466–14472 (2009).
9. Le Ru, E. C., Blackie, E. J., Meyer, M. & Etchegoin, P. G. Surface Enhanced Raman Scattering Enhancement Factors: A Comprehensive Study. *J. Phys. Chem. C* **111**, 13794–13803 (2007).
10. Roelli, P., Galland, C., Piro, N. & Kippenberg, T. J. Molecular cavity optomechanics as a theory of plasmon-enhanced Raman scattering. *Nat. Nanotechnol.* **11**, 164–169 (2016).
11. Schmidt, M. K., Esteban, R., González-Tudela, A., Giedke, G. & Aizpurua, J. Quantum Mechanical Description of Raman Scattering from Molecules in Plasmonic Cavities. *ACS Nano* **10**, 6291–6298 (2016).
12. Benz, F. *et al.* Single-molecule optomechanics in "picocavities". *Science* **354**, 726–729 (2016).
13. Schmidt, M. K., Esteban, R., Benz, F., Baumberg, J. J. & Aizpurua, J. Linking classical and molecular optomechanics descriptions of SERS. *Faraday Discuss.* **205**, 31–65 (2017).
14. Lombardi, A. *et al.* Pulsed Molecular Optomechanics in Plasmonic Nanocavities: From Nonlinear Vibrational Instabilities to Bond-Breaking. *Phys. Rev. X* **8**, 011016 (2018).
15. Kamandar Dezfouli, M. & Hughes, S. Quantum Optics Model of Surface-Enhanced Raman Spectroscopy for Arbitrarily Shaped Plasmonic Resonators. *ACS Photonics* **4**, 1245–1256 (2017).
16. Baumberg, J. J., Aizpurua, J., Mikkelsen, M. H. & Smith, D. R. Extreme nanophotonics from ultrathin metallic gaps. *Nat. Mater.* **18**, 668–678 (2019).
17. Zhang, Y. *et al.* Addressing Molecular Optomechanical Effects in Nanocavity-Enhanced Raman Scattering beyond the Single Plasmonic Mode. *Nanoscale* **13**, 1938–1954 (2021).
18. Zhang, Y., Aizpurua, J. & Esteban, R. Optomechanical Collective Effects in Surface-Enhanced Raman Scattering from Many Molecules. *ACS Photonics* **7**, 1676–1688 (2020).
19. Vries, P. De, Coevorden, D. V. Van & Lagendijk, A. Point scatterers for classical waves. *Rev. Mod. Phys.* **70**, 447–466 (1998).
20. Chaumet, P. C., Sentenac, A. & Rahmani, A. Coupled dipole method for scatterers with large permittivity. *Phys. Rev. E* **70**, 036606 (2004).
21. Vlack, C. Van & Hughes, S. Finite-difference time-domain technique as an efficient tool for calculating the regularized Green function: applications to the local-field problem in quantum optics for inhomogeneous lossy materials. *Opt. Lett.* **37**, 2880–2882 (2012).
22. Frenkel, J. On the transformation of light into heat in solids. i. *Phys. Rev.* **37**, 17–44 (1931).

23. Pouthier, V. Vibrational exciton mediated quantum state transfer: Simple model. *Phys. Rev. B - Condens. Matter Mater. Phys.* **85**, 214303 (2012).
24. Muller, E. A. *et al.* Vibrational exciton nanoimaging of phases and domains in porphyrin nanocrystals. *Proc. Natl. Acad. Sci. U. S. A.* **117**, 7030–7037 (2020).
25. Crampton, K. T. *et al.* Ultrafast Coherent Raman Scattering at Plasmonic Nanojunctions. *J. Phys. Chem. C* **120**, 20943–20953 (2016).
26. Pozzi, E. A. *et al.* Operational Regimes in Picosecond and Femtosecond Pulse-Excited Ultrahigh Vacuum SERS. *J. Phys. Chem. Lett.* **7**, 2971–2976 (2016).
27. Gruenke, N. L. *et al.* Ultrafast and nonlinear surface-enhanced Raman spectroscopy. *Chem. Soc. Rev.* **45**, 2263–2290 (2016).
28. Zhang, R. *et al.* Chemical mapping of a single molecule by plasmon-enhanced Raman scattering. *Nature* **498**, 82–86 (2013).
29. Haslett, T. L., Tay, L. & Moskovits, M. Can surface-enhanced Raman scattering serve as a channel for strong optical pumping? *J. Chem. Phys.* **113**, 1641–1646 (2000).
30. Li, K. *et al.* Large-area, reproducible and sensitive plasmonic MIM substrates for surface-enhanced Raman scattering. *Nanotechnology* **27**, 495402 (2016).
31. Lee, J., Tallarida, N., Chen, X., Jensen, L. & Apkarian, V. A. Microscopy with a single-molecule scanning electrometer. *Sci. Adv.* **4**, eaat5472 (2018).
32. Ventalon, C. *et al.* Coherent vibrational climbing in carboxyhemoglobin. *Proc. Natl. Acad. Sci.* **101**, 13216–13220 (2004).
33. Witte, T., Yeston, J. ., Motzkus, M., Heilweil, E. . & Kompa, K.-L. Femtosecond infrared coherent excitation of liquid phase vibrational population distributions (v>5). *Chem. Phys. Lett.* **392**, 156–161 (2004).
34. Morichika, I., Murata, K., Sakurai, A., Ishii, K. & Ashihara, S. Molecular ground-state dissociation in the condensed phase employing plasmonic field enhancement of chirped mid-infrared pulses. *Nat. Commun.* **10**, 3893 (2019).
35. Crampton, K. T., Fast, A., Potma, E. O. & Apkarian, V. A. Junction Plasmon Driven Population Inversion of Molecular Vibrations: A Picosecond Surface-Enhanced Raman Spectroscopy Study. *Nano Lett.* **18**, 5791–5796 (2018).
36. Park, K. D. *et al.* Variable-Temperature Tip-Enhanced Raman Spectroscopy of Single-Molecule Fluctuations and Dynamics. *Nano Lett.* **16**, 479–487 (2016).
37. Artur, C., Le Ru, E. C. & Etchegoin, P. G. Temperature dependence of the homogeneous broadening of resonant Raman peaks measured by single-molecule surface-enhanced Raman spectroscopy. *J. Phys. Chem. Lett.* **2**, 3002–3005 (2011).
38. Harris, C. B., Shelby, R. M. & Cornelius, P. A. Effects of energy exchange on vibrational dephasing times in Raman scattering. *Phys. Rev. Lett.* **38**, 1415–1419 (1977).
39. Zhang, K. & Chen, X.-J. Identification of the incommensurate structure transition in biphenyl by Raman scattering. *Spectrochim. Acta Part A Mol. Biomol. Spectrosc.* **206**, 202–206 (2019).
40. Mei, H.-Y. *et al.* Study of melting transition on biphenyl by Raman scattering. *AIP Adv.* **9**, 095049 (2019).
41. Johnson, P. B. & Christy, R. W. Optical Constants of the Noble Metals. *Phys. Rev. B* **6**, 4370–4379 (1972).
42. Cheetham, M. R. *et al.* Out-of-Plane Nanoscale Reorganization of Lipid Molecules and Nanoparticles Revealed by Plasmonic Spectroscopy. *J. Phys. Chem. Lett.* **11**, 2875–2882 (2020).
43. Kamp, M. *et al.* Cascaded nanooptics to probe microsecond atomic-scale phenomena. *Proc. Natl. Acad. Sci. U. S. A.* **117**, 14819–14826 (2020).
44. Huang, J. *et al.* Tracking interfacial single-molecule pH and binding dynamics via vibrational spectroscopy. *Sci. Adv.* **7**, eabg1790 (2021).

45. Carnegie, C. *et al.* Room-Temperature Optical Picocavities below 1 nm3 Accessing Single-Atom Geometries. *J. Phys. Chem. Lett.* **9**, 7146–7151 (2018).
46. Griffiths, J., De Nijs, B., Chikkaraddy, R. & Baumberg, J. J. Locating Single-Atom Optical Picocavities Using Wavelength-Multiplexed Raman Scattering. *ACS Photonics* **8**, 2868–2875 (2021).
47. Brinks, D. *et al.* Ultrafast dynamics of single molecules. *Chem. Soc. Rev.* **43**, 2476–2491 (2014).
48. Shalabney, A. *et al.* Coherent coupling of molecular resonators with a microcavity mode. *Nat. Commun.* **6**, 5981 (2015).
49. George, J. *et al.* Multiple Rabi Splittings under Ultrastrong Vibrational Coupling. *Phys. Rev. Lett.* **117**, 183601 (2016).
50. Aspelmeyer, M., Kippenberg, T. J. & Marquardt, F. Cavity optomechanics. *Rev. Mod. Phys.* **86**, 1391–1452 (2014).
51. Hegner, M., Wagner, P. & Semenza, G. Ultralarge atomically flat template-stripped Au surfaces for scanning probe microscopy. *Surf. Sci.* **291**, 39–46 (1993).

# Giant optomechanical spring effect in plasmonic nano- and picocavities probed by surface-enhanced Raman scattering

# Supplementary Information

Lukas A. Jakob[1], William M. Deacon[1], Yuan Zhang[2*], Bart de Nijs[1], Elena Pavlenko[1], Shu Hu[1], Cloudy Carnegie[1], Tomas Neuman[3], Ruben Esteban[3], Javier Aizpurua[3*], Jeremy J. Baumberg[1*]

[1] Nanophotonics Centre, Cavendish Laboratory, University of Cambridge, Cambridge CB3 0HE, UK.
[2] Henan Key Laboratory of Diamond Optoelectronic Materials and Devices, Key Laboratory of Material Physics, Ministry of Education, School of Physics and Microelectronics, Zhengzhou University, Zhengzhou 450052, China.
[3] Institute of Quantum Materials and Physics, Henan Academy of Sciences, Zhengzhou 450046, China
[4] Center for Material Physics (CSIC - UPV/EHU and DIPC), Paseo Manuel de Lardizabal 5, Donostia-San Sebastian Gipuzkoa 20018, Spain.

## Theory and Simulation

In the detailed theory worked out below, we show the extended molecular optomechanical model and describe what approximations have been used in these derivations (Section S1). We then calculate the specific vibrations modes for the molecules probed in the experiments (Section S2), and the specific plasmon modes in the nanoparticle-on-mirror (NPoM) structures used (Section S3). Using these, we then calculate the parameters for our optomechanical model within the experimental situation here (Section S4), and finally compare the simulations to the experiment (Section S5).

To help the reader, we provide some additional information, starting with a direct comparison of the single-mode optomechanical model with the continuum-field extension used here (Section S6). We then provide a simplified formula to understand the Raman lineshifts that are produced, as well as what intensity thresholds can be identified (Section S7). Finally, we also show that these can be understood in terms of local dipoles (Section S8), and confirm that introducing vibrational anharmonicity does not account for the results (Section S9).

### S1. Molecular Optomechanical Theory

In a previous study[1], the molecular optomechanical theory of non-resonant Surface-Enhanced Raman Scattering (SERS) by Dezfouli and Hughes[2], which considered a single molecule and a single vibration, was extended to a single molecule with multiple vibrations. In such a theory, the electromagnetic field of the nanocavity (plasmonic response) was quantized in a dispersive and lossy medium using a classical dyadic Green's function[3,4], and this continuum of quantized electric field was coupled with the molecular vibrations via the optomechanical interaction. Finally, the electromagnetic field degree of freedom was eliminated to achieve an effective master equation for the molecular vibrations.

This approach, applied in the work here, accounts for the full complexity of the plasmonic response via continuum-field theory. In contrast, the simpler molecular optomechanics description considers a single plasmonic mode. We sketch in Figure S1 these two conceptual optomechanical frameworks, which show their essence and details.

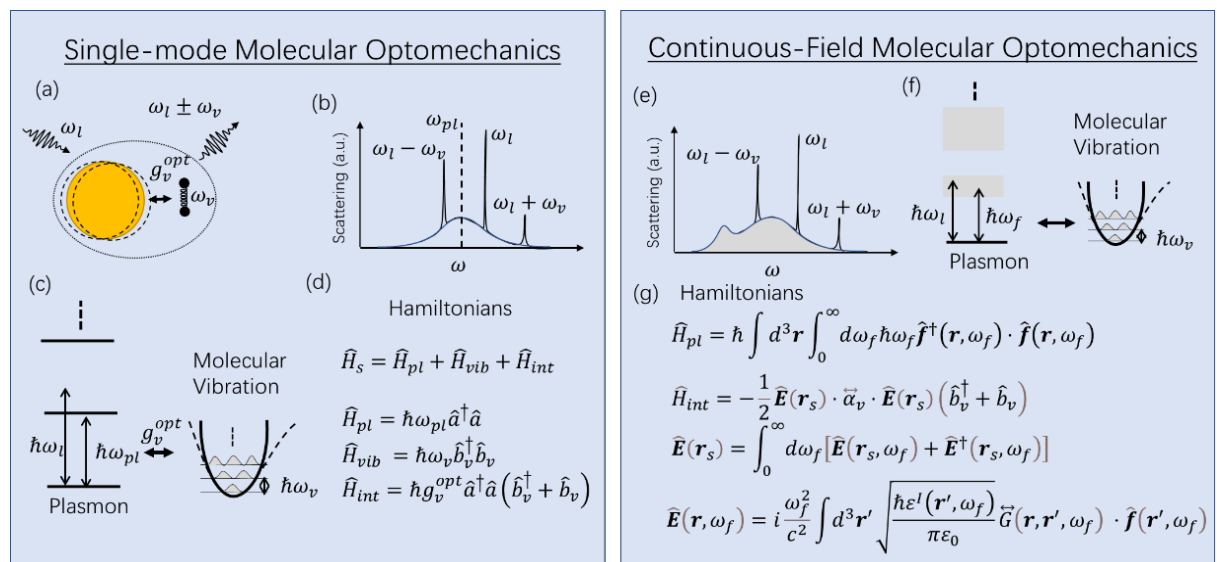

**Figure S1.** Theoretical frameworks for molecular optomechanics within (a-d) cavity single-mode (on left side) and (e-g) continuum-field (on right side) descriptions.

Figure S1 (a) shows that, in the situation of a single plasmonic mode of (angular) frequency $\omega_{pl}$, this mode couples with a molecular vibration of much lower frequency $\omega_v$. The strength of the interaction is given by the optomechanical coupling $g_v^{opt}$. As a consequence of this coupling, the laser excitation at frequency $\omega_l$ results in Stokes and anti-Stokes Raman lines at frequencies $\omega_l - \omega_v$ and $\omega_l + \omega_v$ (Figure S1 b). The spring shift effect can induce changes in the vibrational frequency so that emission now occurs at modified $\omega_l - (\omega_v - \Delta\omega)$, and $\omega_l + (\omega_v - \Delta\omega)$ frequencies. This shift is important to understand the results at large laser intensities used in the experiments. Figure S1 (c) illustrates the corresponding energy scheme for the plasmon mode and the molecular vibration, which are both treated as quantized harmonic oscillators. For the molecule, this model corresponds to approximating the full potential energy surface for the electronic ground state as parabolic. The Hamiltonians of the plasmon $\widehat{H}_{pl}$, the molecular vibration $\widehat{H}_{vib}$ and the optomechanical interaction $\widehat{H}_{int}$ are explicitly written in Figure S1 (d), where $\hat{a}^\dagger, \hat{a}$ ($\hat{b}_v^\dagger, \hat{b}_v$) are the creation and annihilation operators of the plasmon (vibrational) quanta. We notice in particular that the optomechanical coupling term is given by $\hbar g_v^{opt} \hat{a}^\dagger \hat{a} \left( \hat{b}_v^\dagger + \hat{b}_v \right)$ (note, in the main text, this optomechanical coupling is written for concision as $g$). Losses are included via Lindblad operators and the master equation (not written explicitly in the panel for simplicity).

For the NPoM system studied in this work on the other hand, describing the plasmonic system as a single mode is insufficient and it is necessary to incorporate the full modal structure[1–3], as depicted in Figure S1 (e-g). The complex plasmonic response (Figure S1 e) requires a continuum-field description corresponding to a continuum of electromagnetic modes (annihilation operator $\boldsymbol{f}(\boldsymbol{r}, \omega_f)$ of frequency $\omega_f$ and at position $\boldsymbol{r}$), which couple with the molecular vibration (again treated as a harmonic oscillator). The resulting Hamiltonian is given in Figure S1 (g) where $\overleftrightarrow{\alpha}_v$ and $\widehat{\boldsymbol{E}}(\boldsymbol{r}_s)$ are the Raman polarizability tensor and the quantized electric field operator at the molecular position $\boldsymbol{r}_s$. Here, this total electric field operator is the integral of field operators $\widehat{\boldsymbol{E}}(\boldsymbol{r}_s, \omega_f)$ associated with different frequencies $\omega_f$ and is a function of the $\widehat{\boldsymbol{f}}(\boldsymbol{r}, \omega_f)$ operators and of the dyadic Green's function $\overleftrightarrow{G}(\boldsymbol{r}, \boldsymbol{r}', \omega_f)$ [together with a prefactor that depends on the permittivity $\varepsilon_0$ and light speed $c$ in vacuum and the imaginary part of permittivity $\varepsilon^I(\boldsymbol{r}', \omega_f)$]. This description already contains the plasmonic losses so that only vibrational losses need to be incorporated via Lindblad operators.

In principle, we can easily generalize the molecular optomechanical theory to the more complex situation involving multiple molecules and multiple vibrations. In this case, the different vibrational modes of the same molecule or different molecules can be correlated with each other. As shown in our previous study[1], this kind of correlation is small if the different vibrational modes are strongly off-resonant with each other.

In contrast, the same vibrational mode of identical molecules becomes strongly correlated due to their degenerate tuning. Thus, in the following we focus on the particular case of a single vibrational mode and multiple molecules. To describe this situation, we modify the theory presented in Ref. [1] with the following procedures: (i) we replace the index $v$ of the vibrational modes of a single molecule with the molecular index $s$; and (ii) we extend the dyadic Green's functions $\overleftrightarrow{G}(\boldsymbol{r}_m, \boldsymbol{r}_m; \omega)$, $\overleftrightarrow{G}(\boldsymbol{r}_d, \boldsymbol{r}_m; \omega)$ for the single molecule at position $\boldsymbol{r}_m$ and the detector at position $\boldsymbol{r}_d$, to the more general ones $\overleftrightarrow{G}(\boldsymbol{r}_{s'}, \boldsymbol{r}_s; \omega)$, $\overleftrightarrow{G}(\boldsymbol{r}_d, \boldsymbol{r}_s; \omega)$ for the $s$-th and the $s'$-th molecules at positions $\boldsymbol{r}_s$ and $\boldsymbol{r}_{s'}$, and the detector at position $\boldsymbol{r}_d$. As the derivation is a straightforward extension of the results in Ref. [1], we just present the final results and main ideas in Sections S1.2 and S1.3. Additionally, we summarize in Section S1.1 all the approximations involved in the development of the molecular optomechanical framework used in this work.

## S1.1 Approximations in the Description of Molecular Optomechanical Interactions

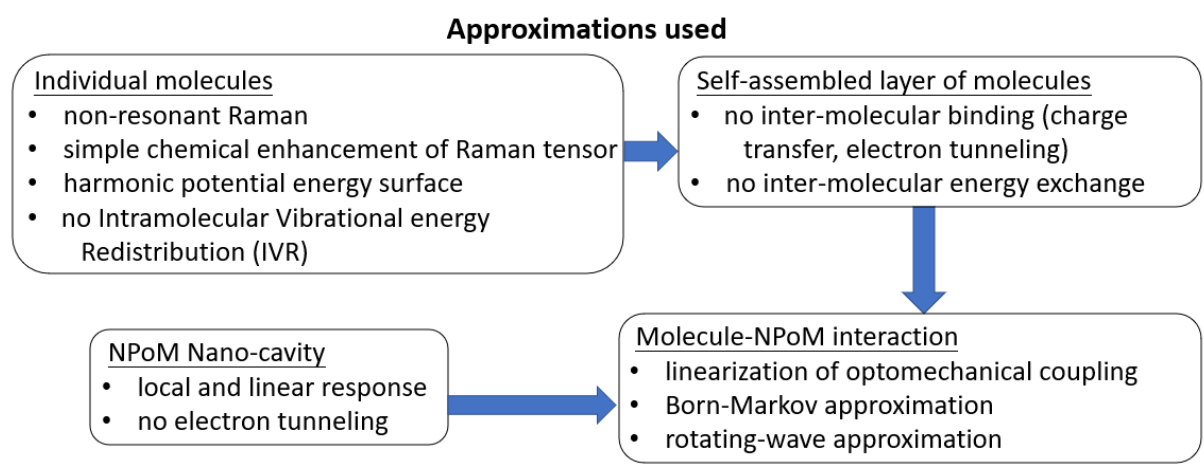

**Figure S2.** Summary of the approximations used in the molecular optomechanical framework here.

In the theory of molecular optomechanics used in this paper, a number of approximations are adopted as summarized in Figure S2, and their validity is justified below:

- Electronic excited states of the molecules (not populated due to the large laser detuning) do not need to be considered. We thus follow the standard theory of non-resonant Raman scattering, in which the effect of excited states is properly included indirectly through the Raman tensor of the molecular vibrations. This key parameter is calculated using rigorous atomistic DFT calculation that includes (partially) the chemical effects (Section S2). We assume a parabolic potential energy surface for the electronic ground state, so that the vibrational modes can be modeled as harmonic oscillators - the validity of this approximation is discussed in detail in Section S9. Lastly, we do not consider intra-molecular vibrational energy redistribution (IVR) processes[5,6], which we expect to be more important for the low-energy vibrational modes, instead of the 1586 cm$^{-1}$ mode which is the focus here. Furthermore, these IVR processes should not affect the emergence of collective modes that explain the behavior of the system.

- The Raman tensor is obtained for isolated molecules, and thus charge transfer, electron-tunneling or energy exchange between molecules are not included. For the biphenyl-4-thiol (BPT) molecules (and in contrast to redox molecules recently studied[7]), charge transfer is not seen (or expected). Previous DFT calculations[8] show that the Raman polarizability of individual molecules thiol-attached to a gold atom incorporates the effect of partial charge transfer from the substrate and are in good agreement with the low-power Raman spectra of molecular monolayers. DFT also confirms that electronic inter-molecular interactions are very weak contributing <5 cm$^{-1}$ shifts. Electron tunneling is not a factor until >10$^9$ Vm$^{-1}$ electrical fields are applied, as recently confirmed experimentally for these molecules[9].

- The optical response of the NPoM nanocavity is modeled by using linear and local classical electromagnetic theory. Non-locality modifies slightly the effective gap size[10] and the exact contribution of higher order modes to the self-interaction Green's function[11]. These effects (which only become strong for gaps much narrower than those considered here) might introduce a correction to the calculated values, but do not affect the observed trends. Similarly, quantum effects such as electron-tunneling only become important for considerably narrower gaps[12]. Crucially, our model incorporates the full plasmonic response via the dyadic Green's function, instead of the contribution from a finite number of modes[1].

- To solve the resulting optomechanical Hamiltonian describing the non-resonant Raman process, the field exciting the molecular vibrations is considered to be classical[2]. This neglects a small optomechanically-induced energy shift of the plasmonic response and effectively linearizes the Hamiltonian. The linearization affects the optomechanical response only weakly for typical coupling strengths[13], introducing a significant error only when the optomechanical coupling strength is comparable or larger than the plasmonic losses. Here, the coupling of the 1586 cm$^{-1}$ vibration of a single molecule with the (20) plasmonic mode is 0.06 meV, which is three orders of magnitude smaller than the plasmonic losses 135 meV (Section S6). The Hamiltonian is then solved using the Born-Markovian approximation according to the standard open quantum system approach, which is well justified for our system because the optomechanical coupling and vibrational dynamics are significantly slower than the plasmonic dynamics. During the theoretical derivation, the rotating wave-approximation (RWA) has been applied in several places, which ignores fast oscillating components. Normally, the RWA fails only for significantly coupling strengths, which do not occur in our system. In addition, the validity of the RWA for our system has also been checked numerically.

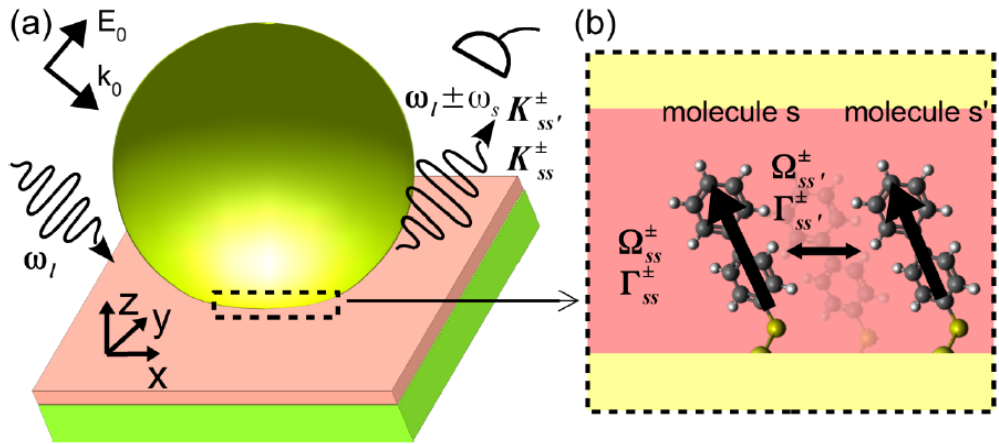

**Figure S3. SERS from many molecules inside a NPoM nanocavity.** (**a**) Illumination of the NPoM by a plane wave (laser) with frequency $\omega_l$, propagation wave-vector $\boldsymbol{k}_0$ and amplitude $\boldsymbol{E}_0$, and the detection of the Stokes and anti-Stokes Raman scattering photons at frequencies $\omega_l - \omega_s$, $\omega_l + \omega_s$ by the detector in the far-field. The exact geometry of the NPoM system is given in Section S3. (**b**) Zoom-in of the gap region, where the BPT molecules are treated as point-like dipoles, and their vibrational modes are modified by the parameters $\Omega_{ss}^\pm$, $\Gamma_{ss}^\pm$, and the molecules are coupled by $\Omega_{ss'}^\pm$, $\Gamma_{ss'}^\pm$ (with $s \neq s'$). In addition, the propagation of the Raman photons of individual molecules is captured by $K_{ss}^\pm$ while the interference of the scattering of different molecules is given by $K_{ss'}^\pm$ (with $s \neq s'$).

## S1.2 Expressions for the SERS Spectra

We now present a detailed derivation of the inelastic response. In the generalized theory, the differential power of the Raman scattering is determined by

$$\frac{dP}{d\Omega} = \sum_{s=1}^{N_m} \text{Re}\{ K_{ss}(\omega)[ S_{ss}^{st} (\omega - \omega_l) + S_{ss}^{as} (\omega - \omega_l)]\}$$
$$+ \sum_{s=1}^{N_m} \sum_{s \neq s}^{N_m} \text{Re} \{K_{s's}(\omega)[ S_{ss'}^{st} (\omega - \omega_l) + S_{ss'}^{as} (\omega - \omega_l)]\} \ . \tag{S1}$$

In this equation, the first line describes the Raman scattering of the $N$ individual molecules, while the second line captures the interference of the Raman scattering of different molecules[14,15]. The factors $K_{ss'}(\omega)$ are defined as

$$K_{ss'}(\omega) = \frac{cr^2}{2\pi\varepsilon_0} \left[\frac{\omega^2}{c^2} \overleftrightarrow{G}^*(\mathbf{r}_d,\mathbf{r}_{s'};\omega) \cdot \mathbf{p}_{s'}^*\right] \cdot \left[\frac{\omega^2}{c^2} \overleftrightarrow{G}(\mathbf{r}_d,\mathbf{r}_s;\omega) \cdot \mathbf{p}_s\right] \ . \tag{S2}$$

Here, the induced Raman dipole $\mathbf{p}_s = \overleftrightarrow{\alpha}_v \mathbf{E}(\mathbf{r}_s,\omega_l)$ is determined by the Raman polarizabilitly $\overleftrightarrow{\alpha}_v$ of the $v$-th vibrational mode and the local electric field $\mathbf{E}(\mathbf{r}_s,\omega_l)$ at the molecular position $\mathbf{r}_s$ induced by the laser of frequency $\omega_l$. $\overleftrightarrow{G}(\mathbf{r}_d,\mathbf{r}_s;\omega)$ is the classical dyadic Green's function, which relates the electric field at the detector position $\mathbf{r}_d$ with the dipole point source at the molecular position $\mathbf{r}_s$ as $\mathbf{E}(\mathbf{r}_d,\omega) = \omega^2/c^2\varepsilon_0 \overleftrightarrow{G}(\mathbf{r}_d,\mathbf{r}_s;\omega) \cdot \mathbf{p}_s$. In addition, $c, \epsilon_0, r$ are the speed of light in vacuum, the vacuum permittivity, and the molecule-detector distance, respectively. Here, we have updated the definition of the propagation factor compared to Ref. [1] in order to write the differential scattering power for many molecules in a compact and matrix-form, see Eq. (S16) and (S17).

Importantly, the terms in the brackets of Eq. (S1) describe the frequency- and excitation-dependence of the Stokes ($st$) and anti-Stokes ($as$) Raman scattering spectral density component $S_{ss'}^{st}(\omega) = \int_{-\infty}^{\infty} d\tau \, e^{-i\omega\tau}\theta(\tau)\langle\hat{b}_s(\tau)\hat{b}_{s'}^\dagger(0)\rangle$, $S_{ss'}^{as}(\omega) = \int_{-\infty}^{\infty} d\tau \, e^{-i\omega\tau}\theta(\tau)\langle\hat{b}_s^\dagger(\tau)\hat{b}_{s'}(0)\rangle$ (for $s = s'$ and $s \neq s'$). These dependencies are defined by the Fourier transform of the two-time correlations $\langle\hat{b}_s(\tau)\hat{b}_{s'}^\dagger(0)\rangle$, $\langle\hat{b}_s^\dagger(\tau)\hat{b}_{s'}(0)\rangle$ with $\tau$ the time difference from the steady state labeled by 0 and $\langle\cdot\rangle$ indicating the average. Here, $\hat{b}_s^\dagger, \hat{b}_s$ are the creation and annihilation operators of the $v$-th vibrational mode of the $s$-th molecule, and $\theta(\tau)$ accounts for the causality of the two-time correlations. The equations for the two-time correlations can be derived from the equations for the vibrational amplitudes $\langle\hat{b}_s\rangle, \langle\hat{b}_s^\dagger\rangle$ by applying the quantum regression theorem[16]. The equations for $\beta_s \equiv \langle\hat{b}_s\rangle$ are given by Eq. (1) in the main text. Using these equations and carrying out the Fourier transform, we obtain the following equations

$$i\left[\left(\omega + \omega_v - i\frac{\gamma_v}{2}\right) - v_{ss}^{(1)}\right] S_{ss}^{st}(\omega) - i\sum_{s''\neq s}^{N_m} v_{s''s}^{(1)} S_{s''s}^{st}(\omega) = 1 + \langle\hat{b}_s^\dagger\hat{b}_s\rangle_{ste} \ , \tag{S3}$$

$$i\left[\left(\omega + \omega_v - i\frac{\gamma_v}{2}\right) - v_{ss}^{(1)}\right] S_{ss'}^{st}(\omega) - i\sum_{s''\neq s}^{N_m} v_{s''s}^{(1)} S_{s''s'}^{st}(\omega) = \langle\hat{b}_{s'}^\dagger\hat{b}_s\rangle_{ste} \ , \tag{S4}$$

$$i\left[\left(\omega - \omega_v - i\frac{\gamma_v}{2}\right) + v_{ss}^{(2)}\right] S_{ss}^{as}(\omega) + i\sum_{s''\neq s}^{N_m} v_{ss''}^{(2)} S_{s''s}^{as}(\omega) = \langle\hat{b}_s^\dagger\hat{b}_s\rangle_{ste} \ , \tag{S5}$$

$$i\left[\left(\omega - \omega_v - i\frac{\gamma_v}{2}\right) + v_{ss}^{(2)}\right] S_{ss'}^{as}(\omega) + i\sum_{s''\neq s}^{N_m} v_{ss''}^{(2)} S_{s''s'}^{as}(\omega) = \langle\hat{b}_s^\dagger\hat{b}_{s'}\rangle_{ste} \ . \tag{S6}$$

Equation (S3) describes the Stokes Raman densities of individual molecules. Here, $\omega_\nu$ and $\gamma_\nu$ are the frequency, and the intrinsic decay of the $\nu$-th vibrational mode. $v_{ss}^{(1)}$ is defined as $v_{ss}^{(1)} = v_{ss} = (S_{ss}^+)^* + S_{ss}^-$ (with the spectral densities $S_{s's}^\pm$ given by Eq. (2) in the main text), and its real part $\mathrm{Re}\,v_{ss} = \frac{1}{2}(\Omega_{ss}^+ + \Omega_{ss}^-)$ and imaginary part $\mathrm{Im}\,v_{ss}^{(1)} = -\frac{1}{2}(\Gamma_{ss}^+ - \Gamma_{ss}^-)$ describe the reduction of the vibrational frequency and the change of the vibrational decay, respectively, that occur due to the optomechanical coupling. $v_{s's}^{(1)} = \frac{1}{2}(\Omega_{s's}^+ + \Omega_{ss'}^-) - i\frac{1}{2}(\Gamma_{s's}^+ - \Gamma_{ss'}^-)$ (with $s \neq s'$) describes the coupling with other molecules. $\langle \hat{b}_s^\dagger \hat{b}_s \rangle_{ste}$ describes the vibrational population of the $s$-th molecule at steady-state (indicated by the sub-index "$ste$"). Equation (S4) describes the coupling of the individual Stokes densities of the $s$-th and $s'$-th molecule. This equation has a similar structure as Eq. (S3) except that it depends on the correlation $\langle \hat{b}_{s'}^\dagger \hat{b}_s \rangle_{ste}$ of the $s'$-th and $s$-th molecule. Since the correlation is usually smaller than 1 (see below), $S_{ss'}^{st}(\omega)$ contributes less to the final Stokes density than $S_{ss}^{st}(\omega)$.

From the spectral density $S_{s's}^\pm$ given by Eq. (2) in the main text, we can obtain the parameters $\Omega_{s's}^\pm$ and $\Gamma_{s's}^\pm$ (with $s = s'$ and $s \neq s'$):

$$\Omega_{s's}^\pm = \frac{1}{2\hbar\varepsilon_0}\left(\frac{\omega_l \mp \omega_\nu}{c}\right)^2 \boldsymbol{p}_{s'}^* \cdot \mathrm{Re}\,\overleftrightarrow{G}(\boldsymbol{r}_{s'},\boldsymbol{r}_s;\omega_l \mp \omega_\nu) \cdot \boldsymbol{p}_s \ , \tag{S7}$$

$$\Gamma_{s's}^\pm = \frac{1}{2\hbar\varepsilon_0}\left(\frac{\omega_l \mp \omega_\nu}{c}\right)^2 \boldsymbol{p}_{s'}^* \cdot \mathrm{Im}\,\overleftrightarrow{G}(\boldsymbol{r}_{s'},\boldsymbol{r}_s;\omega_l \mp \omega_\nu) \cdot \boldsymbol{p}_s \ , \tag{S8}$$

which depend on the real and imaginary part of the dyadic Green's function $\overleftrightarrow{G}(\boldsymbol{r}_{s'},\boldsymbol{r}_s;\omega)$ between the $s$-th molecule (at the position $\boldsymbol{r}_s$) and the $s'$-th molecule (at the position $\boldsymbol{r}_{s'}$), respectively. The parameters with the superscript "$+$" ("$-$") are evaluated at the Stokes frequency $\omega_l - \omega_\nu$ (anti-Stokes frequency $\omega_l + \omega_\nu$). Since we focus on the optomechanical effects due to the plasmonic field, we only use in the calculations the contribution of the scattered field of the NPoM nanocavity to the dyadic Green's function[1].

In the same way, we can understand Eqs. (S5) and (S6) for the anti-Stokes Raman density components of individual molecules and pairs of molecules. In these equations, $v_{ss}^{(2)}$ and $v_{ss'}^{(2)}$ (with $s \neq s'$) are defined as $v_{ss}^{(2)} = (v_{ss}^{(1)})^*$ and $v_{ss'}^{(2)} = (v_{s's}^{(1)})^*$, respectively. The anti-Stokes density of individual molecules depends explicitly on the vibrational population. Since the correlation can become comparable with the vibrational population for moderate laser illumination (see below), $S_{ss'}^{as}(\omega)$ contributes to the final anti-Stokes density almost as much as $S_{ss}^{as}(\omega)$. Additionally, since the populations and correlations are normally much smaller than 1, the anti-Stokes spectral density is much smaller than the Stokes density [as seen by comparing Eq. (S5) to Eq. (S3)].

To obtain the Stokes and the anti-Stokes Raman densities, we need to solve the coupled equations (S3) and (S4). Similarly, the corresponding anti-Stokes results are obtained from Eqs. (S5) and (S6). To this end, we introduce the matrices

$$M = \left(\omega_\nu - i\frac{\gamma_\nu}{2}\right)I - \begin{bmatrix} v_{11}^{(1)} & \cdots & v_{N1}^{(1)} \\ \vdots & \vdots & \vdots \\ v_{1N}^{(1)} & \cdots & v_{NN}^{(1)} \end{bmatrix}, \quad S^{i=st,as}(\omega) = \begin{bmatrix} S_{11}^i(\omega) & \cdots & S_{1N}^i(\omega) \\ \vdots & \vdots & \vdots \\ S_{N1}^i(\omega) & \cdots & S_{NN}^i(\omega) \end{bmatrix},$$

$$B = \begin{bmatrix} \langle \hat{b}_1^\dagger \hat{b}_1 \rangle & \cdots & \langle \hat{b}_N^\dagger \hat{b}_1 \rangle \\ \vdots & \vdots & \vdots \\ \langle \hat{b}_1^\dagger \hat{b}_N \rangle & \cdots & \langle \hat{b}_N^\dagger \hat{b}_N \rangle \end{bmatrix} \tag{S9}$$

to rewrite the coupled equations in a matrix form

$$i\omega S^{st}(\omega) + iMS^{st}(\omega) = I + B \quad , \tag{S10}$$

$$i\omega S^{as}(\omega) - i(M^*)^T S^{as}(\omega) = B^T \quad . \tag{S11}$$

where $I$ is an identity matrix. To solve these equations, we first solve the eigenvalue problem

$$Mc_\phi = \lambda_\phi c_\phi \tag{S12}$$

with the eigenvalue $\lambda_\phi$ and the $\phi$-th eigenvector $c_\phi = \{\ldots c_{\phi s} \ldots\}$. These eigenvectors correspond to the collective vibrational modes, where the squared module $|c_{\phi s}|^2$ describes the contribution of the $s$-th molecule, and the real and imaginary part of $\lambda_\phi$ describe the frequency shift relative to $\omega_v$ and the decay rate of these collective modes. From the eigen values and vectors, we can construct the singular matrix $S$ and the diagonal matrix $\Lambda$:

$$S = \begin{bmatrix} c_{11} & \cdots & c_{N1} \\ \vdots & \vdots & \vdots \\ c_{1N} & \cdots & c_{NN} \end{bmatrix} \quad , \quad \Lambda = \begin{bmatrix} \lambda_1 & 0 & 0 \\ 0 & \ddots & 0 \\ 0 & 0 & \lambda_N \end{bmatrix} \tag{S13}$$

With these matrices, we obtain the following solutions of Eq. (S10) and (S11)

$$S^{st}(\omega - \omega_l) = S \frac{1}{i((\omega - \omega_l)I + \Lambda)} S^{-1}(I + B) \quad , \tag{S14}$$

$$S^{as}(\omega - \omega_l) = S \frac{1}{i((\omega - \omega_l)I - \Lambda^*)} S^{-1} B^T \quad . \tag{S15}$$

Inserting the above solutions into Eq. (S1) and then rearranging the matrix multiplication, we obtain the Stokes Raman spectrum

$$\frac{dP^{st}}{d\Omega} = \text{Re} \sum_{\phi=1}^{N} \frac{1}{i(\omega - \omega_l + \lambda_\phi)} [S^{-1}(I + B)K^+ S]_{\phi\phi} \quad , \tag{S16}$$

and the anti-Stokes Raman spectrum

$$\frac{dP^{as}}{d\Omega} = \text{Re} \sum_{\phi=1}^{N} \frac{1}{i(\omega - \omega_l - \lambda_\phi^*)} [S^{-1} B^T K^- S]_{\phi\phi} \quad . \tag{S17}$$

Here, $K^\pm$ is the matrix form of the propagation factors $K_{s's}^\pm = K_{s's}(\omega_l \mp \omega_v)$. We see that the Stokes (anti-Stokes) spectrum is constituted by Lorentzian functions with the frequencies $\omega_l - \text{Re}\lambda_\phi$ and $\omega_l + \text{Re}\lambda_\phi$ and the linewidths $2|\text{Im}\lambda_\phi|$, and the amplitude of the different collective modes is given by the diagonal elements of the term in the brackets.

Since the vibrational population and the correlation are typically much smaller than 1, i.e. $\langle \hat{b}_s^\dagger \hat{b}_s \rangle$, $|\langle \hat{b}_s^\dagger \hat{b}_{s'} \rangle| \ll 1$ (see below), we can approximate Eq. (S16) as

$$\frac{dP^{st}}{d\Omega} \approx \text{Re} \sum_{\phi=1}^{N} \frac{1}{i(\omega - \omega_l + \lambda_\phi)} [S^{-1} K^+ S]_{\phi\phi} \quad . \tag{S18}$$

This indicates that the collective modes are essentially caused by the coupling of the vibrational modes.

## S1.3 Expressions for the Vibrational Population and Correlation

To compute the Stokes and anti-Stokes spectrum with Eqs. (S16) and (S17), we have to first compute the vibrational population $\langle \hat{b}_s^\dagger \hat{b}_s \rangle$ and correlation $\langle \hat{b}_s^\dagger \hat{b}_{s'} \rangle$ (with $s \neq s'$). The equation for $\langle \hat{b}_s^\dagger \hat{b}_s \rangle$ reads

$$\frac{\partial}{\partial t} \langle \hat{b}_s^\dagger \hat{b}_s \rangle = -\gamma_v \langle \hat{b}_s^\dagger \hat{b}_s \rangle + n_v^{\text{th}} \gamma_v$$
$$+ \Gamma_{ss}^+ + (\Gamma_{ss}^+ - \Gamma_{ss}^-)\langle \hat{b}_s^\dagger \hat{b}_s \rangle \quad (S19)$$
$$+ i \sum_{s' \neq s}^{N_m} \left( \langle \hat{b}_s^\dagger \hat{b}_{s'} \rangle v_{s's}^{(1)} - v_{ss'}^{(2)} \langle \hat{b}_{s'}^\dagger \hat{b}_s \rangle \right) .$$

The first line describes the intrinsic vibrational decay with rate $\gamma_v$ and the thermal excitation at temperature $T$ with the thermal population $n_v^{\text{th}} = [\exp\{\hbar \omega_v / k_B T\} - 1]^{-1}$ ($k_B$ is the Boltzmann constant). The second line describes the optomechanical coupling-induced vibrational pumping and damping with the rates $\Gamma_{ss}^+$, $\Gamma_{ss}^-$ for the individual molecules. The third line describes the coupling to the correlations with other molecules. The equation for $\langle \hat{b}_s^\dagger \hat{b}_{s'} \rangle$ between the $s$-th and $s'$-th molecule ($s \neq s'$) has the following form:

$$\frac{\partial}{\partial t} \langle \hat{b}_s^\dagger \hat{b}_{s'} \rangle = -i(\omega_{s'} - \omega_s)\langle \hat{b}_s^\dagger \hat{b}_{s'} \rangle - \frac{1}{2}(\gamma_{s'} + \gamma_s)\langle \hat{b}_s^\dagger \hat{b}_{s'} \rangle$$
$$+ i \left( v_{s's'}^{(1)} - v_{ss}^{(2)} \right) \langle \hat{b}_s^\dagger \hat{b}_{s'} \rangle + \Gamma_{ss'}^+$$
$$+ i \left( \langle \hat{b}_s^\dagger \hat{b}_s \rangle v_{ss'}^{(1)} - v_{ss'}^{(2)} \langle \hat{b}_{s'}^\dagger \hat{b}_{s'} \rangle \right) \quad (S20)$$
$$+ i \sum_{s'' \neq s, s'}^{N_m} \left( \langle \hat{b}_s^\dagger \hat{b}_{s''} \rangle v_{s''s'}^{(1)} - v_{ss''}^{(2)} \langle \hat{b}_{s''}^\dagger \hat{b}_{s'} \rangle \right) .$$

The first line indicates that the correlation depends on the vibrational frequency difference $\omega_{s'} - \omega_s$ (and thus is a resonant phenomenon) and the dephasing rate $(\gamma_{s'} + \gamma_s)/2$ of the different molecules. In our study, we focus on the same vibrational mode for the molecules, and thus have $\omega_{s'} - \omega_s = 0$ and $(\gamma_{s'} + \gamma_s)/2 = \gamma_v$. The second line describes the change of the frequency and the dephasing rate as given by $\text{Re}(v_{s's'}^{(1)} - v_{ss}^{(2)})$ and $\text{Im}(v_{s's'}^{(1)} - v_{ss}^{(2)})$, respectively, and the pumping of the correlation by the rate $\Gamma_{ss'}^+$. The third line couples the correlation with the vibrational population of individual molecules. The last line couples the correlation with the correlations between other pairs of molecules.

To solve the coupled equations (S19) and (S20), we introduce the abbreviation $\eta_{ss'} = \Gamma_{ss'}^+ + \delta_{ss'} n_v^{\text{th}} \gamma_v$ to rewrite them in a compact form

$$\frac{\partial}{\partial t} \langle \hat{b}_s^\dagger \hat{b}_{s'} \rangle = -\gamma_v \langle \hat{b}_s^\dagger \hat{b}_{s'} \rangle + \eta_{ss'} + i \sum_{s''=1}^{N_m} \left( \langle \hat{b}_s^\dagger \hat{b}_{s''} \rangle v_{s''s'}^{(1)} - v_{ss''}^{(2)} \langle \hat{b}_{s''}^\dagger \hat{b}_{s'} \rangle \right) . \quad (S21)$$

Then, we introduce the vectors and matrices with elements $x_\alpha = \langle \hat{b}_s^\dagger \hat{b}_{s'} \rangle$, $\lambda_\alpha = \eta_{ss'}$, $\Gamma_{\alpha\beta} = \delta_{\alpha\beta}\gamma_v$, $V_{\alpha\beta}^{(1)} = \delta_{ss''} v_{s'''s'}^{(1)}$, $V_{\alpha\beta}^{(2)} = \delta_{s's'''} v_{ss''}^{(2)}$, where the labels $\alpha, \beta$ are defined as $\alpha = s \times N_m + s'$ and $\beta = s'' \times N_m + s'''$ with $N_m$ the number of molecules. With these vectors and matrices, we can rewrite Eq. (S21) in matrix form

$$\frac{\partial}{\partial t} x = -[\Gamma - i(V^{(1)} - V^{(2)})]x + \lambda . \quad (S22)$$

The steady-state solution is simply $x_{ste} = [\Gamma - i(V^{(1)} - V^{(2)})]^{-1} \lambda$, from which we obtain the steady-state values of the vibrational population $\langle \hat{b}_s^\dagger \hat{b}_s \rangle_{ste}$ and the correlations $\langle \hat{b}_s^\dagger \hat{b}_{s'} \rangle_{ste}$ (with $s \neq s'$).

## S2. DFT Calculations of Raman-active Molecular Vibrational Modes

We obtain the vibrational properties of the BPT molecule binding to a single gold atom (see Figure S3 b) by carrying out DFT calculations with the Gaussian 16 package[17], where we utilize the B3LYP hybrid functional, 6-31 G(d,p) basis set for the carbon, sulphur and hydrogen atoms, and the LANL2DZ basis set for the gold atom. Here, we consider the binding with a single gold atom to partially account for the enhancement of the Raman tensor due to charge transfer between the molecule and the gold substrate (known as chemical enhancement of SERS). From the calculations, we identify the three main Raman-active vibrational modes as the dominant peaks measured in the experiment, and summarize their wavenumber and Raman tensor in Table S1.

**Table S1. Wavenumber and Raman tensor components $R_{ij}$** (with $i,j = x,y,z$) in units $\varepsilon_0 \text{Å}^2 \text{ amu}^{-1/2}$ of three Raman-active vibrational modes of the BPT molecule binding with a single gold atom. The wavenumber $\omega_v$ has been scaled by a factor 0.967 for better match with the experiments.

| $\omega_v$ | $R_{xx}$ | $R_{xy} = R_{yx}$ | $R_{xz} = R_{zx}$ | $R_{yy}$ | $R_{yz} = R_{zy}$ | $R_{zz}$ |
|---|---|---|---|---|---|---|
| 1066 cm$^{-1}$ | 3.9 | -0.3 | 11.8 | 11.5 | 16.0 | 141.5 |
| 1269 cm$^{-1}$ | 3.9 | -0.5 | -15.4 | 8.2 | 5.9 | 127.9 |
| 1586 cm$^{-1}$ | 2.6 | 2.7 | -7.5 | -6.5 | 22.9 | 277.1 |

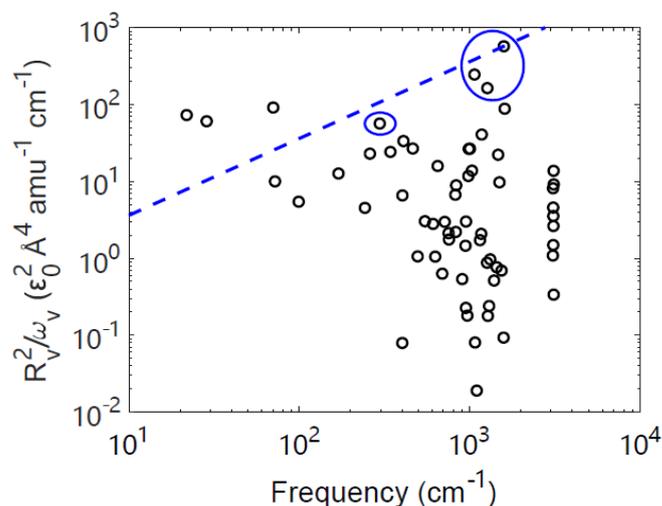

**Figure S4. Raman activity for different vibrational modes in BPT,** with those studied here circled. We note the scaling (Eq. (S34)) of the frequency shifts as $R^2/\omega_v$ which is shown by the dashed line. Note that DFT of the low frequency modes is unreliable, but gives an indication that low frequency modes may also be prone to light-induced shifts.

## S3. Plasmonic Response of the NPoM Nanocavity

We simulate the NPoM nanocavity shown in Figure S3 with the Boundary Element Method (BEM)[18,19] as implemented in the Metal Nanoparticle BEM toolkit[20,21], and present the results in Figure S5. The NPoM nanocavity is formed by a truncated gold nanosphere of diameter 90 nm (slight deviation from experimental size to match resonance wavelength) and a bottom facet of 16 nm radius separated from a gold substrate by a dielectric layer of 1.3 nm thickness and dielectric constant $\varepsilon_g = 2.1$. The dielectric $\varepsilon_{Au}$ of gold is taken from Johnson-Christy[22], and the full system is placed in air. The NPoM nanocavity is illuminated by a plane-wave with an incident angle of 55° with respect to the normal of the substrate, and the scattered Raman photons are collected along the reflected direction.

We observe in Figure S5 (a) two peaks around 830 nm and 670 nm in the computed far-field scattering spectrum (black line), which match those measured in the experiments, and two peaks at the same wavelengths in the near-field spectrum (blue line). We identify these peaks as the (10) and (20) plasmon modes according to the nomenclature in Ref. [23], whose near-field distributions in the nanocavity show anti-nodes in the radial direction as shown in Figure S5(b,c), respectively. The (20) mode is also seen in the experiments as a weak scattering peak [Figure 3(a) in the main text]. Recent work shows that the out-coupling efficiency of the (20) mode is a strong function of the facet shape, and triangular facets (as most prevalent in nanoparticles used here[24]) depress it still further compared to the (10) peak than for circular facets. However, as the simulations show in Figure S5(a), the near field is still large at (20) compared to (10), which is why it is important in the optomechanical coupling below.

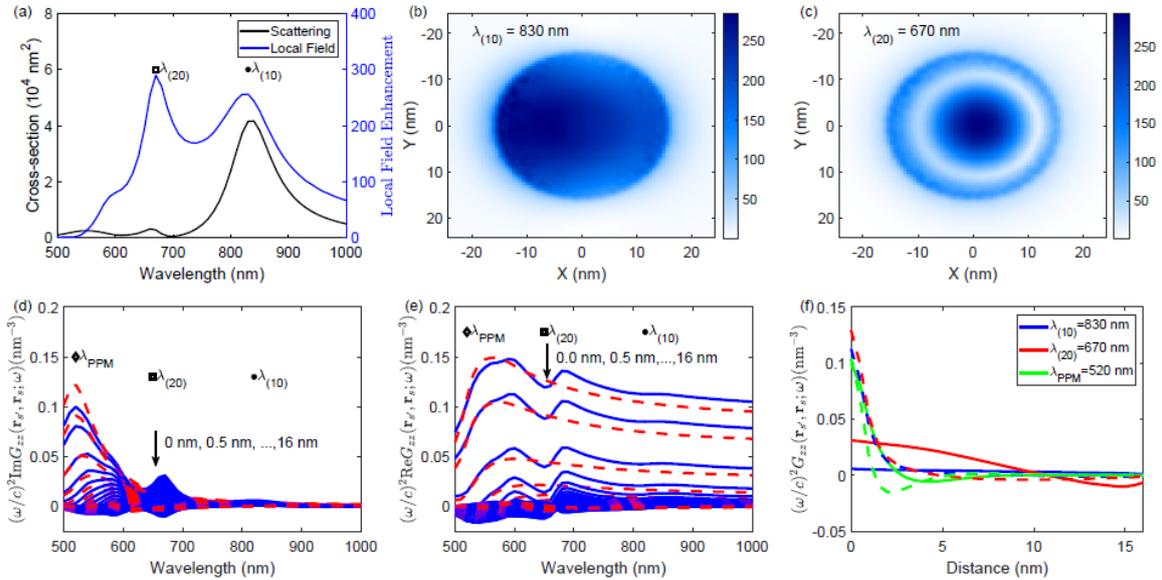

**Figure S5. Simulated electromagnetic response of the NPoM nanocavity.** (**a**) Far-field scattering spectrum (black line, left axis) and absolute value of the near-field enhancement $|E_z/E_0|$ (blue line, right axis) at the centre of the nanocavity. The peaks in the spectra are identified as the (10) plasmonic mode (at $\lambda_{(10)} = 830$ nm) and the (20) plasmonic mode (at $\lambda_{(20)} = 670$ nm). (**b,c**) Absolute value of the electric field enhancement $|E_z/E_0|$ mapping in the $x-y$ plane through the nanocavity centre for the wavelengths $\lambda_{(10)}, \lambda_{(20)}$, respectively. (**d,e**) Imaginary part (d) and real part (e) of the scattered dyadic Green's function $zz$-component $(\omega/c)^2 G_{zz}(\mathbf{r}_{s'}, \mathbf{r}_s; \omega)$ of the NPoM nanocavity (blue solid lines) and of the MIM structure [red dashed lines, obtained with Eq.(S23)] as a function of wavelength $\lambda$ for increasing separation $|\mathbf{r}_{s'} - \mathbf{r}_s|$ in steps of 0.5 nm (with $\mathbf{r}_s$ corresponding to the middle point of the nanogap $(x,y,z)=(0,0,d/2)$ and $\mathbf{r}_{s'}$ situated in the middle plane $z=d/2$). The largest peak in the imaginary part is identified as the Plasmonic Pseudo-Mode (PPM) (at $\lambda_{PPM} = 520$ nm) formed by the overlapping higher order plasmonic modes. (**f**) Real part (dashed lines) and imaginary part (solid lines) of the dyadic Green's function as a function of the separation $|\mathbf{r}_{s'} - \mathbf{r}_s|$ for $\lambda_{(10)}$ (blue lines), $\lambda_{(20)}$ (red lines), and $\lambda_{PPM}$ (green lines).

We see in Figure S5(d) an additional peak at 520 nm in the imaginary part of the dyadic Green's function besides the peaks at the (10) and (20) plasmon modes (note that we always consider the scattered dyadic Green's function). We identify this extra peak as the Plasmon Pseudo-Mode (PPM) formed by the overlapping of many higher order plasmonic modes[23]. Further, the contribution of all modes leads to a large real part of the dyadic Green's function [Figure S5(e)] that varies slowly across a large wavelength range, except for Fano-like features at the wavelengths of the resonant (10) and (20) plasmon modes. Note that Figure S5(d,e) show the evolution of the real (d) and imaginary (e) parts of the Green's function between one molecule at the nanocavity centre and another molecule moving away from the centre. In general, the Green's function decreases with increasing inter-molecular distance.

To better observe this distance dependence, we show in Figure S5(f) the evolution of the real (dashed lines) and imaginary (solid lines) part of the dyadic Green's function as a function of the position of the second molecule, at the wavelengths of the (10) mode (blue lines), (20) mode (red lines), and the PPM (green lines). The real part drops to zero in a short distance ~2.5 nm, while the qualitative behaviour of the imaginary part depends more strongly on the wavelength. The imaginary part for the wavelength of the (10) mode is small and decreases monotonously (blue solid line), while that for the wavelength of the (20) mode is larger but becomes negative for distances larger than 10 nm (red solid line). Both results approximately follow the near-field distribution of the corresponding modes, as shown in Figure S5(b,c), so that their variation with distance is slower than for the real part. By contrast, for the frequency corresponding to the PPM the imaginary part drops to zero in a short distance, ≈2.5 nm, in a similar way as the real part.

In Section S3.1, we derive a semi-analytical expression of the dyadic Green's function of the corresponding Metal-Insulator-Metal (MIM) structure by following the rigorous approach in Ref. [25] [Eq.(S23)] and the image charges method [Eq. (S25)]. The imaginary and real part of Green's function computed with Eq. (S23) are shown with the dashed lines in Figure S5 (d) and (e), respectively, which show great agreement with those for the NPoM nanocavity except for the features around the (10) and (20) plasmon mode. This agreement gives us the opportunity to unravel the physics leading to the large optical spring effect.

## S3.1 Dyadic Green's Function of Metal-Insulator-Metal Structure

To understand the different behaviour of the real and imaginary parts of the dyadic Green's function, we compute it for a metal-insulator-metal (MIM) structure, where the metal is made of gold with permittivity $\varepsilon_{Au}$, and the gap insulator has a permittivity $\varepsilon_g = 2.1$ and thickness of $d = 1.3$ nm (i.e. the gap of the NPoM structure). The two metal-insulator interfaces are located at $z = 0$ and $z = d$, respectively. We are interested in the $zz$-component of the dyadic Green's function $G_{zz}(\boldsymbol{r}_s, \boldsymbol{r}_{s'}, \omega)$ for two molecules located at the middle of the gap $z_s = z_{s'} = d/2$ and separated from each other by a distance $\rho = |\boldsymbol{r}_s - \boldsymbol{r}_{s'}|$. We follow the rigorous procedure in Refs. [25,26] to compute this component

$$G_{zz}(\boldsymbol{r}_s, \boldsymbol{r}_{s'}, \omega) = \frac{i}{4\pi} \int_0^\infty dk_\rho \left[ \frac{k_\rho^3}{k_g^2 k_{gz}} \frac{2 F \exp(ik_{gz} d)}{1 - F \exp(ik_{gz} d)} J_0(k_\rho \rho) \right] . \quad (S23)$$

Here, $F = \frac{\varepsilon_{Au} k_{gz} - \varepsilon_g k_{Auz}}{\varepsilon_{Au} k_{gz} + \varepsilon_g k_{Auz}}$ is the Fresnel reflection coefficient, and $k_j = \sqrt{\varepsilon_j} \omega / c$ is the wave-vector in the gold ($j = Au$) and gap insulator ($j = g$). In addition, $k_\rho$, $k_{jz} = \sqrt{k_j^2 - k_\rho^2}$ are the projection of the wavevector in the radial direction and the z-axis, respectively. By analysing the kernel of the integral in Eq. (S23) in the same way as in Ref. [1], we identify that the Green's function is mainly determined by (i) the propagating surface plasmon mode with radial wavevector $k_\rho$ much larger than $k_{Au}, k_g$, and (ii) the surface wave mode with even much larger wavevector $k_\rho$, which both can be related to the so-called plasmon pseudo-modes[23,27]. Since the radial wavevectors of these modes are much larger than the wavevectors of the metal and the insulator, i.e. $k_\rho \gg k_{Au}, k_g$, we can account for their contribution to the dyadic Green's function by approximating $k_{jz} = \sqrt{k_j^2 - k_\rho^2} \approx ik_\rho$ and carrying out the integral in Eq. (S23) to obtain the approximate analytical expression

$$\left(\frac{\omega}{c}\right)^2 G_{zz}(\boldsymbol{r}_s, \boldsymbol{r}_{s'}, \omega) \approx \frac{1}{2\pi \varepsilon_g} \sum_{j=1}^{J_m} \left(\frac{\varepsilon_{Au} - \varepsilon_g}{\varepsilon_{Au} + \varepsilon_g}\right)^j \frac{2(jd)^2 - \rho^2}{[(jd)^2 + \rho^2]^{5/2}} . \quad (S24)$$

This equation can diverge for $J_m \to \infty$ but we verify that the expression for $J_m = 4$ reprises the full behaviour of the dyadic Green's function of the MIM for distance shorter than $\sim 2.5$ nm (not shown).

A similar expression can also be obtained using an image charge (dipole) method to compute $G_{zz}(\boldsymbol{r}_s, \boldsymbol{r}_{s'}, \omega)$. A vertical dipole $p$ inside the insulator (with dielectric $\varepsilon_g$) near the Au metal substrate ($\varepsilon_{Au}$) at distance $d$ creates two image dipoles inside and outside the metal with amplitudes $\beta p$ and $(1 - \beta) p$,[25] where $\beta = \frac{\varepsilon_g - \varepsilon_{Au}}{\varepsilon_g + \varepsilon_{Au}}$ is the Fresnel reflection coefficient of the dielectric-metal interface in the quasi-static limit. These two image dipoles are required to generate the correct field outside and inside the metal, respectively. Since we are interested in the field outside the metal, we focus below on the first image dipole.

We can apply this image charges method to study the induced field for a vertical dipole in the middle of the MIM gap of thickness $d$. For simplicity, we assume that the initial dipole is located at $z = z_s = d/2$. In this case, the vertical dipole of amplitude $p$ induces two image dipoles inside the upper and lower metal at positions $z - z_s = \pm d$ with amplitude $\beta p$. These image dipoles then create another pair of dipoles inside the metal on the opposite side with amplitude $\beta^2 p$ at positions $z - z_s = \pm 2d$. Following this logic, we get a series of image dipoles with amplitudes $\beta^j p$ at positions $z - z_s = \pm jd$ with $j = 1, 2, \dots \infty$. For the vertical dipole with amplitude $p$ at origin, the z-component of the generated electric field at position $\rho \boldsymbol{e}_x + z \boldsymbol{e}_z$ is given by[28] $E_z = \frac{p}{4\pi\varepsilon_0 \varepsilon_g} \frac{3(z-z_s)^2 - (\rho^2 + (z-z_s)^2)}{[\rho^2 + (z-z_s)^2]^{5/2}}$. Thus, for the image dipoles with amplitude $\beta^j p$ at positions $z - z_s = \pm jd$, we obtain electric fields $E_z^{\pm j} = \frac{\beta^j p}{4\pi\varepsilon_0 \varepsilon_g} \frac{2(jd)^2 - \rho^2}{[\rho^2 + (jd)^2]^{5/2}}$. From this expression, we obtain the total electric field of all the image dipoles $E_z^{\text{tot}} = \sum_{j=1}^{\infty}(E_z^{+j} + E_z^{-j})$, and the $zz$-component of the dyadic Green's function

$$\left(\frac{\omega}{c}\right)^2 G_{zz} = \frac{\varepsilon_0 E_z^{\text{tot}}}{p} = \frac{1}{2\pi\varepsilon_g} \sum_{j=1}^{\infty} \left(\frac{\varepsilon_{Au} - \varepsilon_g}{\varepsilon_{Au} + \varepsilon_g}\right)^j \frac{2(jd)^2 - \rho^2}{[\rho^2 + (jd)^2]^{5/2}} . \tag{S25}$$

Since Eq. (S25) is identical to Eq. (S24), we have proven here the equivalence of the plasmon pseudo-mode (full MIM expression) and the image charges.

We have applied Cauchy's integral theorem to transform the integral along the real radial wave-number[25,29] to evaluate numerically Eq. (S23) and obtain the dyadic Green's function for a MIM structure, see the dashed lines in Figure S5 (d,e). We found that the result of Eq. (S23) agrees both qualitatively and quantitatively with that for the corresponding NPoM nanocavity, see the solid lines in Figure S5 (d,e), except for small features near the (10) and (20) plasmon modes. In addition, we have also verified that the approximated Green's functions [Eqs. (S24) and (S25)] reproduce correctly the wavelength-dependence of the full expression [Eq. (S23)] for zero-distance ($\rho = 0$) and its real part for large distance, but strongly underestimates the imaginary part of the dyadic Green's function for large distance. These results verify that the real part of the Green's function and thus the optomechanical coupling-induced vibrational frequency shift (optical spring effect) can be understood with the plasmon pseudo-mode and, more intuitively, the image charges.

## S4. Optomechanical Parameters: Vibrational Frequency Shift, Damping, Pumping, and Coupling Parameters

We can now combine the calculations in the previous sub-sections to compute the key optomechanical parameters. Figure S6 shows the evolution of $\Omega_{ss'}^{\pm}$ (a-c) and $\Gamma_{ss'}^{\pm}$ (d-f) as functions of the laser wavelength $\lambda_l$ and the inter-molecular distance for the reference laser intensity $I_l = 1\,\mu\text{W}/\mu\text{m}^2$ (note the smaller intensity used here as compared to the main text). Since all these parameters are linearly proportional to the laser intensity, it is straightforward to calculate their values for other laser intensities. Here, we do not analyse the parameters $K_{ss'}^{\pm}$ (with $s = s'$ and $s \neq s'$) since they do not affect the vibrational broadening and shifts.

According to Eq. (S7) and (S8), $\Omega_{ss'}^{\pm}$ and $\Gamma_{ss'}^{\pm}$ are determined by the product of the local electric field at the laser wavelength $\lambda_l = 2\pi c/\omega_l$ (via the induced Raman dipoles) and the dyadic Green's function at the wavelengths $\lambda_{st} = 2\pi c/(\omega_l - \omega_v), \lambda_{as} = 2\pi c/(\omega_l + \omega_v)$ of the Stokes and anti-Stokes lines. Since these quantities reach global or local maxima at the wavelengths $\lambda_{(10)}, \lambda_{(20)}$ of the (10) and (20) plasmon modes, the optomechanical parameters should achieve large values when either $\lambda_l$ or $\lambda_{st}, \lambda_{as}$ match these wavelengths. By matching $\lambda_{st}, \lambda_{as}$ with $\lambda_{(10)}, \lambda_{(20)}$, we can define the on-resonance laser wavelengths $\lambda_{(10)}^{st} = 2\pi c/(\omega_{(10)} + \omega_v), \lambda_{(20)}^{st} = 2\pi c/(\omega_{(20)} + \omega_v)$, and $\lambda_{(10)}^{as} = 2\pi c/(\omega_{(10)} - \omega_v), \lambda_{(20)}^{as} = 2\pi c/(\omega_{(20)} - \omega_v)$.

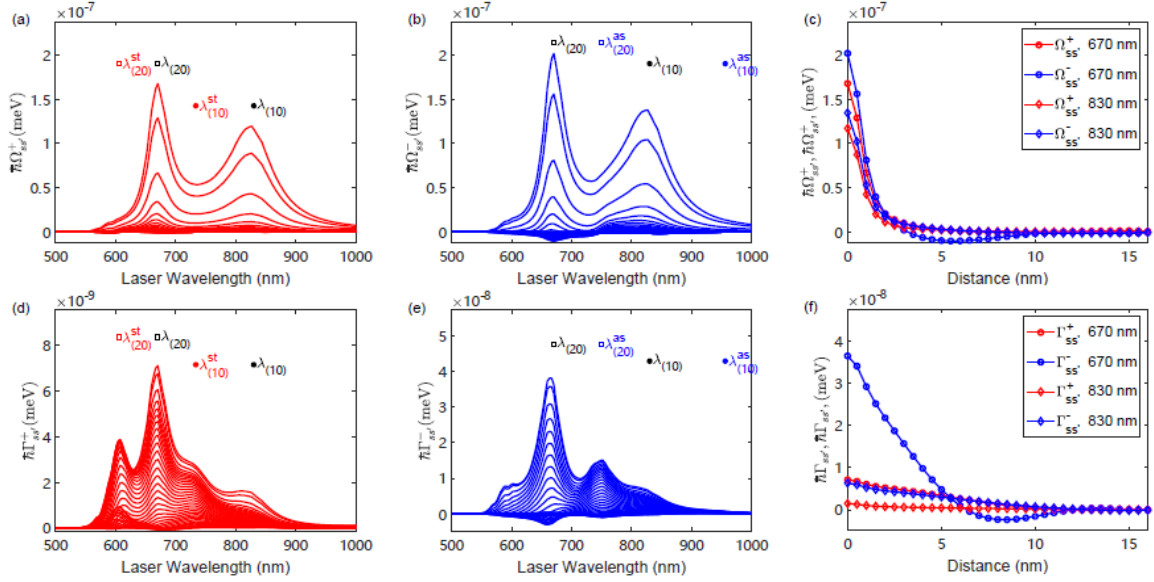

**Figure S6. Key optomechanical parameters for two molecules in the cavity.** (a,b,c) Vibrational frequency shift parameters $\Omega_{ss}^{\pm}$ and coherent coupling $\Omega_{ss'}^{\pm}$, and (d,e,f) vibrational pumping $\Gamma_{ss}^{+}$, damping rate $\Gamma_{ss}^{-}$ and incoherent coupling $\Gamma_{ss'}^{\pm}$, as functions of the laser wavelength for reference intensity $I_l = 1$ µW.µm$^{-2}$ and increasing inter-molecule distance from 0 to 16 nm in steps of 1 nm (from upper to lower lines). Note that the situation of $\Omega_{ss}^{\pm}$, $\Gamma_{ss}^{+}$, and $\Gamma_{ss}^{-}$ (zero inter-molecule distance) thus corresponds to the upper lines in panels a,b,d,e. One of the two molecules is placed at the centre of the gap. Quantities related to the Stokes/ anti-Stokes scattering (with superscripts "+/−") are plotted with red/ blue lines, respectively. (c,f) show these parameters as functions of inter-molecular distance for wavelengths 830 nm and 670 nm corresponding to the (10) and (20) plasmon modes. Results shown here are for the most Raman-active vibrational mode with wavenumber 1586 cm$^{-1}$. Other parameters are specified in the text.

Figure S6 (a,b) indicate that the parameters $\Omega_{ss}^{\pm}$ determining the vibrational frequency shift of individual molecules, show two maxima around the wavelength $\lambda_{(20)} = 670$ nm and $\lambda_{(10)} = 830$ nm, and the dependence of these parameters on $\lambda_l$ resembles that of the local electric field as shown by the blue solid line in Figure S5 (a). This resemblance originates from $\Omega_{ss}^{\pm}$ being determined by the real part of the reflected Green's function $\text{Re}\overleftrightarrow{G}(r_s, r_s; \omega)$ (at the same location) and the square of local electric field, with the former largely insensitive to the wavelength [see Figure S5(e)][1]. Figure S6 (c) shows that the parameters $\Omega_{ss'}^{\pm}$ decrease dramatically in a short inter-molecular distance of 2-3 nm, and become relatively small for larger distance. This dependence resembles that of the real part of the dyadic Green's function shown in Figure S5 (f).

Figure S6 (d,e) indicate that the vibrational damping rate $\Gamma_{ss}^{-}$ shows mainly three peaks at the wavelengths $\lambda_{(20)}$, $\lambda_{(10)}$ and $\lambda_{(20)}^{as}$, where the former two appear since the local field is enhanced by the (20) and (10) plasmon mode, and the latter because the anti-Stokes radiation is resonant with this plasmonic mode. The vibrational pumping rate $\Gamma_{ss}^{+}$ is generally significantly smaller than $\Gamma_{ss}^{-}$ because the dyadic Green's function is much smaller at the longer wavelength $\lambda_{st}$ than at $\lambda_{as}$. Furthermore, $\Gamma_{ss}^{+}$ shows two peaks around the wavelengths $\lambda_{(20)}^{st}$, $\lambda_{(20)}$, which are caused by the maximized Stokes radiation and local field due to the (20) plasmonic mode. The incoherent coupling $\Gamma_{ss'}^{\pm}$ decreases with increasing intra-molecular distance. In particular, $\Gamma_{ss'}^{-}$ reduces much faster than $\Gamma_{ss'}^{+}$ [see Figure S6(f)]. This occurs since the plasmonic modes at shorter wavelength (determining $\Gamma_{ss'}^{-}$) have a narrower spatial distribution than the modes at longer wavelength (dominating $\Gamma_{ss'}^{+}$).

## S5.  Simulations of the Experimental Results

In this section, we analyse in detail the simulations of the Raman signal emitted by the molecules in the NPoM gap. Before discussing the results, we first list the simulation details for the analysis in Figures 3 and 4 of the main text.  The same system is also considered in the next subsections, except when noted otherwise.

As already described in Section S1, we consider the plasmonic structure of a NPoM nanocavity formed by a gold[22] nanoparticle situated a short distance above a gold substrate. The 1.3 nm separation between the two is imposed by the thickness of a monolayer of BPT molecules, characterized by dielectric constant $\varepsilon_g$ = 2.1. The nanoparticle is a truncated 45 nm-radius gold nanosphere that presents a 16 nm radius flat facet at the bottom (i.e. in contact with the molecules). The centre of the gap corresponds to $x = y = 0$ and $z = 0.65$ nm $= d/2$, with $z$ the direction perpendicular to the substrate (see sketch in Figure S3). We characterized the plasmonic response of this cavity (near-field enhancement and dyadic Green's function) in Section S3 and computed the molecular optomechanical parameters in Section S4. For the latter, we consider a continuous-wave laser incoming at 55° angle to the $z$ axis. To introduce the broadening of the Raman lines induced by the spectral width of the pulsed illumination used in the experiments, we set the vibrational decay rate to $\hbar\gamma_v = 2.5$ meV, a significantly larger value than the intrinsic vibrational losses.

The optomechanical parameters serve as the key input to simulate the Raman spectrum. However, while these parameters can be obtained by modelling the molecules as a homogeneous layer of dielectric constant $\varepsilon_g$ = 2.1, the calculation of the Raman spectrum according to the optomechanical theory developed in Section S1 requires explicitly considering an ensemble of individual BPT molecules, which are treated as point-like. Specifically, we place 100 molecules in the intermediate position ($z$ = 0.65 nm plane) between the substrate and the NPoM interfaces, forming a square lattice centred in the middle of the gap $x = y = 0$ [Figure S7(a) inset]. The Raman tensor of the BPT molecules is given in Section S2 and the separation between the molecules is 0.58 nm, as taken from Ref. [30]. The effect of the position and number of the molecules is discussed is Sections S5.3 and S5.4, respectively.

The Raman scattering is calculated by solving Eq. (S16) and (S17). We consider the scattering detected by an infinitesimally small detector situated at the same 55° angle as the illumination, in the reflected direction. Furthermore, we usually analyse i) the integrated 'Raman' signal $S^{int}(\omega_{1586})$ for the 1586 cm$^{-1}$ vibrational mode, defined as the integral of scattering between 1566 cm$^{-1}$ and 1606 cm$^{-1}$, and ii) the integrated 'background' $S^{int}(\omega_{bkg})$ obtained by integrating the signal between 1350 cm$^{-1}$ and 1500 cm$^{-1}$. Notice that this terminology is chosen only to follow the usual distinction made when analysing experimental measurements. However, as discussed in the main text and more in detail below, to understand the behaviour of the system it is crucial to consider that the background also contains a contribution from the fundamental bright collective Raman mode.

## S5.1 Collective Vibrational Modes in SERS Spectrum

Using the SERS expressions explained in Section S1.2 and the calculated key optomechanical parameters in Section S4, we can now compute the SERS spectrum for the system analogous to the one investigated in the experiment. Figure S7(a) shows the Stokes SERS spectrum of the 1586 cm$^{-1}$ vibrational mode for laser illumination at 633 nm of intensity $I_l = 5 \times 10^7$ µW.µm$^{-2}$, which shows two peaks at 1460 cm$^{-1}$ and 1550 cm$^{-1}$. To identify the origin of these peaks, we solve the eigen-problems given by Eq. (S12) to compute the collective vibrational modes. Figure S7(b) shows the wavenumber and the damping (linewidth) of these modes. Using Eq. (S16), we evaluate the contribution of the different collective modes to the Stokes SERS spectrum, and found that the spectrum is mainly determined by the first and sixth collective modes. Thus, we identify these modes as the bright collective modes, and the remaining modes as the dark collective modes. Notably, one of these two bright modes ($\phi$ = 1) experience much larger broadening and shift than the other ($\phi$ = 6). We refer to this $\phi$ = 1 mode as the fundamental collective Raman bright mode. The contribution of the different molecules to this fundamental collective Raman bright mode is illustrated in Figure S7(c).

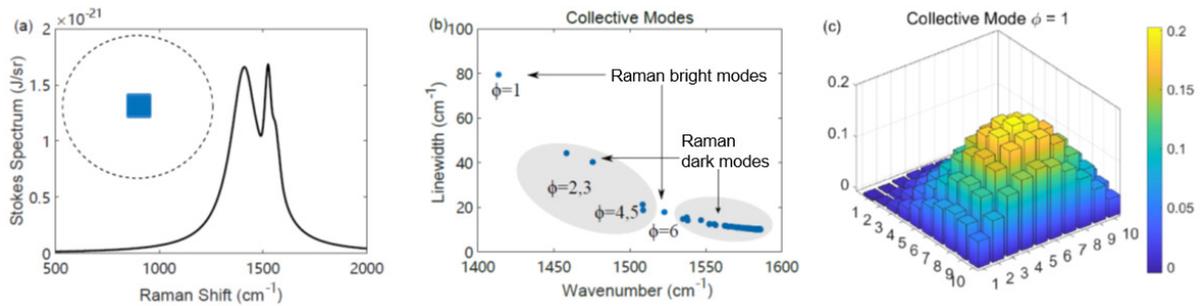

**Figure S7. Collective SERS response** for continuous-wave laser illumination of wavelength 633 nm and intensity $I_l = 5 \times 10^7$ µW.µm$^{-2}$. (**a**) Stokes SERS spectrum and the square lattice of 100 molecules (inset) inside the plasmonic nanocavity facet (dashed circle) with inter-molecular distance 0.58 nm. [30] (**b**) Wavenumber and damping rates of collective vibrational modes $\phi_i$, with classification into bright and dark modes. (**c**) Spatial distribution of the contribution of each molecule in the cavity to the fundamental collective Raman bright mode $\phi_1$ leading to the broad peak in (a). Here, we consider the intrinsic vibrational decay rate $\hbar\gamma_v$ = 2.5 meV, and the temperature T = 293 K (room temperature). Other parameters are specified in the text.

## S5.2 Evolution of the SERS signal with Increasing Laser Intensity

After understanding the SERS response for a given laser intensity, we can now analyse the evaluation of SERS under laser illumination at 633 nm with increasing intensity from 1 µW µm$^{-2}$ to 10$^8$ µW.µm$^{-2}$, as shown in Figure S8. We see that that the Stokes spectrum in Figure S8(a) initially shows a strong increase of the single peak at the vibrational frequency 1586 cm$^{-1}$ of the individual molecules. For even larger intensities, this single peak evolves into two peaks. Both peaks are increasingly red-shifted and broadened, but this effect is considerably stronger for the one at lower Raman shift. After normalizing with the laser intensity, the 1586 cm$^{-1}$ peak actually becomes weaker with increasing laser intensity, as shown in Figure S8(b). We also find qualitatively similar behaviours for the anti-Stokes spectrum (not shown).

The integrated SERS intensity as shown in Figure S8(c) increases linearly with laser intensity for the Stokes scattering (blue solid line), but increases firstly super-linearly and then linearly again for the anti-Stokes scattering (red dashed lines). We note that, in contrast with the analysis of the experimental signal in Figures 2 and 4 of the main text, in Figure S8(c) we are integrating over a larger spectral range, to account for the energy scattered by all collective vibrational modes. These results show that the plasmon-induced vibrational frequency shift and plasmon-mediated coupling re-distribute the SERS spectrum, but do not modify the total integrated intensity. The super-linear increase of the anti-Stokes spectrum for moderate laser intensity is analogous to the situation of vibrational pumping for single molecules, but in this case the strong correlations shown in Figure S8(d) play a key role[31]. Similarly, the linear scaling of the anti-Stokes scattering for large laser intensity is caused by the saturation of the correlations (Eq. (S17)). Thus, following previous work[32] we distinguish three regimes according to the evolution of the average population and correlation: (i) the thermal regime for relatively weak laser intensity (light blue region), where the vibrational population is determined by the thermal value; (ii) the vibrational pumping regime for moderate laser intensity (light red region), where the average population increases and $N_m$ times the average correlation becomes larger than the average population; and (iii) the strong intensity regime (light green region), where the correlation saturates. The Stokes scales linearly because the population and correlations remain much smaller than one.

To compare our numerical results with the experimental results, we proceed as described in the beginning of Section S5 and separate the Raman spectrum into two portions (similar to the treatment of the experimental data) with $S^{int}(\omega_{1586})$ in the wavenumber range 1586 ± 20 cm$^{-1}$ while $S^{int}(\omega_{bkg})$ spans the range 1425 ± 75 cm$^{-1}$ (the 'background'). The integrated intensity of the Raman signal in these two ranges reflects the redistribution of energy between the sharp peak and the broad background, as observed in the experiment. Figure S8(e) shows that the intensity of the sharp peak is much larger than that of the background for relatively small laser intensity, but the situation reverses for laser intensities above 10$^7$ µW/µm$^2$. Figure S8(f) shows similar behaviour for the intensity normalized to the laser intensity, except that (i) the normalized intensity for the Stokes scattering near 1586 cm$^{-1}$ and the background is constant in the vibrational pumping regime while that for the anti-Stokes scattering increases with laser intensity, and (ii) for large intensities the normalized Stokes scattering near 1586 cm$^{-1}$ becomes weaker while the normalized background increases and finally decreases (due to the large spectral shift). These behaviours are precisely what is observed in the experiment. Interestingly, we find similar laser thresholds for the correlation saturation and the qualitative change of the SERS spectrum shape.

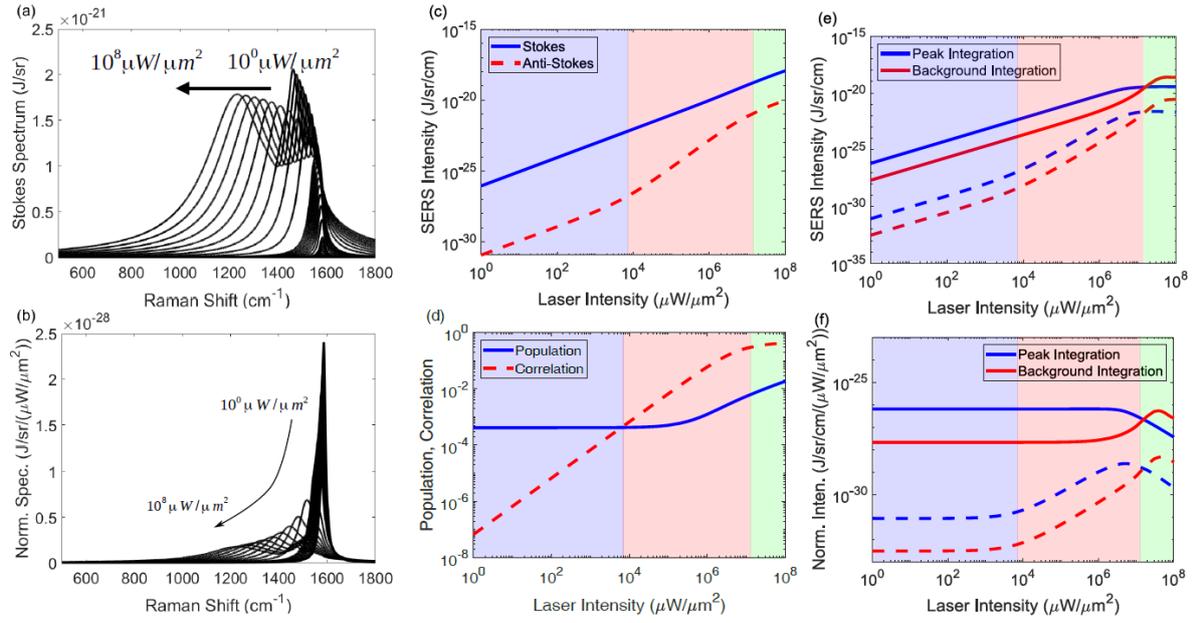

**Figure S8. Evolution of SERS with increasing laser intensity $I_l$** from 1 µW.µm$^{-2}$ to $10^8$ µW.µm$^{-2}$ for a continuous-wave laser of wavelength 633 nm. (**a,b**) Stokes SERS spectrum, and spectrum normalized by $I_l$, for increasing $I_l$. Anti-Stokes spectrum shows qualitatively similar behaviour (not shown). (**c**) Evolution of the integrated Stokes intensity (blue solid line) and anti-Stokes intensity (red dashed line). (**d**) Evolution of the average vibrational population $\sum_{s=1}^{N_m}\langle\hat{b}_s^\dagger\hat{b}_s\rangle/N_m$ (blue solid line) and $N_m$ times the average correlation (red dashed lines) $N_m\left[\sum_{s=1}^{N_m}\sum_{s'\neq s}^{N_m}\langle\hat{b}_s^\dagger\hat{b}_{s'}\rangle/N_m(N_m-1)\right]$. (**e**) Evolution of the integrated SERS signal in the range $1586 \pm 20$ cm$^{-1}$ [blue line, 'peak integration' $S^{int}(\omega_{1586})$] and $1425 \pm 75$ cm$^{-1}$ [red lines, 'background integration' $S^{int}(\omega_{bkg})$], for Stokes (solid lines) and anti-Stokes (dashed lines) signals. (**f**) shows the equivalent results for the SERS signal normalized to the laser intensity. Regions of light blue, red, green backgrounds indicate the thermal regime, the vibrational pumping regime and the correlation saturation regime, respectively. Here, we consider the 1586 cm$^{-1}$ vibrational mode, intrinsic vibrational decay rate $\hbar\gamma_v = 2.5$ meV, and temperature of 293 K (room temperature). Other parameters are same as in Figure S7.

## S5.3 Dependence of the SERS signal on Molecular Positions

After understanding the evolution of the SERS response with increasing laser intensity when the molecules are placed in the centre of the NPoM nanocavity, we now study the effect of the molecular positions (Figure S9). We consider again a patch of $N_m = 100$ molecules in a 10x10 array with 0.58 nm spacing[30], and compare the SERS evolution with laser intensity when the patch is moved from the centre to halfway-out and finally to the edge of the NPoM nanocavity facet. This change reduces the local electric field exciting the molecules, and also reduces the imaginary part of dyadic Green's function at the frequency near to the (20) and (10) modes, but has little effect on this quantity at frequencies near the plasmon pseudo mode, and has almost no effect on the real part of the dyadic Green's function (which instead is mostly affected by the local MIM configuration). Therefore, we expect that the optomechanical parameters (which depend on both the Green's function and the local field) reduce for the patch moved further away from the nanocavity centre. As a result, the laser intensity threshold to reach the saturation of the integrated SERS signal $S^{int}(\omega_{1586})$ and superlinear scaling of the integrated background $S^{int}(\omega_{\text{bkg}})$ (both normalized by the excitation intensity) increases, as shown in Figure S9(b). The change is small when comparing the centre and half-way out position, but very significant for the edge.

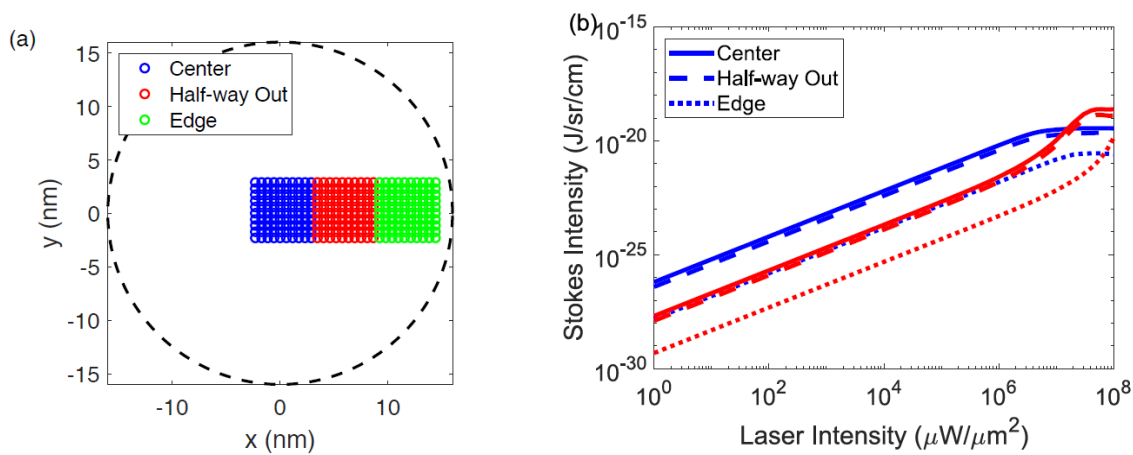

**Figure S9. Dependence of the SERS response on the molecular positions.** (**a**) Effect of changing the position of a square lattice of molecules (points), which is placed around the centre (blue), half-way out (red), or at the edge (green) of the NPoM nanocavity facet (dashed circle). (**b**) Evolution of the integrated Stokes intensity of the sharp peak [blue lines, $S^{int}(\omega_{1586})$) and of the background [red lines, $S^{int}(\omega_{\text{bkg}})$], which are defined in the same way as in Figure S8(e). Solid, dashed and dotted lines are the results for molecules in the centre, halfway-out, and at the edge of the nanocavity facet. Other parameters are as used in Figure S8.

## S5.4 Dependence of the SERS signal on the Number of Molecules

In Figure S10, we study the evolution of the SERS peaks associated with the fundamental (lowest energy) collective Raman bright mode with increasing number of molecules $N_m$ in the middle of the NPoM nanocavity. We see from Figure S10(a) that the Stokes and anti-Stokes lines are blue- and red-shifted, respectively (corresponding to a decrease in vibrational energy), and the shift increases linearly with $N_m$. The periodicity of 10 in small fluctuations of the curves occurs because we add the molecules by following the lattice column from left to right (see inset of Figure S7a) in our simulation. In addition, from this figure we also observe that the linewidth of the Stokes and anti-Stokes lines are identical, comparable to the shift of the Raman line, and also increase linearly with $N_m$. The shift and broadening for 100 molecules are about 32 and 82 times larger than for the single molecule at the edge of the molecular patch, respectively. Furthermore, we study the evolution of the Raman intensity (integrated over the whole spectrum) in Figure S10(b), and observe that the Stokes intensity increases firstly super-linearly and then linearly with $N_m$, and the anti-Stokes saturates with $N_m$.

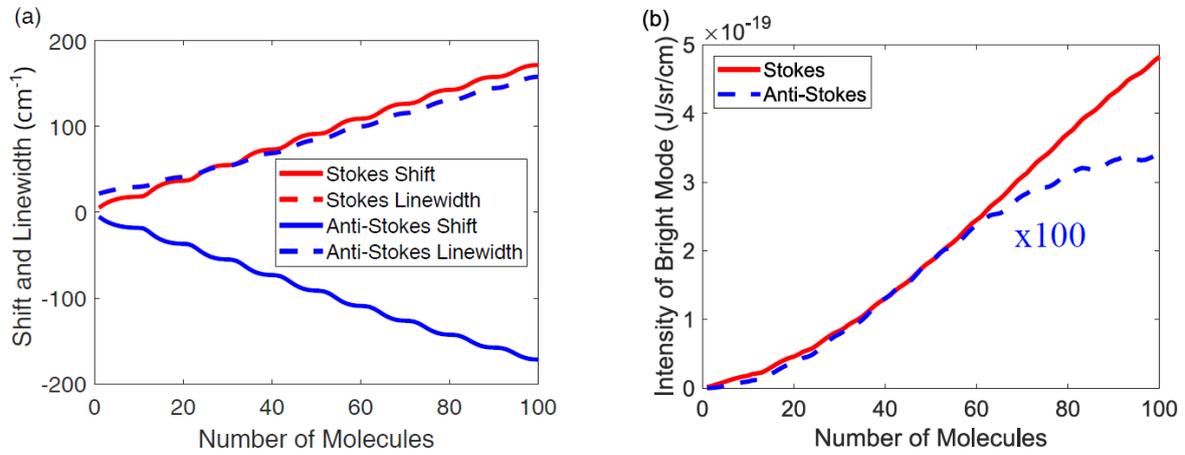

**Figure S10. Influence of the number of molecules $N_m$ on the SERS peak associated with the fundamental (broad) collective Raman bright mode** for a lattice of molecules in the middle of gap. (**a**) Frequency shift $\omega^{st}_{\phi=1} - (\omega_l - \omega_v)$ [$\omega^{as}_{\phi=1} - (\omega_l + \omega_v)$] (solid lines) and the linewidth (dashed lines) of the fundamental (broad) collective bright Stokes (red) and anti-Stokes peak (blue) with $N_m$, while (**b**) shows the evolution of their intensity. In (a) the red dashed line overlaps with the blue dashed line, and thus is invisible. In (b), the blue line is amplified by 100 times for clarity. Other parameters are the same as used in Figure S7.

## S6. Comparison of Continuum-field Model with Single-mode Model

To more intuitively understand the origin of the large optical spring effect, we compare the optical spring effect predicted by the continuum-field and single-mode models, see Figure 1c (main text). To make a fair comparison, we extract the parameters for the single plasmonic mode by comparing the simulated local field enhancement and near-field dyadic Green's function as shown in Figure S5(a,d,e) with those expected from a single-mode model. The prescription for such a comparison has been already reported in our previous article (see Sect. S2 in Ref. [1] Supp. Inf.). For simplicity, we focus on a single molecule in the middle of an NPoM nanocavity, at $r_s$ = (0,0,0.65 nm). In this case, the key optomechanical parameters can be extracted from Eqs. (S7) and (S8) for a single molecule and read

$$\Omega_{ss}^{\pm} = \frac{1}{2\hbar\varepsilon_g}\left(\frac{\omega_l \mp \omega_v}{c}\right)^2 \boldsymbol{p}_s^* \cdot \text{Re}\overleftrightarrow{G}(\boldsymbol{r}_s, \boldsymbol{r}_s; \omega_l \mp \omega_v) \cdot \boldsymbol{p}_s \ , \tag{S26}$$

$$\Gamma_{ss}^{\pm} = \frac{1}{2\hbar\varepsilon_g}\left(\frac{\omega_l \mp \omega_v}{c}\right)^2 \boldsymbol{p}_s^* \cdot \text{Im}\overleftrightarrow{G}(\boldsymbol{r}_s, \boldsymbol{r}_s; \omega_l \mp \omega_v) \cdot \boldsymbol{p}_s \ . \tag{S27}$$

with the induced Raman dipole $\boldsymbol{p}_s = \overleftrightarrow{\alpha}_v \cdot \boldsymbol{E}(\boldsymbol{r}_s, \omega_l)$ according to the continuum-field model. In the single-mode model, these parameters have the same form except that the electric field takes the form

$$\boldsymbol{E}(\boldsymbol{r}_s, \omega_l) \approx \frac{\kappa}{2}\frac{\boldsymbol{u}(\boldsymbol{r}_s)\,\text{EF}\,E_0}{\omega_c - \omega_l - i\kappa/2} \ , \tag{S28}$$

where $\boldsymbol{u}(\boldsymbol{r}_s)$ is the mode function at the molecular position, EF is the maximal field enhancement factor, $\omega_c$ and $\kappa$ are the frequency and damping rate of the single plasmon mode, and the dyadic Green's function takes the form

$$\left(\frac{\omega_l \mp \omega_v}{c}\right)^2 \overleftrightarrow{G}(\boldsymbol{r}_s, \boldsymbol{r}_s; \omega_l \mp \omega_v) \approx \frac{1}{\varepsilon_g V_{\text{eff}}}\frac{\omega_c}{2}\frac{\boldsymbol{u}(\boldsymbol{r}_s) \otimes \boldsymbol{u}^*(\boldsymbol{r}_s)}{\omega_c - (\omega_l \mp \omega_v) - i\kappa/2} \ , \tag{S29}$$

with $\varepsilon_g, V_{\text{eff}}$ the relative dielectric of the NPoM gap and the effective mode volume, respectively.

In the single-mode description of the plasmon, the effective mode volume is inversely proportional to $\text{EF}^2$. Accounting for the linear dependence of the local field on EF [see Eq. (S28)] then leads to the $\text{EF}^4$ dependence of the optomechanical parameters, in agreement with the previous theory based on the single plasmon mode description.

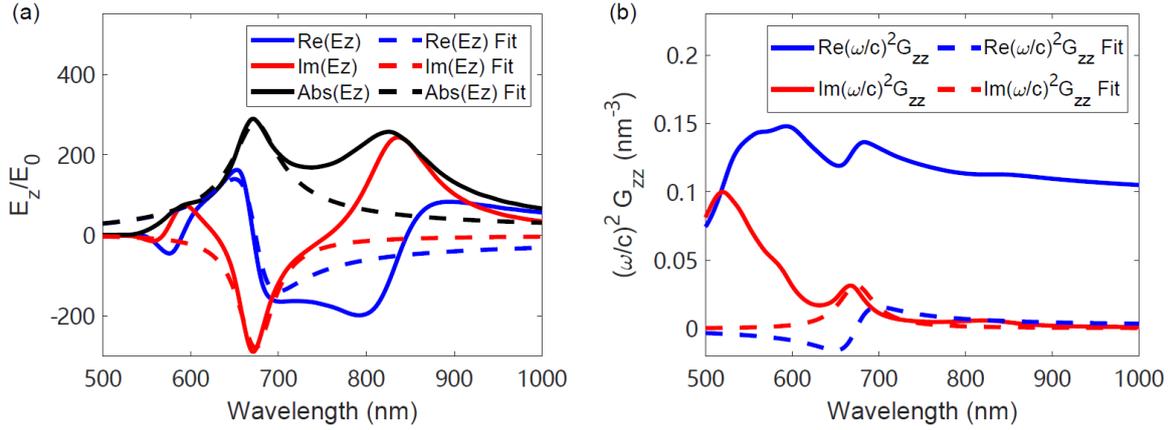

**Figure S11. Local electric field enhancement ($z$-component, a) and dyadic Green's function ($zz$-component, b) for a single molecule in the middle of the NPoM nanocavity.** In (a), the blue, red, black solid lines are the real, imaginary part and absolute value of the field enhancement from the electromagnetic simulations, and the corresponding dashed lines are the single-mode fits according to Eq. (S28). In (b), the blue and red solid line are the real and imaginary parts of the dyadic Green's function obtained from the electromagnetic simulations, and the red and blue dashed lines are the fits according to Eq. (S29) for the single-mode model.

Figure S11 shows the local field enhancement and the dyadic Green's function from the electromagnetic simulations used in the continuum-field model (solid lines) and fits to Eqs. (S28) and (S29) for the single-mode model (dashed lines). For this comparison, we only consider the dominant $z$ and $zz$ components. Here, we focus on the (20) plasmon mode because this mode dominates the simulated local field enhancement of the NPoM under 633 nm and 658 nm excitation. Because of this choice, we do not reproduce the features in the local-field enhancement around 830 nm associated with the (10) plasmonic mode. Importantly, Eq. (S29) captures the peak of the imaginary part of the Green's function around $\lambda \sim 670$ nm corresponding to the (20) plasmon mode, but is unable to correctly capture the real part which is strongly underestimated. From these fits, we extract the parameters related to the single plasmon mode: the plasmon energy $\hbar\omega_c \approx 1.84$ eV (or wavelength $\lambda_c \approx 674$ nm), the plasmon damping rate $\hbar\kappa_c = 135$ meV, the maximal local field enhancement EF $\approx 280$, the plasmon-laser coupling $\hbar\Omega \approx 1.3$ meV (for laser intensity 1 µW.µm$^{-2}$), and the single-plasmon optomechanical parameter $\hbar g \approx 0.06$ meV for the 1586 cm$^{-1}$ vibrational mode. These fitting parameters are also utilized in Section S9 to estimate the impact of the anharmonicity on the molecular optomechanics.

**Table S2. Parameters extracted by fitting the continuum-field model with the single-mode model for the (20) plasmon.**

| $\hbar\omega_c$ for (20) plasmonic mode | $\hbar\kappa_c$ | $\hbar\Omega$ for laser at 1 µW/µm$^2$ | $\hbar g$ for 1586 cm$^{-1}$ |
|---|---|---|---|
| 1.84 eV | 135 meV | 1.3 meV | 0.06 meV |

Notice that although these parameters are useful to better situate molecular optomechanics in the context of traditional cavity optomechanical setups, they only give a limited view of the system under study. To fully understand the emitted Raman signal (and in particular the large value of the spring shift) it is necessary to include the full modal structure of the plasmonic NPoM.

Figure S12 shows the key optomechanical parameters $\Omega_{ss}^{\pm,s}$, $\Gamma_{ss}^{\pm,s}$ vs laser wavelengths $\lambda_l = 2\pi/\omega_l$ in the single-mode model (dashed lines), and their comparison with $\Omega_{ss}^{\pm,c}$, $\Gamma_{ss}^{\pm,c}$ in the continuum-field model (solid lines). The values of $\Omega_{ss}^{\pm,c}$ are about a factor 10 larger than $\Omega_{ss}^{\pm,s}$ because of the larger value of the real part of the Green's function (Figure S11b). Further, we see that $\Omega_{ss}^{\pm,s}(\omega_l)$ both approach a maximum around $\lambda_c$ but with opposite sign, which results in a small total frequency shift $[\Omega_{ss}^{+,s} + \Omega_{ss}^{-,s}]/2$ due to cancelation of frequency shifts from combining both Stokes and anti-Stokes scattering. In contrast, we see that $\Omega_{ss}^{\pm,c}$ both approach a maximum around $\lambda_c$ but have the same sign, which results in a much larger total frequency shift than in the single cavity mode description, due to the addition of frequency shifts produced from Stokes and anti-Stokes scattering in this case. Extra peaks are also seen around 810 nm for $\Omega_{ss}^{\pm,c}$ arising from the (10) plasmon mode, which is not accounted for in the single-mode model.

We also see that $\Gamma_{ss}^{+,s}$ approaches a maximum around $\lambda_c \approx 664$ nm and another for $2\pi/(\omega_c + \omega_v) \approx 601$ nm, which are mainly due to the maximal local field enhancement and a maximum of the enhancement of the Stokes emission. In a similar manner, $\Gamma_{ss}^{-,s}$ approaches a maximum around $2\pi/(\omega_c - \omega_v) \approx 742$ nm, which is mainly due to the maximum in the enhancement of the anti-Stokes emission. Importantly, $\Gamma_{ss}^{+,c}$ resembles $\Gamma_{ss}^{+,s}$ for wavelengths shorter than 600 nm, but is much larger than $\Gamma_{ss}^{+,s}$ for laser wavelengths larger than 700 nm, which is due to the absence of the (10) mode in the single-mode model. In contrast, $\Gamma_{ss}^{-,c}$ is much larger than $\Gamma_{ss}^{-,s}$ for all laser wavelengths considered, which for low wavelengths is due to the strong influence of the plasmon pseudo-mode.

Overall, the above results indicate the importance of accounting for the full plasmon response while dealing with the optomechanical effects from molecules in the NPoM nanocavity. Only the continuum-field model is reliable to capture the full response and thus is suitable for the quantitative study of molecular optomechanics.

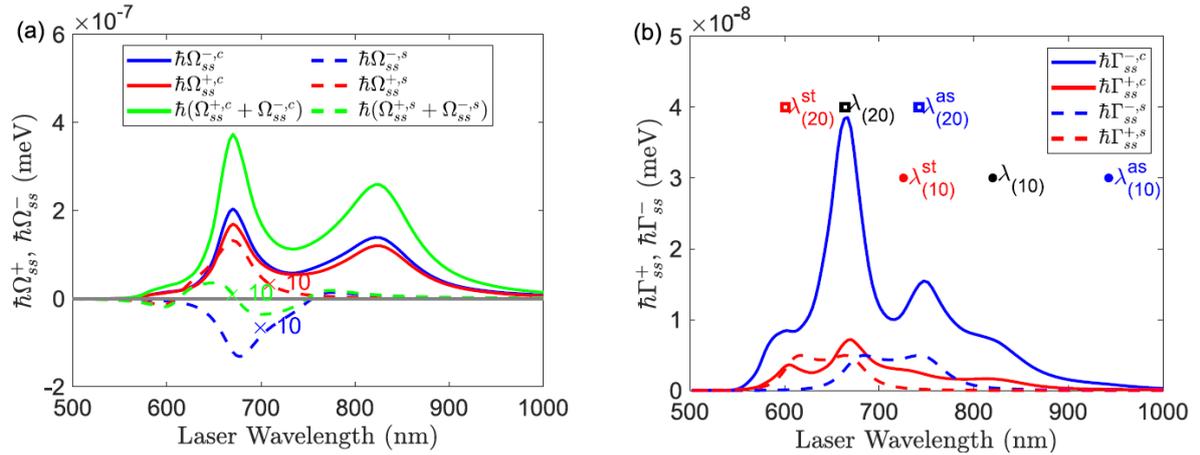

**Figure S12. Vibrational frequency shift parameters $\Omega_{ss}^{\pm}$ (a) and vibrational pumping $\Gamma_{ss}^{+}$/damping $\Gamma_{ss}^{-}$ rate (b) for a single molecule in the middle of the NPoM nanocavity.** In (a), the blue and red lines are $\Omega_{ss}^{-}$, $\Omega_{ss}^{+}$ associated with the anti-Stokes and Stokes scattering, respectively, and the green lines are the sum $\Omega_{ss}^{-} + \Omega_{ss}^{+}$. In (b), the blue and red lines are $\Gamma_{ss}^{-}$, $\Gamma_{ss}^{+}$, respectively, and the wavelengths $\lambda_k, \lambda_k^{st}, \lambda_k^{as}$ indicate the peaks or shoulders caused by the maximal local field enhancement, and the enhancement of the Stokes, anti-Stokes emission due to the $k = (10)$ or (20) plasmon mode. In all panels, the solid and dashed lines are the results according to the continuum-field model and the single-mode model, which are labelled by "c" and "s", respectively.

## S7. Effective Description of Raman Lineshift and Analytic Estimates

The results in Section S5.4 indicate that the Raman lineshift and broadening are linearly proportional to the number of molecules $N_m$ for a given laser intensity, at least up to relatively large $N_m$. Based on this observation, in this section, we derive simple expressions for the Raman lineshift and broadening to help understand the evolution of these quantities. Our starting point is the Raman lineshift $\Delta_N$ and broadening $\Gamma_N$ for a single molecule $N = 1$. The former can be derived from Eq. (S7) as

$$\Delta_1 = \frac{1}{2}(\Omega_{ss}^+ + \Omega_{ss}^-) = \frac{1}{4\hbar\varepsilon_0} \, \boldsymbol{p}_s^* \cdot \left[ \begin{array}{l} \left(\frac{\omega_l - \omega_v}{c}\right)^2 \mathrm{Re}\overleftrightarrow{G}(\boldsymbol{r}_s, \boldsymbol{r}_s; \omega_l - \omega_v) \\ + \left(\frac{\omega_l + \omega_v}{c}\right)^2 \mathrm{Re}\overleftrightarrow{G}(\boldsymbol{r}_s, \boldsymbol{r}_s; \omega_l + \omega_v) \end{array} \right] \cdot \boldsymbol{p}_s \, , \quad \text{(S30)}$$

In our system, the local field is mostly polarized along the z-direction, and the Raman-active vibrational modes have the largest value for the zz-component. Thus, the induced Raman dipoles have also their largest value along the z-direction, i.e. $|\boldsymbol{p}_s| \approx |p_{sz}| = |\alpha_{v,zz} E_z(\omega_l)|$ with Raman polarizability $\alpha_{v,zz} = \sqrt{\hbar/2\omega_v} \, R_{v,zz}$ and the local field component $E_z(\omega_l) = \mathrm{EF} \, E_0$ where EF is the enhancement factor. In this case, in the above expression, only the zz-component of the Raman tensor $G_{zz}$ is important.

Using the fact that $(\omega/c)^2 \, \mathrm{Re} G_{zz}(\boldsymbol{r}_s, \boldsymbol{r}_s; \omega)$ only varies slightly around a large positive number [see Figure 1(c) in the main text, and Figure S5(e) for molecular distance zero] and applying Eq. (S24) or (S25) with the dominant term $j = 1$ and $\rho = 0$, we arrive at

$$\Delta_1 \approx \frac{1}{\omega_v} \frac{1}{2\pi\varepsilon_0^2 c \varepsilon_g d^3} \, \mathrm{Re}\left(\frac{\varepsilon_{Au}(\omega_l) - \varepsilon_g}{\varepsilon_{Au}(\omega_l) + \varepsilon_g}\right) R_{v,zz}^2 \, \mathrm{EF}^2 \, I_l \, , \quad \text{(S31)}$$

where we introduced the laser intensity $I_l = \frac{1}{2}\varepsilon_0 c E_0^2$. For the system with $N_m$ molecules in the centre of the NPoM nanocavity, we thus predict that the shift is

$$\Delta\omega_v = \Delta_N \approx \eta_1 N_m \Delta_1 \approx \frac{\eta_1(\rho) N_m}{\omega_v} \frac{1}{2\pi\varepsilon_0^2 c \varepsilon_g d^3} \, \mathrm{Re}\left(\frac{\varepsilon_{Au}(\omega_l) - \varepsilon_g}{\varepsilon_{Au}(\omega_l) + \varepsilon_g}\right) R_{v,zz}^2 \, \mathrm{EF}^2 \, I_l \, , \quad \text{(S32)}$$

where $\eta_1(\rho)$ is a factor that accounts for the effective coupling to the broad collective Raman bright mode depending on molecular spacing $\rho$ and geometrical factors. This coupling can be determined by fitting the exact results in Figure S10(a) for a given laser wavelength, laser intensity and arrangement of the molecules. To observe the frequency shift $\Delta_N$, we require that $\Delta_N$ is larger than the intrinsic vibrational decay rate $\gamma_v$, from which we can determine the laser threshold intensity $I_c$ to observe the SERS suppression.

We also need to understand how this molecular coupling changes as the intermolecular distance $\rho$ increases. From Figure S5(f), we find an analytic estimate $\mathrm{Re} G_{zz}(\rho) \simeq \mathrm{Re} G_{zz}(0,0) \exp\{-(\rho/\delta)^2\}$ with $\delta = d/1.13$ found by fitting the full computed result. Thus $\eta_1(\rho) \approx \eta_1' \exp\{-(\rho/\delta)^2\}$, giving the simple expression for the spring shift from Eq. (S32) which is used in the main text,

$$\Delta_N \approx e^{-(\rho/\delta)^2} \frac{\eta_1' N_m}{\omega_v} \frac{1}{2\pi\varepsilon_0^2 c \varepsilon_g d^3} \, \mathrm{Re}\left(\frac{\varepsilon_{Au}(\omega_l) - \varepsilon_g}{\varepsilon_{Au}(\omega_l) + \varepsilon_g}\right) R_{v,zz}^2 \, \mathrm{EF}^2 \, I_l \, , \quad \text{(S33)}$$

or explicitly including unit conversions

$$\Delta_N \, [\mathrm{cm}^{-1}] \sim \frac{1}{\omega_v[\mathrm{cm}^{-1}]} \frac{10^{-5}}{m_u(2\pi c)^3} \frac{\eta_1' N_m}{d^3 \varepsilon_g} e^{-(\rho/\delta)^2} \mathrm{Re}\left(\frac{\varepsilon_{Au} - \varepsilon_g}{\varepsilon_{Au} + \varepsilon_g}\right) \mathrm{EF}^2 \frac{R^2}{12} \, I_l[10^6 \mu\mathrm{W}/\mu\mathrm{m}^2] \quad \text{(S34)}$$

for $m_u$ = 1.67x10$^{-27}$ kg, gap size $d$ in nm, and $R = \sqrt{12} R_{v,zz} \sim 960$ (in units of $\epsilon_0 \text{Å}^2/\sqrt{\text{amu}}$) for the Raman activity $R^2$ of the 1586 cm$^{-1}$ line, taking also $\eta_1 \sim 0.12$ from Figure S10(a) (with $\rho$ = 0.58 nm). This can be simply inverted to give the intensity threshold $I_c$ when $\Delta_N > \gamma_v \sim 20$ cm$^{-1}$, given in the main text, where $\gamma_v$ is the Raman spectral resolution beyond which any shift can be observed.

Since the linewidth of the Stokes and anti-Stokes lines also increases linearly with the laser intensity, as shown in Figure S10 (a), and the linewidth of a Lorentzian-shape spectrum is associated with the total decay rate of the vibrational collective mode, the vibrational pumping rate of the collective Raman mode $\Gamma_N^\pm = \eta_2 N_m \Gamma_{N=1}^\pm$ is simply proportional to $N_m$ times the value for single molecule $\Gamma_{N=1}^\pm$. The proportional factor $\eta_2$ plays the same role as $\eta_1$. According to Eq. (S7), we obtain

$$\Gamma_{N=1}^\pm = \frac{1}{2\hbar\varepsilon_0} \boldsymbol{p}_s^* \cdot \left(\frac{\omega_l \mp \omega_v}{c}\right)^2 \text{Im}\overleftrightarrow{G}(\boldsymbol{r}_s, \boldsymbol{r}_s; \omega_l \mp \omega_v) \cdot \boldsymbol{p}_s \ . \tag{S35}$$

Using the same argument for the Raman induced dipoles and the local electric field as well as the Green's function, we can further simplify the above expression and obtain

$$\Gamma_N^\pm \approx \eta_2 \frac{N_m}{2\pi\varepsilon_0^2 c \omega_v \varepsilon_g d^3} \text{Im}\left(\frac{\varepsilon_{Au}(\omega_l \mp \omega_v) - \varepsilon_g}{\varepsilon_{Au}(\omega_l \mp \omega_v) + \varepsilon_g}\right) R_{v,zz}^2 \text{ EF}^2 I_l \ . \tag{S36}$$

We note that this estimate neglects the contribution of the low-energy NPoM modes to the dyadic Green's function, which introduces a larger correction to $\Gamma_N^\pm$ as compared to the spring shift.

If we assume that the collective bright vibrational mode works like a giant vibrational mode with enhanced vibrational pumping and damping rate $\Gamma_N^\pm$, the population of this mode should satisfy the same equation as for a single molecule

$$n = n_v^{\text{th}} + \frac{\Gamma_N^+ - n_v^{\text{th}}(\Gamma_N^- - \Gamma_N^+)}{\gamma_v + (\Gamma_N^- - \Gamma_N^+)} \ . \tag{S37}$$

where the thermal vibrational population and decay $n_v^{\text{th}}, \gamma_v$ are assumed to be the same as those for single molecules (see Section S4 in Ref. [31]). For moderate laser intensity, the second part of the above equation can be simplified as $\Gamma_N^+/\gamma_v$, which describes the population increase due to vibrational pumping. Thus, the vibrational pumping regime is achieved when $\Gamma_N^+/\gamma_v$ becomes comparable with the thermal population $n_v^{th}$. From this condition, we can define a second laser intensity threshold $I_v$ for vibrational pumping.

The ratio between the laser intensity thresholds for SERS saturation and vibrational pumping can be then estimated as

$$\frac{I_v}{I_c} \approx n_v^{th} \frac{\eta_1 \text{Re}\left(\frac{\varepsilon_{Au}(\omega_l) - \varepsilon_g}{\varepsilon_{Au}(\omega_l) + \varepsilon_g}\right)}{\eta_2 \text{Im}\left(\frac{\varepsilon_{Au}(\omega_l - \omega_v) - \varepsilon_g}{\varepsilon_{Au}(\omega_l - \omega_v) + \varepsilon_g}\right)} \ . \tag{S38}$$

For closely-spaced molecules ($\rho < d$) and assuming $\eta_1 \sim \eta_2$, then taking $n_v^{th} \approx 4 \times 10^{-4}$ for the 1586 cm$^{-1}$ vibrational mode, as well as using $\varsigma = \text{Re}\left(\frac{\varepsilon_{Au}(\omega_l) - \varepsilon_g}{\varepsilon_{Au}(\omega_l) + \varepsilon_g}\right) / \text{Im}\left(\frac{\varepsilon_{Au}(\omega_l - \omega_v) - \varepsilon_g}{\varepsilon_{Au}(\omega_l - \omega_v) + \varepsilon_g}\right) \sim 68$, we obtain that $I_v \sim 3 \times 10^{-2} I_c$. In other words, the laser threshold to observe the vibrational pumping is about two orders of magnitude smaller than that to reach the SERS saturation. This is verified by the results in Figure S8.

Since in this regime $n \sim \Gamma_N^+/\gamma_v$ and at $I_c$, $\Delta_N \sim \varsigma \Gamma_N^+ \sim \gamma_v$, we find that at the spring-shift threshold $n_c \sim 1/\varsigma$. As a result, even at these large spring-shifts, there is still not a significant population in the vibrational mode, which is less than 2% occupied. Equating this to an equivalent thermal population, $\hbar\omega_v = k_b T \ln(\varsigma)$ shows that for mode frequencies $\omega_v < \ln(\varsigma) \times 202$ cm$^{-1} \sim 850$ cm$^{-1}$, this model shows that Raman frequency shifts could occur even before vibrational pumping above the thermal background is observed (depending on the size of $R$).

Note that these estimates do not include the nanocavity plasmonic resonances which give peaks in the broadening term Im$G$, but they capture the main MIM nanocavity effects which result from molecular dipoles very close to the metal interfaces. Examining the simple expressions for these shows how the Raman lineshifts dominate increasingly for off-resonant tuning (Figure S13).

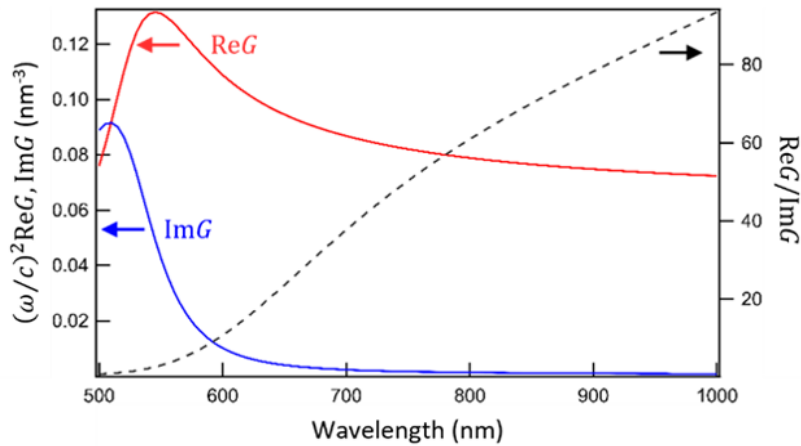

**Figure S13. Comparison of real (red) and imaginary (blue) part of the Green's function without nanocavity plasmon contribution.** The ratio of real to imaginary part is shown in black, dotted, for low frequency vibrations.

The optomechanical parametric instability threshold has been shown to occur only in very specific conditions in this system, as typically $\Gamma_v^{opt} < 0$ for only some $v$ and only for specific pump wavelengths[1]. By contrast, the optical spring shifts are broadband. In general, the ratio $\rho = \Delta\omega / \Gamma_v^{opt}$ between the optomechanical shifts $\Delta\omega$ and broadening/narrowing $\Gamma_v^{opt}$ depends additionally on contributions from the extra resonances of the resonant plasmonic modes, but the underlying dependence without these follows (in the single molecule approximation for low frequency vibrations $\omega_v$)

$$\rho = \frac{\text{Re}\{\varsigma\}}{2\omega_v \, \text{Im}\{d\varsigma/d\omega\}} \quad . \tag{S39}$$

Even below 550 nm, the shift is 10-fold greater than the broadening for $\omega_v = 100$ cm$^{-1}$ (see Figure S14). We note that the shift dominates proportionately more for low wavenumber modes.

Properly including optomechanical vibrational mode shifts shows that conventional optomechanical parametric instabilities are difficult to reach in plasmonic systems. Overall thus the hierarchy of thresholds for the different regimes observed in the experiments is vibrational pumping first, and then Raman frequency shifting, at hundred-fold higher power but with vibrational occupations of only 1%.

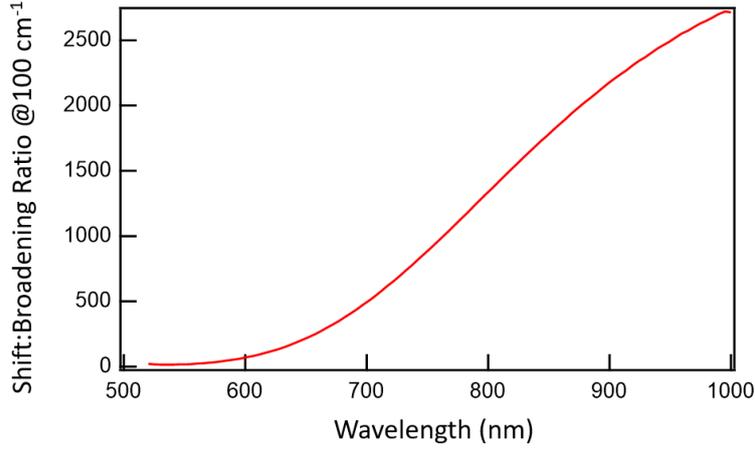

**Figure S14. Ratio of optomechanical shift $\Delta\omega$ to broadening $\Gamma_v^{opt}$.** From Eq. (S39), which is derived in a simple approximation for only very low wavenumber modes.

## S8. Raman Redshift Contributions from Local Dipoles and NPoM Modes

It is helpful to estimate the relative ratio of the local image dipole contributions compared to the nanocavity plasmon contributions to the red-shift. Since this is related to dipole-dipole interactions either from local image charges, or via the plasmon nanocavity mode which is spread over the cavity, we can very roughly estimate their relative contributions as being inversely proportional to their volumes. For the local images, the model above gives the lateral radius $\delta \sim d$ and thus

$$V_{local} \sim d.\pi d^2 \tag{S40}$$

Previously[28], we have estimated that the NPoM plasmonic modes have volumes

$$V_{nanocavity} \sim d.\pi \left(\frac{\sqrt{Dd}}{n}\right)^2 \tag{S41}$$

for NP diameter $D$. The ratio of Raman frequency shifts for the light-dressing is thus in this approximation

$$\frac{\Delta\omega_{local}}{\Delta\omega_{nanocavity}} \sim \frac{V_{nanocavity}}{V_{local}} \sim \frac{Dd}{nd^2} \sim \frac{D}{nd} \sim 41 \tag{S42}$$

which indeed shows how redshifts are dominated by the local dipole-dipole interactions. We note that nanoparticle faceting acts to increase the nanocavity mode volume, but also increases the number of molecules which are optomechanically coupled.

## S9. Raman Redshift from Vibrational Anharmonicity

To consider the role of vibrational anharmonicity we next present a numerical model of an optomechanical molecular system considering an anharmonic molecular vibration. We model the vibrational anharmonic potential-energy surface as a Morse potential of the following form:

$$V(q) = D\left(1 - e^{-\alpha(q-q_0)}\right)^2 \ , \tag{S43}$$

where $D$ is the dissociation energy characterising the oscillator, $\alpha$ contains the information about the oscillator stiffness, and $q_0$ is the equilibrium position of the oscillator in the mass-weighted coordinates. The vibrational energy levels of the bound states of a Morse oscillator are

$$E_n = \hbar\omega_v'\left[\left(n + \frac{1}{2}\right) - \chi\left(n + \frac{1}{2}\right)^2\right] \tag{S44}$$

with $= \sqrt{\omega_v'^2/2D}$, $\chi = \hbar\omega_v'/4D$, and $\lambda = \sqrt{2D}/\alpha\hbar$. There is only a finite number of bound states $\max(n) < \lambda - 1/2$. Note that the frequency of the observed vibrational transition at low temperature is thus connected with $\omega_v'$ via $\omega_v = \omega_v'(1 - 2\chi)$. The Morse potential (black) is shown in Figure S15 as a function of the normalized coordinate $q/q_Q$, with $q_Q = \sqrt{\hbar/(2\omega_v')}$, for two different values of the anharmonicity parameter $\chi$: (a) $\chi = 5 \times 10^{-4}$, and (b) $\chi = 5 \times 10^{-3}$. The vibrational wave functions $\psi_i(q)$ are depicted in blue and are vertically offset by their respective eigenenergies.

We obtain the oscillator wave functions by numerically solving the 1D Schrödinger equation for the Morse oscillator defined in the mass-weighted coordinate $q$ of the vibrational mode. Using the wave functions $\psi_i(q)$, we numerically obtain the matrix elements of the position operator $\hat{q}$ as:

$$\langle n|\hat{q}|m\rangle = \int \psi_n^*(q)\, q\, \psi_m(q)\, dq \ , \tag{S45}$$

We consider the optomechanical Hamiltonian containing the anharmonic Morse oscillator in the form:

$$\hat{H}_{omM} = \hat{H}_M + \hbar\omega_c \hat{a}^\dagger \hat{a} + \frac{\hbar\Omega}{2}\left(\hat{a}^\dagger e^{-i\omega_l t} + \hat{a} e^{i\omega_l t}\right) + \hbar g_{om} \hat{a}^\dagger \hat{a} \hat{q}/q_Q \ , \tag{S46}$$

where $\hat{H}_M = \sum_n E_n |n\rangle\langle n|$. We note that the normalized operator $\hat{q}/q_Q$ corresponds to the operator $\hat{b} + \hat{b}^\dagger$ considered for the harmonic vibration.

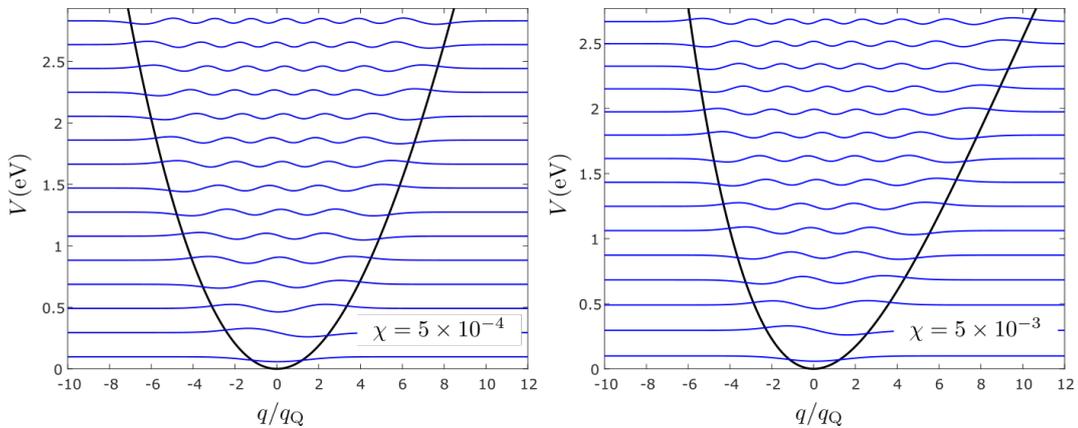

**Figure S15. The Morse potential used to model the anharmonic effects in molecular optomechanics.** The potential (black) and the corresponding vibrational wave functions (blue) offset by their respective vibrational frequencies are shown for (**a**) $\chi = 5 \times 10^{-4}$, and (**b**) $\chi = 5 \times 10^{-3}$.

Plasmonic and vibrational damping is included by the following Lindblad superoperators acting on the density operator $\hat{\rho}$ of the the system:

$$\mathcal{L}_{\hat{a}}^{-}[\hat{\rho}] = \left(1 + n_{th}(\hbar\omega_c;T)\right)\frac{\kappa_c}{2}\left(2\hat{a}\hat{\rho}\hat{a}^{\dagger} - \{\hat{a}^{\dagger}\hat{a},\hat{\rho}\}\right), \tag{S47}$$

$$\mathcal{L}_{\hat{a}^{\dagger}}^{-}[\hat{\rho}] = n_{th}(\hbar\omega_c;T)\frac{\kappa_c}{2}\left(2\hat{a}^{\dagger}\hat{\rho}\hat{a} - \{\hat{a}\hat{a}^{\dagger},\hat{\rho}\}\right), \tag{S48}$$

$$\mathcal{L}_{M}^{-}[\hat{\rho}] = \sum_{i<j}\left(1 + n_{th}(E_j - E_i;T)\right)\frac{\gamma_{ij}}{2}\left(2|i\rangle\langle j|\hat{\rho}|j\rangle\langle i| - \{|j\rangle\langle j|,\hat{\rho}\}\right), \tag{S49}$$

$$\mathcal{L}_{M}^{+}[\hat{\rho}] = \sum_{i>j}n_{th}(E_i - E_j;T)\frac{\gamma_{ij}}{2}\left(2|i\rangle\langle j|\hat{\rho}|j\rangle\langle i| - \{|j\rangle\langle j|,\hat{\rho}\}\right), \tag{S50}$$

where $\gamma_{ij} = \gamma_v|\langle i|\hat{q}/q_Q|j\rangle|^2$, $n_{th}(E;T) = (\exp(E/k_BT) - 1)^{-1}$, with $k_B$ being Boltzmann's constant and $T$ the bath temperature.

The emission spectra are calculated using the quantum regression theorem as:

$$P(\omega) \propto \omega^4 \mathrm{Re}\left\{\int_0^{\infty}\langle \hat{a}^{\dagger}(0)\hat{a}(t)\rangle e^{i\omega t}dt\right\}. \tag{S51}$$

We calculate the SERS spectra considering the vibrational mode of frequency $\omega_{v1} = 1586$ cm$^{-1}$, the optomechanical coupling $\hbar g_{om} = 0.06$ meV, the vibrational damping rate $\hbar\gamma_v = 2.5$ meV, the plasmon mode energy $\hbar\omega_c = 1.84$ eV, the plasmon damping rate $\hbar\kappa_c = 135$ meV, and the laser frequency $\omega_l = \omega_c + \omega_v$. The pumping amplitude $\Omega = 1.3$ meV corresponds to pump intensity $I_l = 1$ µW/µm². Here, the plasmon mode corresponds to the (20) plasmon mode and the related parameters are obtained by fitting the near-field and dyadic Green's function with those expected in the single-mode model, see Sec. S6 for more details. The anharmonicity is parametrized by $\chi$ which we estimate to range between $\chi = 5 \times 10^{-4}$ and $\chi = 5 \times 10^{-3}$. In Figure S16 we show the spectra calculated for illumination intensities linearly increasing from $I_l = 10^5$ µW/µm² to $I_l = 10^7$ µW/µm² for (a) $\chi = 5 \times 10^{-4}$, and (b) $\chi = 5 \times 10^{-3}$. The SERS peak moves linearly with increasing power and shifts by (a) $\sim 10$ cm$^{-1}$ and (b) $\sim 33$ cm$^{-1}$ to lower vibrational frequencies when the intensity is increased. The intensity-dependent shift emerges due to the non-zero diagonal elements of the operator $\hat{q}$.

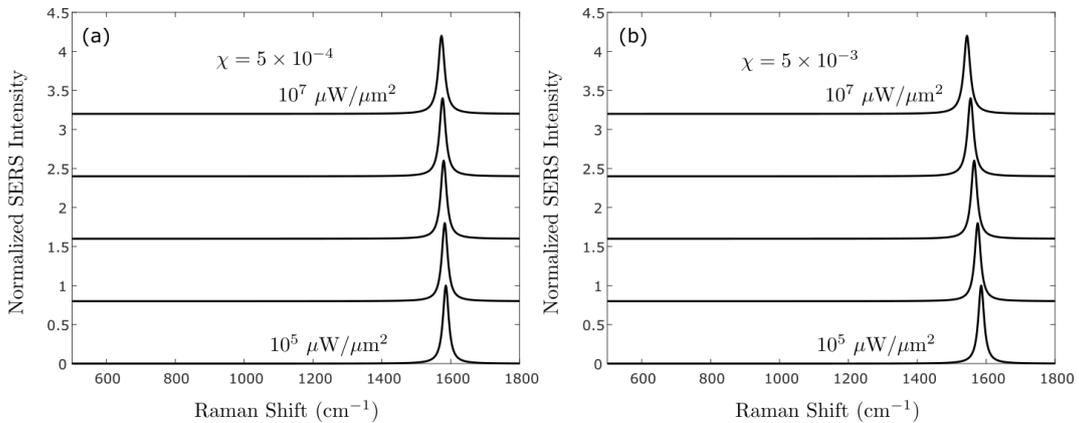

**Figure S16. SERS spectra as a function of illumination intensity $I_{las}$.** (a) The anharmonicity parameter $\chi = 5 \times 10^{-3}$, and (b) $\chi = 5 \times 10^{-4}$. Here we consider $T = 0$ K.

We integrate the total emitted power in the SERS peak and plot the result as a function of the illumination intensity to test whether the total emitted power saturates as the illumination intensity is increased. The result is shown in Figure S17 for $\chi = 5 \times 10^{-4}$ and $\chi = 5 \times 10^{-3}$ in (a) and (b), respectively. In both cases, the intensity-dependent integrated SERS power increases linearly with illumination intensity.

Lastly, we study the effect of temperature on the SERS shift. We fix the illumination intensity $I_l = 10^5\,\mu W/\mu m^2$ and linearly increase the temperature from $T = 0$ K to $T = 1000$ K. We perform the calculation for both $\chi = 5 \times 10^{-4}$ and $\chi = 5 \times 10^{-3}$ and plot the results in Figure S18 (a) and (b), respectively. The peak shift as a function of temperature is negligible.

This shows that transiently exciting molecules in solution to high $n$[33–35] can give increasingly red-shifted Raman peaks[36–39], arising from vibrational anharmonicity. From literature estimates biphenyl peak shifts with temperature[35,40,41] $\hbar\omega_v(T) = \hbar\omega_v - 2\chi k_B T$ we estimate $\chi \sim 5 \times 10^{-3}$, tenfold larger than $\chi \sim 5 \times 10^{-4}$ extracted from DFT of the $\omega_{v1} = 1586$ cm$^{-1}$ ring breathing mode of BPT in solution. Molecular heating would thus only give shifts of $\hbar\Delta\omega = 2\chi k_B T_e < 5$ cm$^{-1}$ (Figure S18), while vibrational pumping can account for a shift of up to 35 cm$^{-1}$ (Figure S16). Both contributions are far smaller than the >150 cm$^{-1}$ observed in the experimental data presented in the main text. This model including anharmonicity thus cannot reproduce our results, instead suggesting that a new interaction must be present with a much larger energy scale, as analysed in the previous sections from local image dipoles at the metal surface.

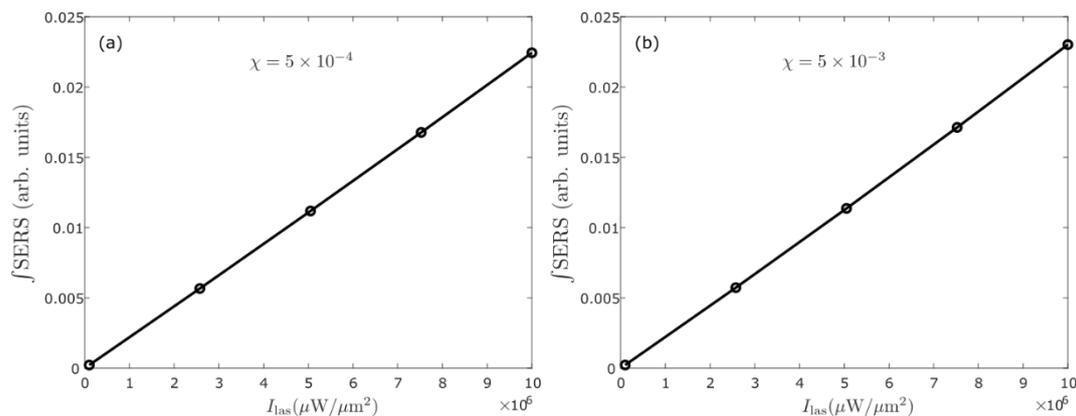

**Figure S17. Intensity dependence of the integrated SERS power.** The dots correspond to the spectra shown in Figure S16. (**a**) $\chi = 5 \times 10^{-3}$, and (**b**) $\chi = 5 \times 10^{-4}$.

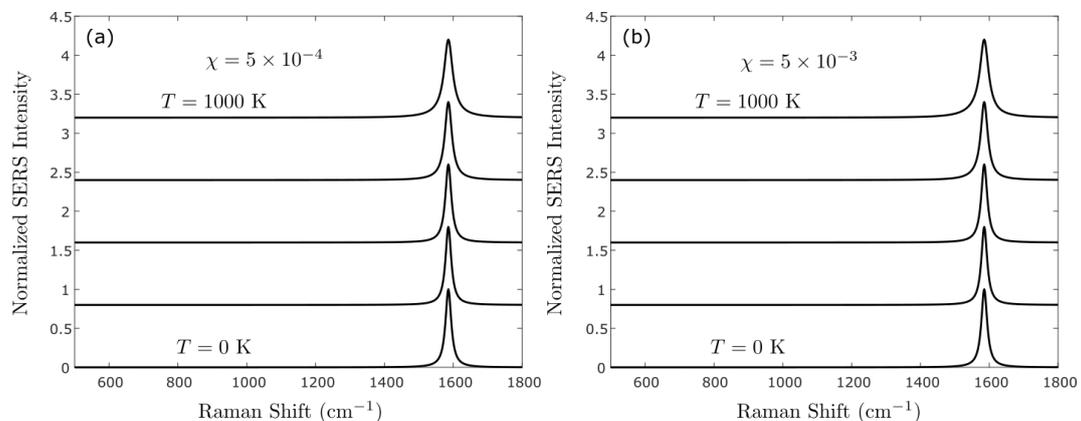

**Figure S18. SERS spectra as a function temperature.** (**a**) $\chi = 5 \times 10^{-3}$, and (**b**) $\chi = 5 \times 10^{-4}$. We considered $I_{las} = 10^5\,\mu W/\mu m^2$ in all calculations.

# Supplementary Experimental Data

## S10. NPoM In-coupling Correction

In the main text, we record power-dependent SERS spectra from hundreds of NPoMs. In order to compare NPoMs ($i$) with slightly different size and shape which varies their excitation and collection efficiencies, we use a normalization based on the integrated SERS counts at $\omega_{v1}$ for the lowest intensity $I_0$, $\eta^i = S^i(\omega_{v1}, I_0)/I_0$. Given average coupling $\bar{\eta} = \sum_i \eta^i$, we then normalize the in-coupled intensity for each NPoM as $I_{in}^i = I_l \cdot \eta^i/\bar{\eta}$ (with units of integrated counts.s$^{-1}$.µW$^{-1}$). This accounts for the in-coupling efficiency so at the same low input power, each NPoM gives the same normalized SERS emission. The effectiveness of this correction is demonstrated in Figure S19. Without correction of in-coupling efficiency, particles show SERS signal varying by up to one order of magnitude at the same laser power. Only by correcting the in-coupled laser intensity, it is possible to compare many different nanostructures and obtain the common trend.

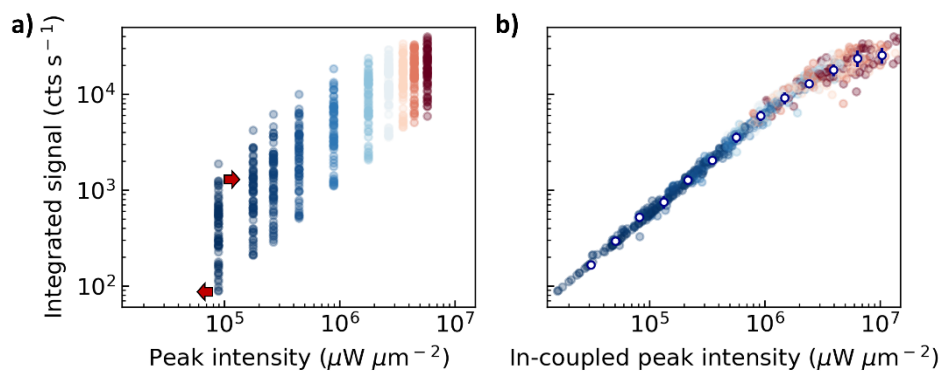

**Figure S19. In-coupling correction.** (**a**) Integrated SERS from $\omega_{v1}$ = 1586 cm$^{-1}$ mode vs. actual laser intensity on sample, excited by pulsed 658 nm laser. Power series run on >60 particles vary by up to one order of magnitude in signal intensity at same laser intensity. Red arrows indicate correction of in-coupling efficiency by renormalising laser intensity for each particle with signal at lowest intensity. (**b**) Integrated SERS vs. in-coupling corrected laser intensity. All particles now coincide to form a linear trend before SERS saturation at high laser intensity. Data as shown in main text Figure 4(e).

## S11. Raw Data of Individual NPoMs

Here, we provide spectral data from individual NPoM structures for the three investigated laser wavelengths. These are part of the dataset of hundreds of NPoMs that is presented in the main text in an averaged way by correcting for the in-coupling efficiency. The optical spring effect can indeed also be observed in data from individual nanostructures. NPoMs shown here are representative for the entire dataset, however particles with high in-coupling efficiency are selected since weakly in-coupling particles do not reach the intensity regime required for a clear optical spring shift.

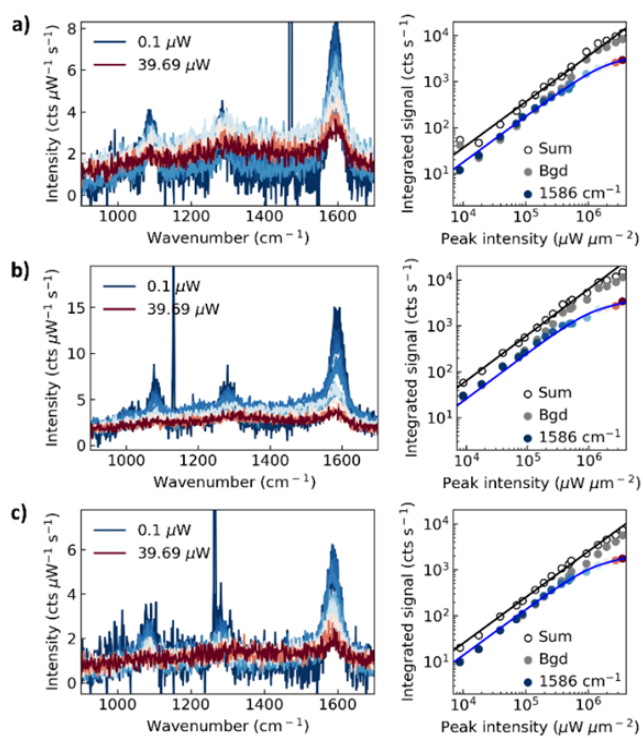

**Figure S20. Power series of individual NPoM structures (633 nm).** (**a-c,** left) Laser power-dependent SERS spectra of three NPoM structures with 633 nm laser. (**a-c,** right) Integrated signal of SERS line (1586 cm$^{-1}$, blue dots), background (grey dots) and sum of SERS line and background (black open dots). SERS line follows optomechanical model (blue line) while sum increases linearly with power (black line).

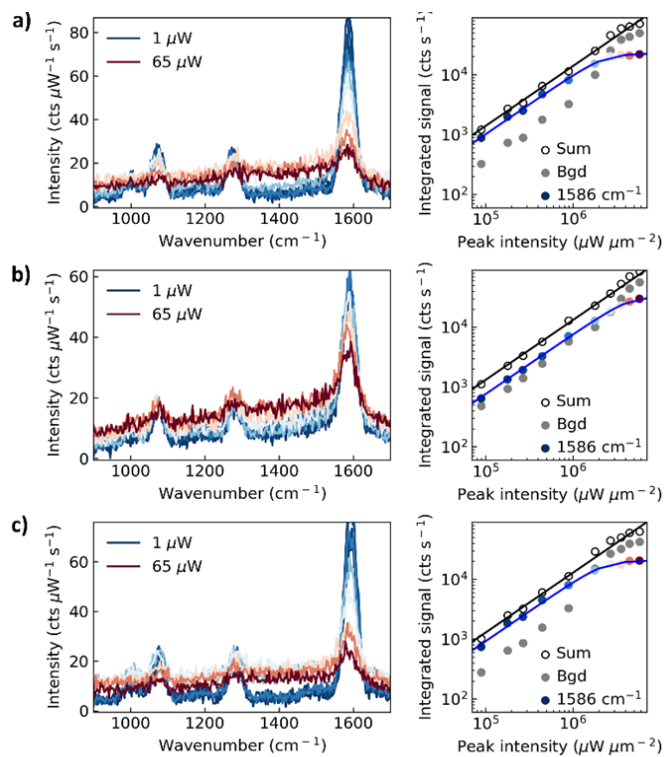

**Figure S21. Power series of individual NPoM structures (658 nm).** (**a-c,** left) Laser power-dependent SERS spectra of three NPoM structures with 658 nm laser. (**a-c,** right) Integrated signal of SERS line (1586 cm$^{-1}$, blue dots), background (grey dots) and sum of SERS line and background (black open dots). SERS line follows optomechanical model (blue line) while sum increases linearly with power (black line).

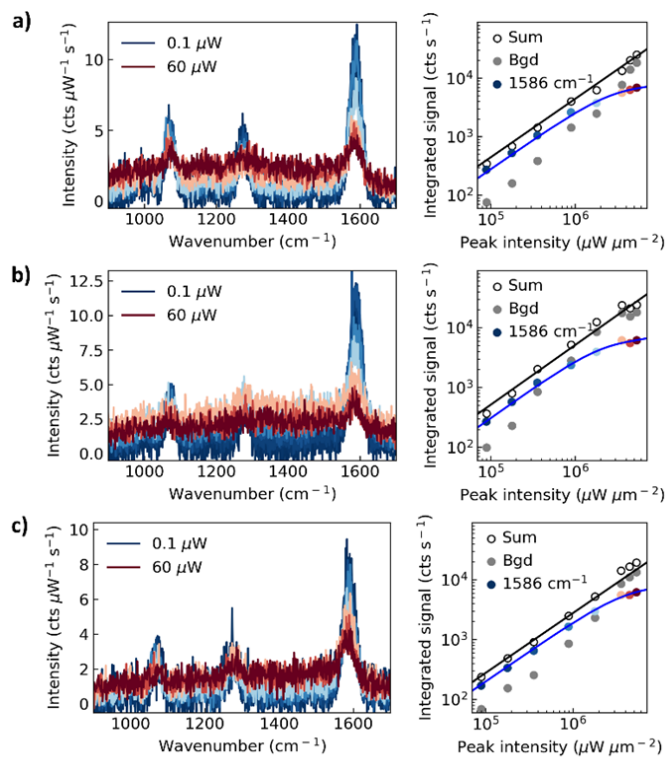

**Figure S22. Power series of individual NPoM structures (700 nm).** (**a-c,** left) Laser power-dependent SERS spectra of three NPoM structures with 700 nm laser. (**a-c,** right) Integrated signal of SERS line (1586 cm$^{-1}$, blue dots), background (grey dots) and sum of SERS line and background (black open dots). SERS line follows optomechanical model (blue line) while sum increases linearly with power (black line).

## S12. Reversibility of Saturation and Damage

To confirm the saturation behaviour and eliminate damage or drift as the cause, two additional experiments are performed. First, power-dependent SERS measurements from low to high power are run twice on each NPoM [Figure S23(a,b)]. Both power sweeps show the same saturation effect caused by the optical spring shift with only a small offset due to damage. Hence, we can confirm the reversibility of the optical spring shift and rule out damage to the nanostructures as an explanation for the SERS saturation.

To quantify the relationship between SERS suppression and damage, measurement series alternating pump intensities between 1 and 20 µW.µm$^{-2}$ are performed [Figure S23(c)] (only the high intensity causes damage). The % suppression in SERS efficiency $\tilde{S}$ from low to high power is much larger than the % damage that reduces recovery after returning from high to low power. The alternating powers show two distinct features of the integrated SERS signal: a small decay (~15%) is caused by accumulating damage [yellow arrows Figure S23(c,d)], while a large suppression (~50%) of the signal comes from SERS saturation (red arrows). The % changes for damage and suppression are estimated from the maximum of the distribution of particles in Figure S23d. (Note: In this experimental procedure, it is not possible to correct for each NPoM's in-coupling efficiency. This leads to a wide range of values for the SERS suppression as molecules experience different effective laser powers. Further, outliers at positive suppression are possible in rare cases due to restructuring of the nanoparticles.)

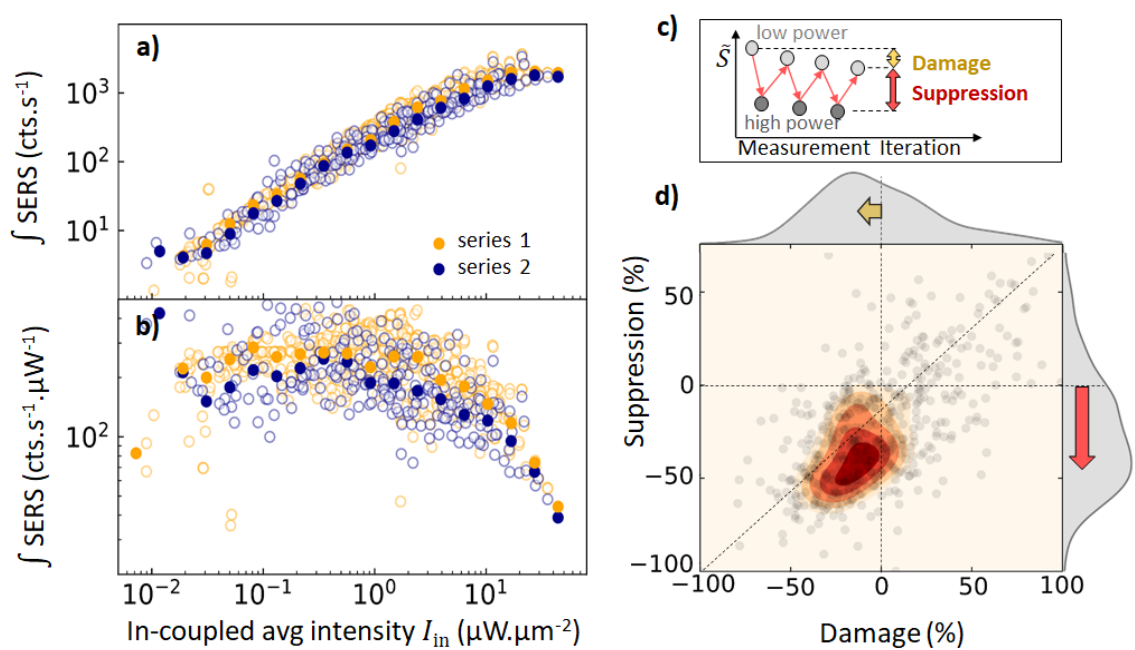

**Figure S23. Repeatability of SERS saturation.** (**a**) Integrated SERS from $v_1$ = 1586 cm$^{-1}$ mode, excited by pulsed 633 nm laser. Power series run twice on each NPoM show saturation in both runs. Filled dots indicate average of measurements on multiple NPoMs (white dots). (**b**) Integrated SERS normalized to in-coupled power and integration time. (**c**) Measurement protocol for SERS efficiency, defining damage vs suppression. (**d**) Extracted suppression of time series on different NPoMs at average power $P_{in}$ = 15 : 100 µW.µm$^{-2}$ (low:high) showing 50% suppression clearly distinguished from damage. We note that outliers here skew the perception of the distribution as most data points overlap in the red shaded area. The histograms on the outside of the plot give a clear representation of the distribution of data points.

# S13. Vibrational Pumping and Non-equilibrium Temperatures

Tuning the pump laser wavelength to 785 nm and using a notch filter, we simultaneously acquired Stokes and anti-Stokes spectra of hundreds of particles [Figure S24(a)]. While the SERS saturation of the 1586 cm$^{-1}$ reported in the main text is observed on the Stokes side (weakly, see below), the anti-Stokes emission scales quadratically before also showing saturation at the same threshold power [Figure S24(c)]. The super-linear scaling of the anti-Stokes scattering demonstrates clearly that vibrational pumping is active at these applied laser powers[42,43]. For a low wavenumber vibration [here 280 cm$^{-1}$, Figure S24(b)], no clear vibrational pumping is seen due to its higher thermal phonon population and the lower Raman cross section. Integrating over the entire spectrum, a linear dependence of the Stokes signal on power without saturation is observed [Figure S24(d)], consistent with the model proposed below (see Figure S8) and again eliminating damage as potential cause of the saturation effect.

Compared to the Stokes spectra presented in the main text, the saturation of the SERS lines and increase in the background can only be observed very weakly. There are two reasons for this observation: firstly, our theory predicts a weaker spring shift at 785 nm laser wavelength compared to 658 nm due to the positions of plasmonic modes [Fig. 2(a) of the main text]. Hence, the power to reach considerable changes in the spectrum will be higher. Secondly, to record anti-Stokes spectra integration times had to be extended fourfold. To avoid damage, the maximum laser intensity in this experiment is thus threefold lower than presented in the main text. Therefore, the intensity regime for SERS saturation is not quite reached in Figure S24.

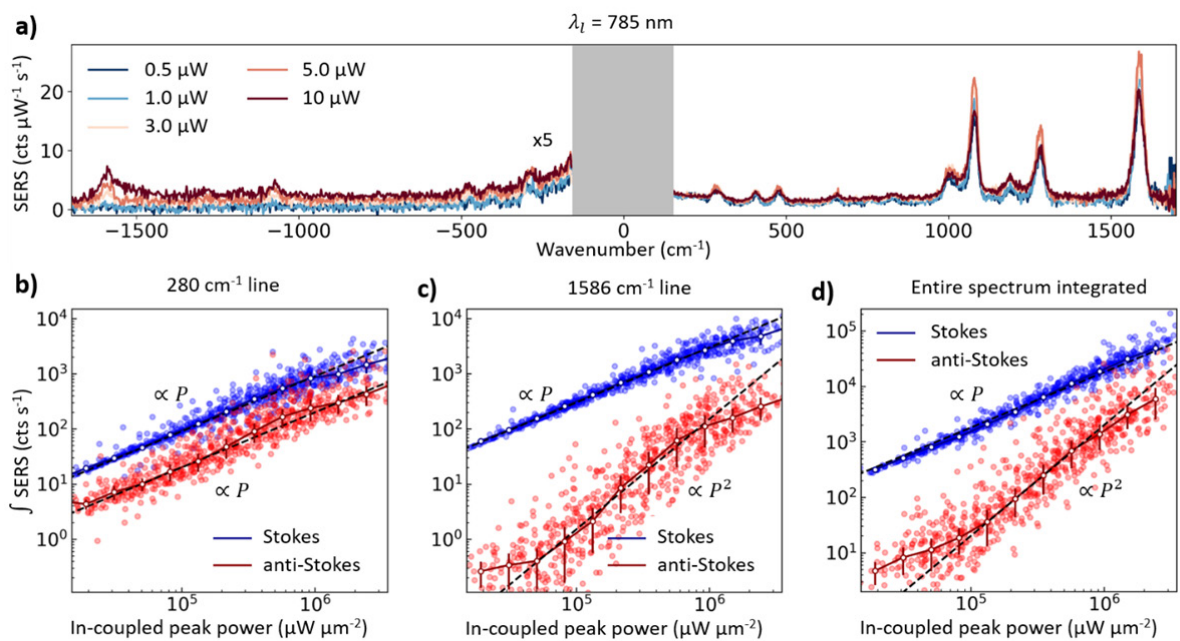

**Figure S24: Comparison of Stokes and anti-Stokes scattering.** (**a**) Power dependent anti-Stokes (x5) and Stokes SERS spectra of one NPoM with pump laser at 785 nm. (**b,c,d**) SERS emission integrated over (b) 280 cm$^{-1}$ line, (c) 1586 cm$^{-1}$ line and (d) entire spectrum for the Stokes (blue) and anti-Stokes (red) sides. Guides to the eye (black, dotted) indicate linear and quadratic scaling with power. White dots and coloured lines indicate average of measurements on multiple NPoMs.

We extract an 'apparent temperature' in the nanostructure during each laser pulse. Note that given the vibrational pumping observed, this is a non-equilibrium temperature. We use the ratio of Stokes to anti-Stokes SERS peak areas, at pump wavelength 633 nm. However, the anti-Stokes signal is very weak (at powers needed to ensure no damage) making only the strongest 1586 cm$^{-1}$ line suitable for estimating the molecular temperature [Figure S25(a,b)]. Due to vibrational pumping, the vibrational population is increased far above the thermal population, from 320 K at 0.44 µW.µm$^{-2}$ to 740 K at 55 µW.µm$^{-2}$. Hence, the estimated temperatures are not representative of the molecular environment temperature. Such excitations similarly affect the apparent (non-equilibrium) temperature of free electrons in the metal leading to nonlinear changes in the SERS background from electronic Raman scattering. Since this study is focused on changes of the molecular SERS signal, we omit these effects in the main text.

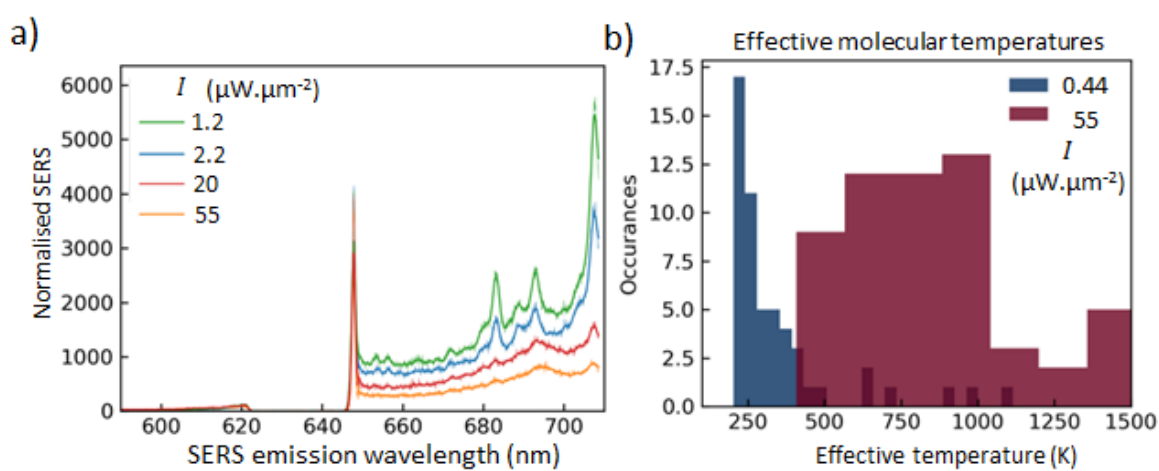

**Figure S25. Apparent molecular temperature evaluation**. (**a**) Power-normalized SERS spectra for increasing average intensities from 1.2 µW.µm$^{-2}$ (green) to 55 µW.µm$^{-2}$ (yellow), laser at 633 nm. (**b**) Histogram of fitted molecular temperatures for low intensity (0.4 µW.µm$^{-2}$) and high intensity (55 µW.µm$^{-2}$) demonstrating the clear increase in apparent molecular temperature for the 1586 cm$^{-1}$ SERS line due to non-equilibrium vibrational pumping.

## S14. SERS Saturation of Other Molecules

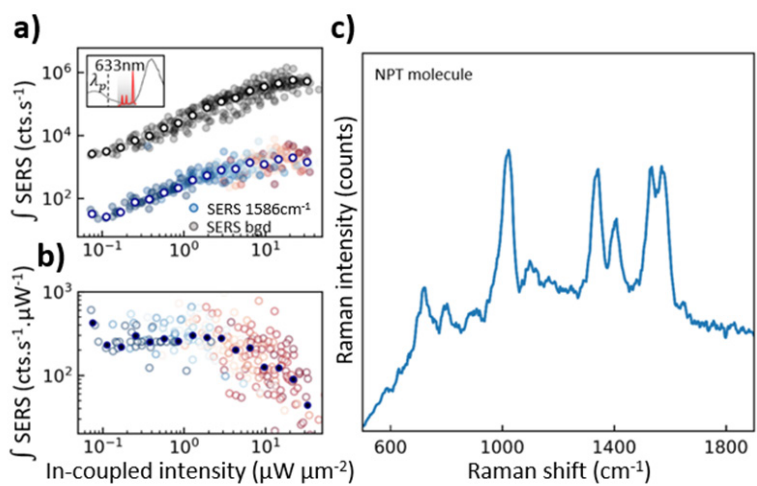

**Figure S26. Saturation of pulsed Stokes Raman scattering from >100 NPoMs containing naphthalene-thiol (NPT).** (**a**) Integrated SERS emission from $\omega_{v_1}$ = 1586 cm$^{-1}$ mode (blue-red) and integrated background (grey,x3), normalised to the integration time. Excitation with pulsed laser tuned to 633 nm. Open points show averaged power dependence. Inset shows relative position of pump wavelength with Stokes lines and plasmon resonances. (**b**) Integrated SERS normalized to excitation power and integration time. (**c**) Example SERS spectrum from single NPoM.

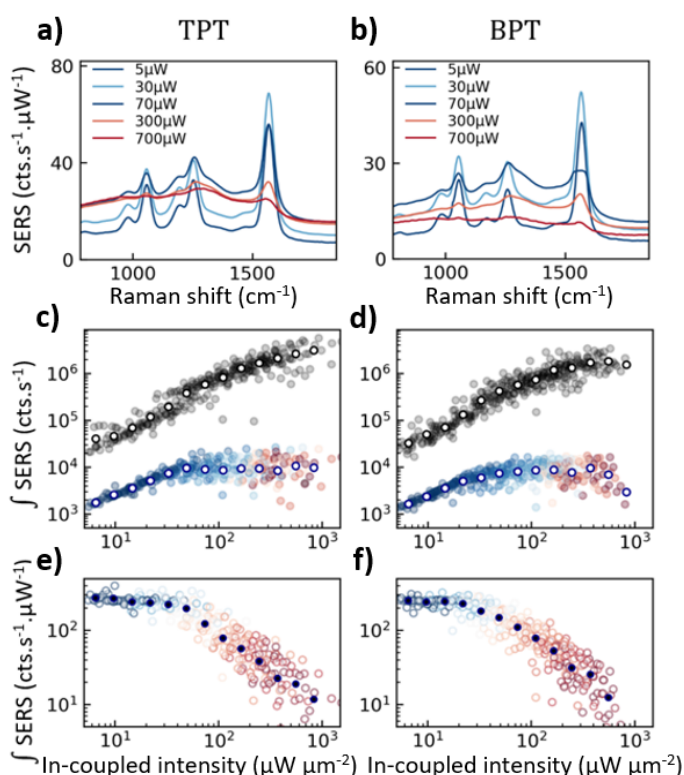

**Figure S27. Saturation of pulsed Stokes Raman scattering from >100 NPoMs containing TPT or BPT.** (**a,b**) Averaged SERS spectra for increasing in-coupled average powers for triphenylthiol (TPT) and biphenylthiol (BPT), recorded in same sample run for direct comparison. (**c,d**) Integrated SERS emission from $\omega_{v_1}$ = 1586 cm$^{-1}$ mode (colors show laser power) and integrated background (grey,x3), excited by pulsed laser at 633 nm. Open points are averages. (**e,f**) Integrated SERS normalized to excitation power and integration time.

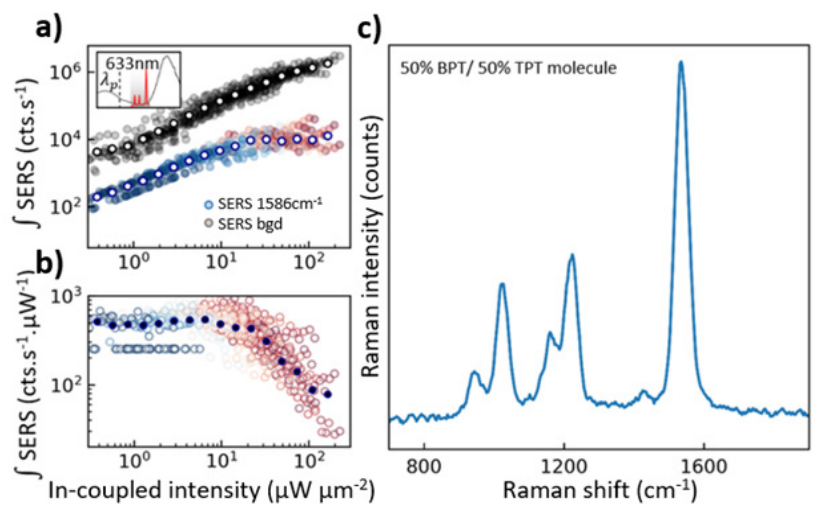

**Figure S28. Saturation of pulsed Stokes Raman scattering from many NPoMs containing 50% BPT/50% TPT mixed SAM.** (**a**) Integrated SERS emission from $\nu_1$ = 1586 cm$^{-1}$ mode (blue-red) and integrated background (grey,x3), normalised to the integration time. Excitation with pulsed laser tuned to 633 nm. Open points show averaged dependence. Inset shows relative position of pump wavelength with Stokes lines and plasmon resonances. (**b**) Integrated SERS normalized to excitation average power and integration time. (**c**) Example SERS spectrum from single NPoM.

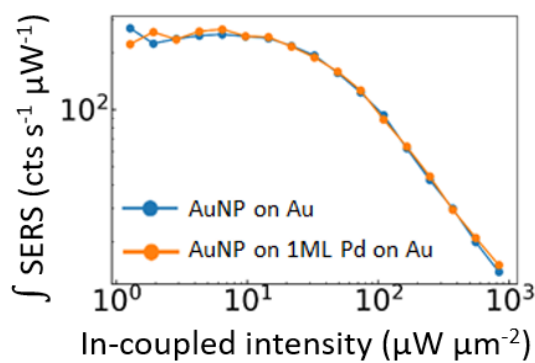

**Figure S29. SERS power dependence normalized to excitation power with different plasmonic confinement.** Data shown is average of power series on many NPoMs averaged with in-coupling correction. No significant difference is observed in the SERS saturation when the lower mirror is pure Au, or Au with a single monolayer of Pd electrochemically deposited on top (spacer: BPT). This suggests that charge transfer between molecule and metal is not playing a key role.

## S15. Additional Picocavity Data

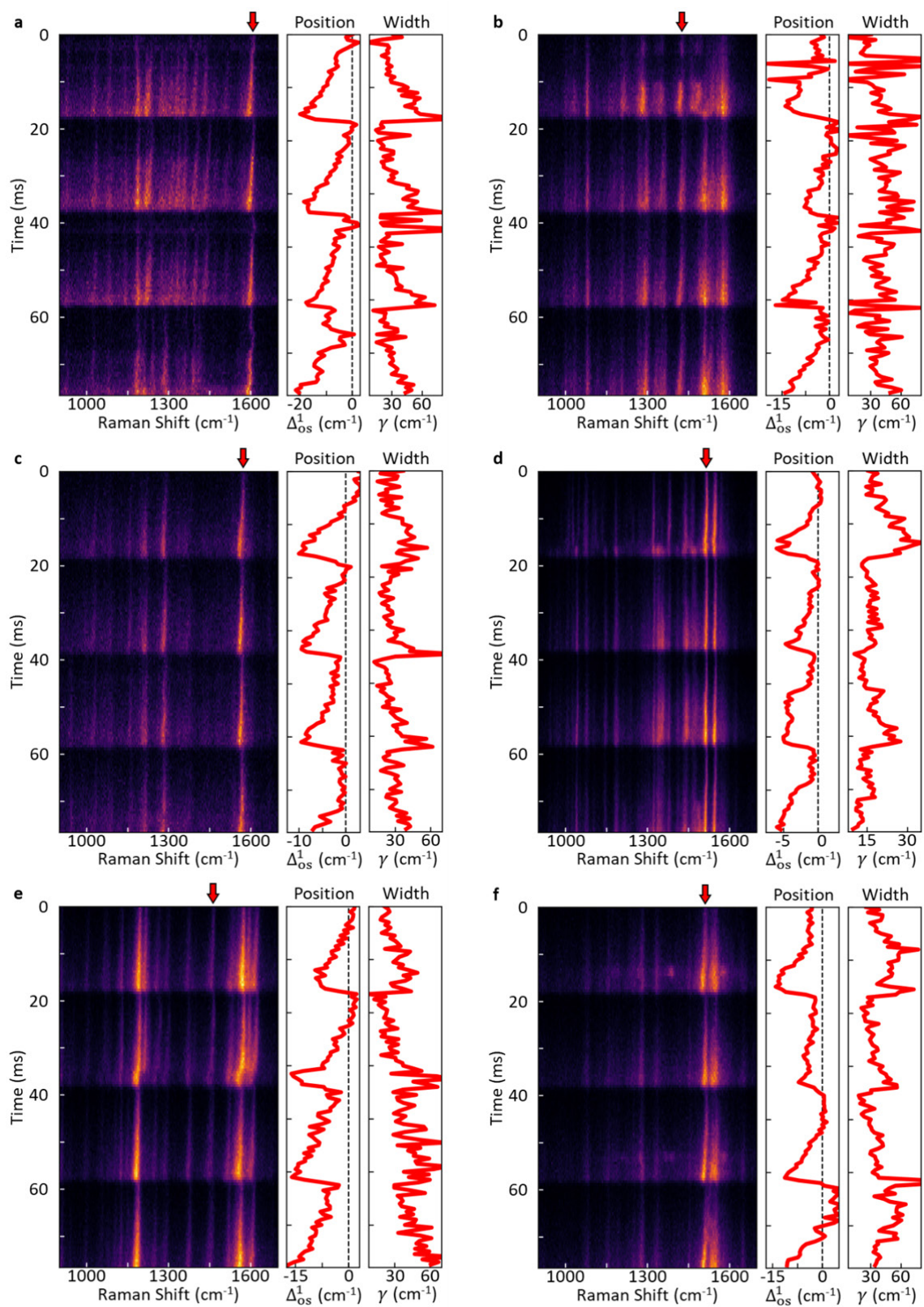

**Figure S30. Additional data from picocavities supporting observation of optomechanical spring shift.** (**a-f**) Fast spectral time scans (left panels) while sweeping laser power on six NPoMs with picocavities, with position and width of SERS line (red arrows) extracted via Lorentzian fits (two right panels). Same experimental conditions as in Figure 6 of the main text. This data underlines the reproducibility of the optical spring effect for different nanostructures and picocavity events.

# References

1. Zhang, Y. *et al.* Addressing Molecular Optomechanical Effects in Nanocavity-Enhanced Raman Scattering beyond the Single Plasmonic Mode. *Nanoscale* **13**, 1938–1954 (2021).
2. Dezfouli, M. K. & Hughes, S. Quantum optics model of Surface Enhanced Raman Spectroscopy for arbitrarily shaped plasmonic resonators. *ACS Photonics* **4**, 1245–1256 (2017).
3. Dung, H. T., Knöll, L. & Welsch, D.-G. Three-dimensional quantization of the electromagnetic fieldin dispersive and absorbing inhomogeneous dielectrics. *Phys. Rev. A* **57**, 3931–3942 (1998).
4. Suttorp, L. G. & van Wonderen, A. J. Fano diagonalization of a polariton modelfor an inhomogeneous absorptive dielectric. *Europhys. Lett.* **67**, 766–772 (2004).
5. Balevičius, V. *et al.* The full dynamics of energy relaxation in large organic molecules: From photo-excitation to solvent heating. *Chem. Sci.* **10**, 4792–4804 (2019).
6. Le Ru, E. C. & Etchegoin, P. G. Vibrational pumping and heating under SERS conditions: Fact or myth? *Faraday Discuss.* **132**, 63–75 (2006).
7. Wright, D. *et al.* Mechanistic study of an immobilized molecular electrocatalyst by in situ gap-plasmon-assisted spectro-electrochemistry. *Nat. Catal.* **4**, 157–163 (2021).
8. Lombardi, A. *et al.* Pulsed Molecular Optomechanics in Plasmonic Nanocavities: From Nonlinear Vibrational Instabilities to Bond-Breaking. *Phys. Rev. X* **8**, 011016 (2018).
9. Kos, D. *et al.* Optical probes of molecules as nano-mechanical switches. *Nat. Commun.* **11**, 5905 (2020).
10. Teperik, T. V., Nordlander, P., Aizpurua, J. & Borisov, A. G. Robust subnanometric plasmon ruler by rescaling of the nonlocal optical response. *Phys. Rev. Lett.* **110**, 263901 (2013).
11. Ciracì, C., Jurga, R., Khalid, M. & Della Sala, F. Plasmonic quantum effects on single-emitter strong coupling. *Nanophotonics* **8**, 1821–1833 (2019).
12. Esteban, R. *et al.* A classical treatment of optical tunneling in plasmonic gaps: Extending the quantum corrected model to practical situations. *Faraday Discuss.* **178**, 151–183 (2015).
13. Aspelmeyer, M., Kippenberg, T. J. & Marquardt, F. Cavity optomechanics. *Rev. Mod. Phys.* **86**, 1391–1452 (2014).
14. Beams, R., Cançado, L. G., Oh, S.-H., Jorio, A. & Novotny, L. Spatial coherence in near-field Raman scattering. *Phys. Rev. Lett.* **113**, 186101 (2014).
15. Maximiano, R. V., Beams, R., Novotny, L., Jorio, A. & Cançado, L. G. Mechanism of near-field Raman enhancement in two-dimensional systems. *Phys. Rev. B* **85**, 235434 (2012).
16. Meystre, P. & Sargent, M. *Elements of quantum optics*. (Springer Science & Business Media, 2007).
17. Frisch, M. J. *et al.* Gaussian 16, Revision B.01. (2016).
18. García de Abajo, F. J. & Howie, A. Retarded field calculation of electron energy loss in inhomogeneous dielectrics. *Phys. Rev. B* **65**, 115418 (2002).
19. García de Abajo, F. J. Optical excitations in electron microscopy. *Rev. Mod. Phys.* **82**, 209–275 (2010).
20. Hohenester, U. & Trügler, A. MNPBEM – A Matlab toolbox for the simulation of plasmonic nanoparticles. *Comput. Phys. Commun.* **183**, 370–381 (2012).
21. Waxenegger, J., Trügler, A. & Hohenester, U. Plasmonics simulations with the MNPBEM toolbox: Consideration of substrates and layer structures. *Comput. Phys. Commun.* **193**, 138–150 (2015).
22. Johnson, P. B. & Christy, R. W. Optical Constants of the Noble Metals. *Phys. Rev. B* **6**, 4370–4379 (1972).
23. Delga, A., Feist, J., Bravo-Abad, J. & Garcia-Vidal, F. J. Quantum Emitters Near a Metal Nanoparticle: Strong Coupling and Quenching. *Phys. Rev. Lett.* **112**, 253601 (2014).
24. Benz, F. *et al.* SERS of Individual Nanoparticles on a Mirror: Size Does Matter, but so Does Shape. *J. Phys. Chem. Lett.* **7**, 2264–2269 (2016).
25. Paulus, M., Gay-Balmaz, P. & Martin, O. J. F. Accurate and efficient computation of the Green's tensor for stratified media. *Phys. Rev. E* **62**, 5797 (2000).
26. Ford, G. W. & Weber, W. H. Electromagnetic Interactions of Molecules with Metal Surfaces. *Phys. Rep.* **113**, 195–287 (1984).

27. Li, R.-Q., Hernángomez-Pérez, D., García-Vidal, F. J. & Fernández-Domínguez, A. I. Transformation Optics Approach to Plasmon-Exciton Strong Coupling in Nanocavities. *Phys. Rev. Lett.* **117**, 107401 (2016).
28. Baumberg, J. J., Aizpurua, J., Mikkelsen, M. H. & Smith, D. R. Extreme nanophotonics from ultrathin metallic gaps. *Nat. Mater.* **18**, 668–678 (2019).
29. Nikitin, A. Y., García-Vidal, F. J. & Martin-Moreno, L. Analytical Expressions for the Electromagnetic Dyadic Green's Function in Grapheneand Thin Layers. *IEEE J. Sel. Top. Quantum Electron.* **19**, 4600611 (2013).
30. Matei, D. G., Muzik, H., Gölzhäuser, A. & Turchanin, A. Structural Investigation of 1,1'-Biphenyl-4-thiol Self-Assembled Monolayers on Au(111) by Scanning Tunneling Microscopy and Low-Energy Electron Diffraction. *Langmuir* **28**, 13905–13911 (2012).
31. Zhang, Y., Aizpurua, J. & Esteban, R. Optomechanical Collective Effects in Surface-Enhanced Raman Scattering from Many Molecules. *ACS Photonics* **7**, 1676–1688 (2020).
32. Schmidt, M. K., Esteban, R., Benz, F., Baumberg, J. J. & Aizpurua, J. Linking classical and molecular optomechanics descriptions of SERS. *Faraday Discuss.* **205**, 31–65 (2017).
33. Ventalon, C. *et al.* Coherent vibrational climbing in carboxyhemoglobin. *Proc. Natl. Acad. Sci.* **101**, 13216–13220 (2004).
34. Witte, T., Yeston, J. , Motzkus, M., Heilweil, E. . & Kompa, K.-L. Femtosecond infrared coherent excitation of liquid phase vibrational population distributions (v>5). *Chem. Phys. Lett.* **392**, 156–161 (2004).
35. Morichika, I., Murata, K., Sakurai, A., Ishii, K. & Ashihara, S. Molecular ground-state dissociation in the condensed phase employing plasmonic field enhancement of chirped mid-infrared pulses. *Nat. Commun.* **10**, 3893 (2019).
36. Crampton, K. T., Fast, A., Potma, E. O. & Apkarian, V. A. Junction Plasmon Driven Population Inversion of Molecular Vibrations: A Picosecond Surface-Enhanced Raman Spectroscopy Study. *Nano Lett.* **18**, 5791–5796 (2018).
37. Park, K. D. *et al.* Variable-Temperature Tip-Enhanced Raman Spectroscopy of Single-Molecule Fluctuations and Dynamics. *Nano Lett.* **16**, 479–487 (2016).
38. Artur, C., Le Ru, E. C. & Etchegoin, P. G. Temperature dependence of the homogeneous broadening of resonant Raman peaks measured by single-molecule surface-enhanced Raman spectroscopy. *J. Phys. Chem. Lett.* **2**, 3002–3005 (2011).
39. Harris, C. B., Shelby, R. M. & Cornelius, P. A. Effects of energy exchange on vibrational dephasing times in Raman scattering. *Phys. Rev. Lett.* **38**, 1415–1419 (1977).
40. Zhang, K. & Chen, X.-J. Identification of the incommensurate structure transition in biphenyl by Raman scattering. *Spectrochim. Acta Part A Mol. Biomol. Spectrosc.* **206**, 202–206 (2019).
41. Mei, H.-Y. *et al.* Study of melting transition on biphenyl by Raman scattering. *AIP Adv.* **9**, 095049 (2019).
42. Kneipp, K. *et al.* Population Pumping of Excited Vibrational States by Spontaneous Surface-Enhanced Raman Scattering. *Phys. Rev. Lett.* **76**, 2444–2447 (1996).
43. Maher, R. C., Etchegoint, P. G., Le Ru, E. C. & Cohen, L. F. A Conclusive Demonstration of Vibrational Pumping under Surface Enhanced Raman Scattering Conditions. *J. Phys. Chem. B* **110**, 11757–11760 (2006).